\documentclass[12pt,english]{article}
\usepackage[T1]{fontenc}
\usepackage[latin9]{inputenc}
\usepackage{float}
\usepackage{rotfloat}
\usepackage{amsmath}
\usepackage{amssymb}
\usepackage{mathrsfs}
\usepackage{graphicx} 
\usepackage{esint}
\usepackage{wasysym}

\usepackage{times}

\newtheorem{theorem}{Theorem}

\newtheorem{lemma}{Lemma}

\newtheorem{proposition}[theorem]{Proposition}

\newtheorem{assumption}{Assumption}

\newcommand{\ackname}{Acknowledgements:}

\usepackage[round,authoryear]{natbib}
\titlepage 

\begin{document}

\title{Latent Factor Analysis in Short Panels}

\author{Alain-Philippe Fortin$^{1}$, Patrick Gagliardini$^{2,3}$, Olivier Scaillet$^{4,3}$\thanks{$^1$Universit\'e de Montr\'eal, $^2$Universit\`a della Svizzera Italiana,  $^3$Swiss Finance Institute, $^4$University of Geneva. {\tiny Acknowledgements: We would like to thank the editor, the associate editor and the two referees for constructive criticism and numerous suggestions which have led to substantial improvements over the previous versions. We are grateful to A.\ Onatski for his very insightful discussion (Onatski (2023)) of our paper Fortin et al.\ (2023) at the 14th Annual SoFiE Conference in Cambridge, which prompted us to exploit factor analysis for estimation in short panels. For their comments, we thank D.\ Amengual, L.\ Barras, S.\ Bonhomme, M.\  Caner, F.\ Carlini,  I.\ Chaieb, F.\ Ghezzi, A.\ Horenstein, G.\ Imbens, S.\ Kim, F.\ Kleibergen, H.\ Langlois, T.\ Magnac, S.\ Ng, E.\ Ossola, H.\ Pesaran, Y.\ Potiron, A.\ Santos, R.\ Sickles, F.\ Trojani, participants at (EC)\^{}2 2022, QFFE 2023, SoFiE 2023, NASM 2023, IPDC 2024, FinEML, 22$^{\text{\`eme}}$ Journ\'ee d'\'Econom\'etrie, HKUST workshop, SFI research days, GIGS conference, EDHEC-ESSEC workshop, FGV conference, Leuven workshop, 41st AFFI meeting, FBA conference, QFDAM, and seminars at UNIMIB, UNIBE, UNIGE, Warwick, Bristol, QMUL, CUHK, Luxembourg, Guanghua B.\ School, UPF, Manchester, and Lancaster. The first and third author also acknowledge financial support by the Swiss National Science Foundation (grants UN11140 and 100018\_215573).} }}

\maketitle
\begin{center}
\textbf{Abstract}
\end{center}
\vspace*{-0.5cm} {We develop a pseudo maximum likelihood method for latent factor analysis in short panels without imposing sphericity nor Gaussianity. We derive an asymptotically uniformly most powerful invariant test for the number of factors. On a large panel of monthly U.S. stock returns, we separate month after month systematic and idiosyncratic  risks in short subperiods of bear vs. bull market. We observe an uptrend in the paths of total and idiosyncratic volatilities. The systematic risk explains a large part of the cross-sectional total variance in bear markets but is not driven by a single factor and not spanned by observed factors. 
} \\

\noindent { \textbf{Keywords:} { Latent factor analysis, uniformly most powerful invariant test, panel data, large $n$ and fixed $T$ asymptotics, equity returns.}
 }

\newpage

\section{Introduction}

Latent variable models have been used for decades in econometrics (Aigner et al.\ (1984)). 
In this paper, we study large cross-sectional latent factor models for panels of equity returns with small time dimension. Latent factors capture unobserved sources of dependence across the sample units and are largely adopted in multiple applications in finance and economics as well as in social sciences. In asset pricing, they underly the celebrated Arbitrage Pricing Theory (APT) of Ross (1976); see also Chamberlain and Rothschild (1983), Giglio and Xiu (2021), and Zaffaroni (2025).  Omitted latent factors are also called interactive fixed effects (Pesaran (2006), Bai (2009), Moon and Weidner (2015), Gobillon and Magnac (2016), Freyberger (2018)) or time-varying individual effects (Ahn et al.\ (2001))  in the panel literature. We find them in asset embeddings, i.e., numerical representations of asset or firm characteristics in a vector space (Gabaix et al.\ (2023)).

Two common methods, sometimes put under the same terminology of ``factor analysis", for estimation of latent factor spaces are principal component analysis (PCA) and factor analysis (FA), see Anderson (2003) Chapters 11 and 14, and Magnus and Neudecker (2007) Chapter 17, for a clear distinction between the two approaches. PCA relies on minimizing a least squares  objective function (leading to a convenient characterization in terms of an eigenvalue-eigenvector decomposition of a variance-covariance matrix) while FA relies on maximizing a Gaussian (pseudo) likelihood function (related to minimizing a weighted least squares objective function as in a weighted PCA) to estimate the latent factors. 
Notably, FA boils down to PCA under the sphericity condition, i.e., when the $T \times T$ covariance matrix of error terms is a multiple of the identity matrix. Otherwise, PCA and FA estimates generally differ - even asymptotically, when one panel dimension (in our case, $T$) is fixed. In fact, error sphericity is both necessary and sufficient for consistency of PCA latent factor estimates with small $T$ (Theorem 4 of Bai (2003)) and we cannot get rid of this restrictive assumption if we choose to apply PCA to short panels. In recent work, Fortin, Gagliardini and Scaillet (FGS, 2023)  show how we can use PCA  to conduct inference on the number of latent factors  without making Gaussian assumptions. 
FGS provides a discussion of the (in)consistency of the PCA estimator with fixed $T$ from the vantage point of the well-known incidental parameter problem of the panel data literature (Neyman and Scott (1948); see Lancaster (2000) for a review). In PCA, sphericity allows to identify the number $k$ of factors from the first $k$ eigenvalue spacings being larger than zero, and being zero for the subsequent ones.  If sphericity fails, all eigenvalue spacings are different from zero, and identification fails. On the contrary, the FA strategy does not exploit eigenvalue spacings and does not require sphericity. However, FA inference  with small $T$ up to now mostly relies on (often restrictive) assumptions such as Gaussian variables (with a notable exception by Anderson and Amemiya (1988)) and error homoskedasticity across sample units. Those are untenable assumptions in our application to stock returns. The strong assumption of sphericity might also fail to hold in some samples, e.g., due to common ARCH effects in idiosyncratic errors (see Section \ref{sectest}). If it happens, our Monte Carlo experiments with non-Gaussian errors show that the eigenvalue spacing test of FGS exhibits size distortions of over 80 percentage points for a nominal size of 5\%. That  massive over-rejection translates into an average estimated number of factors often above 10 instead of 2 (true value) in our simulation design  when $T = 24$ (a sample size close to $T=20$ in our empirics). We get similar Monte Carlo results for a constrained Likelihood Ratio test when sphericity does not hold (see Section \ref{spheri_imp}). Our Monte Carlo experiments also reveal that the chi-square test of the classical FA theory obtained under  cross-sectionally homoskedastic Gaussian errors (Anderson (1963)) and implemented in most numerical packages suffers from massive over-rejection by around 80 percentage points for a nominal size of 5\% when $T=24$.

A central and practical issue in applied work with latent factors is to determine their number. For models with unobservable   factors only, Connor and Korajczyk (1993) are the first to develop a test for the number of factors for large balanced panels of individual stock returns under covariance stationarity and homoskedasticity. Unobservable factors are estimated by the method of asymptotic principal components developed by Connor and Korajczyk (1986) (see also
Stock and Watson (2002)). In heteroskedastic settings, a first strand of that literature focuses on consistent estimation (but not testing) procedures for the
number of factors. Bai and Ng (2002) and Bai (2009) introduce a penalized least-squares strategy to estimate the number of factors. 
Ando and Bai (2015) extend that approach when explanatory variables are present in the linear specification. 
Onatski (2010) looks at the behavior of differences in adjacent eigenvalues to determine the number of factors when  $n$ and  $T$ are both large and comparable, while Ahn and Horenstein (2013) opt for a similar strategy based on eigenvalue ratios. 
Caner and Han (2014) propose an estimator with a group bridge penalization. 
Gagliardini, Ossola and Scaillet (2019) build a simple diagnostic criterion for approximate factor structure in large panels. Given observable
factors, the criterion checks whether the errors are weakly cross-sectionally correlated, or share one or more unobservable common factors (interactive effects), and selects their number; see Gagliardini, Ossola and Scaillet (2020) for a survey of estimation of large dimensional conditional factor models in finance. 
A second strand of that literature develops inference procedures for hypotheses on the number of latent factors. Onatski (2009) deploys a characterization of the largest eigenvalues of a Wishart-distributed covariance matrix with large dimensions in terms of the Tracy-Widom Law. To get a Wishart distribution, Onatski (2009) assumes either Gaussian errors, or $T$ much larger than $n$. Under non-Gaussian errors or when $T$ is much lower than $n$, we also observe notable size distortions for such a test in our Monte Carlo simulations. The test described in Section 4 of Onatski (2009) yields overrejection often largely above 10 to 15 percentage points for a nominal size of 5\% when $T=12, 24$ for the data generating process used in Section \ref{Montecarlo}. Kapetanios (2010) uses subsampling to estimate the limit distribution of the adjacent eigenvalues. Any of the above methods fail with $T$ small and non-Gaussian, non-spherical errors. In a fixed $T$ setting, Ahn et al.\ (2013) develop a test for the number of factors for a panel data model with random interactive effects and i.i.d.\ errors. The test statistic corresponds to the estimated value of the optimal Generalized Method of Moments (GMM) criterion (the usual $J$-test statistic of Hansen (1982)). Hayakawa et al.\ (2023) develop a likelihood ratio test in a transformed Gaussian quasi maximum likelihood approach  for the number of factors for a dynamic panel data model with random interactive effects and i.i.d.\ spherical errors.

This paper puts forward methodological and empirical contributions that complement the above literature. (i) On the methodological side, we extend the inferential tools of FA to non-Gaussian and non-i.i.d.\ settings. First, we characterize the asymptotic distribution of FA estimators obtained under a Pseudo Maximum Likelihood (PML) approach where the time-series dimension is held fixed while the cross-sectional dimension diverges. Hence, the asymptotic analysis targets short panels, and allows for cross-sectionally heteroskedastic and weakly dependent errors. Cochrane (2005, p.\ 226) argues in favour of the development of appropriate large-$n$ small-$T$ tools for evaluating  asset pricing models, a problem only partially addressed in finance. 
In a short panel setting, Zaffaroni (2025) considers  inference for latent factors in conditional linear asset pricing models under sphericity based on PCA, including estimation of the number of factors. Raponi, Robotti and Zaffaroni (2020) develop tests of beta-pricing models and a two-pass methodology to estimate the ex-post risk premia (Shanken (1992)) associated to observable factors (see Kleibergen and Zhan (2025) for robust-identification inference based a continuous updating GMM). Kim and Skoulakis (2018) deals with the error-in-variable problem of the two-pass methodology with small $T$ by regression-calibration under sphericity and a block-dependence structure. The small $T$ setting mitigates concerns for panel unbalancedness (outside the straightforward missing-at-random mechanism) and corresponds to a locally time-invariant factor structure accommodating globally time-dependent features of general forms. It is also appealing to macroeconomic data observed quarterly. For the sake of space, we put part of the theory, namely inference for FA estimates, in the Online Appendix (OA). We refer to Bai and Li (2016)  for inference when $n$ and $T$ are both large (see Bai and Li (2012) for the cross-sectional independent case). Second, we build on our new theoretical results for FA to develop testing procedures for the number of latent factors in a short panel which rely on neither sphericity nor Gaussianity nor cross-sectional independence, thereby extending tests based on eigenvalues, as in Onatski (2009) under a spectral analysis targeting a generalized dynamic factor structure, to small $T$, and as in FGS, to non-spherical errors, thanks to an FA device. Such testing procedures have the practical advantage of not requiring a choice of penalization hyperparameters and of providing p-values to gauge statistical significance. Here, we deliver asymptotic feasible distributional theory, namely implementable theory based on estimated quantities instead of population (unknown) ones, even if FA residuals are not consistent estimates of the true errors under fixed $T$. We further derive the Asymptotically Uniformly Most Powerful  Invariant  (AUMPI) property of the FA likelihood ratio (LR) test statistic in the non-Gaussian case under inequality restrictions on the DGP parameters, and cover inference with weak factors. The AUMPI  property is rare and sought-after in testing procedures (see Engle (1984) for a discussion and Romano, Shaikh and Wolf (2010) for a survey of optimality approaches in testing problems), and often holds only under restrictive assumptions such as Gaussianity. For the user of the test, it provides a theoretical guarantee of a statistical procedure with better power properties w.r.t.\ another choice of testing.
We show that the AUMPI property can hold even if the asymptotic distribution of the LR statistic is driven by a weighted sum of independent chi-square variates in our context instead of a single chi-square variate. We achieve that by providing novel sufficient conditions so that the ratio of the density under local alternative hypotheses and the null hypothesis satisfies the Monotone Likelihood Ratio (MLR) property. Hence, the FA theory gathered in the below delivers a body of new results including their proofs, and differs completely from PCA theory.  (ii) On the empirical side, we apply our FA methodology to panels of monthly U.S.\ stock returns with large cross-sectional and small time-series dimensions, and investigate how the number of driving factors changes over time and particular periods. Furthermore, month after month, we provide a novel separation of the risk coming from the systematic part and the risk coming from the idiosyncratic  part of returns in short subperiods of bear vs.\ bull market based on the selected number of  factors. Such a decomposition with PCA estimates is invalid without sphericity because of inconsistency of factor estimates. We observe an uptrend in the estimated paths of total and idiosyncratic volatilities (see also Campbell et al.\ (2023)) while the systematic risk explains a large part of the cross-sectional total variance  in bear markets but is not driven by a single latent factor. 
We also investigate whether standard observed factors span the estimated latent factors using rank tests that suit our fixed $T$ setting. Observed factors, even beyond the traditional ones, struggle spanning latent factors with a discrepancy between the dimensions of the two factor spaces decreasing over time.

The outline of the paper is as follows. In Section \ref{sectest}, we consider a linear latent factor model  and introduce test statistics on the number of latent factors based on FA. Section \ref{regularity} lists the regularity assumptions underlying the theoretical results. Section \ref{asymexp} presents a feasible asymptotic distributional theory for inference in short panels under a block-dependence structure to allow for weak dependence in the cross-section. Sections \ref{discu} and \ref{local:sec} develop further theory supporting the advantages of our method. Section \ref{discu} discusses three special cases, i.e., Gaussian errors (yielding the classical chi-square test of FA theory in the cross-sectionally homoskedastic case), non-Gaussian settings where the same asymptotic distribution as under Gaussian errors still holds for the test statistics, and spherical errors (leading to PCA). It considers the impact for test validity of wrongly assuming Gaussian errors or sphericity. Section \ref{local:sec} is dedicated to local asymptotic power and AUMPI properties. Section \ref{Montecarlo} gives a Monte Carlo assessment of size and power and selection procedure for the number of
factors for the LR test. We provide our empirical application in Section \ref{appli}. We put our concluding remarks in Section \ref{section:Concluding remarks}. In the Online Appendix (OA), we gather  proofs of the main theoretical results in Appendix A as well as proofs of Lemmas supporting them. We place  additional theory and numerical checks in Appendices B and C. Appendix D collects  the maximum value of $k$ as a function of $T$.
We gather all explicit formulas not listed in the core text but useful for coding  in an online ``Supplementary Materials for Coding'' (SMC) 
attached to the replication files. We also put there  other numerical checks and additional Monte Carlo results  to assess the impact of non-sphericity for the eigenvalue spacing test of FGS and the constrained LR test, as well as of using the classical chi-square test on the size and selection procedure for the number of factors. 

\section{Latent factor model and likelihood ratio test} \label{sectest}

We consider the linear Factor Analysis (FA) model (e.g.\ Anderson (2003)):
\begin{equation} \label{model}
y_i = \mu + F \beta_i + \varepsilon_i, \qquad i=1,...,n,
\end{equation}
where $y_i=(y_{i,1},...,y_{i,T})'$ and $\varepsilon_i = (\varepsilon_{i,1},...,\varepsilon_{i,T})'$ are $T$-dimensional vectors of observed data  and unobserved error terms for individual $i$. The $k$-dimensional vectors $\beta_i = (\beta_{i,1}, ..., \beta_{i,k})'$ are latent individual effects (non-random incidental parameters), while $\mu$ and $F$ are a $T \times 1$ vector and a $T \times k$ matrix of unknown  parameters. The number of latent factors $k$ is an unknown integer smaller than $T$.  In matrix notation, model (\ref{model}) reads $Y = \mu 1_n' + F \beta' + \varepsilon$, where $Y$ and $\varepsilon$ are $T \times n$ matrices, $\beta$ is the $n \times k$ matrix with rows $\beta_i'$, and $1_n$ is a $n$-dimensional vector of ones.
\begin{assumption} \label{Ass1} The $T \times T$ matrix $V_{\varepsilon} = \underset{n \rightarrow \infty}{\lim} E[ \frac{1}{n}\varepsilon \varepsilon']$ is diagonal.
\end{assumption}
Matrix $V_{\varepsilon}$ is the limit cross-sectional average of the unconditional covariance matrix of the - possibly heterogeneous - errors. The diagonality condition in Assumption \ref{Ass1} is standard in FA (in the more restrictive formulation involving i.i.d.\ data).  Assumption \ref{Ass1} allows for serial dependence in idiosyncratic errors in the form of martingale difference sequences, like individual GARCH and Stochastic Volatility (SV) processes, as well as weak cross-sectional dependence (see Assumption \ref{Ass2} below). It also accommodates common time-varying components in idiosyncratic volatilities by allowing different entries along the diagonal of $V_{\varepsilon}$; see  Renault, Van Der Heijden and Werker (2023) for arbitrage pricing in  such settings. When there is a common random component in idiosyncratic volatilities, i.e., an idiosyncratic volatility factor, we have $V_{\varepsilon} = \underset{n \rightarrow \infty}{plim} \frac{1}{n}\varepsilon \varepsilon'$ by a suitable version of the Law of Large Number (LLN) conditional on the sigma-field generated by this common component. With fixed $T$, we treat the sample realizations of the common component in idiosyncratic volatilities as unknown time fixed effects (the diagonal elements of matrix $V_{\varepsilon}$), which yields time heterogeneous distributions for the errors. It is how the unconditional expectation in Assumption \ref{Ass1} has to be understood. The assumption of zero off-diagonal elements in matrix $V_\varepsilon$ rules out pervasive serial autocorrelation in the errors, i.e., $\lim_{n\to\infty} \frac{1}{n}\sum_{i=1}^n E[\varepsilon_{i,t}\varepsilon_{i,s}]=0$ for $t\neq s$.  Generalizations to non-diagonal $V_{\varepsilon}$ are possible (J{\"o}reskog (1970)), but stronger restrictions are required to maintain nonnegative degrees of freedom (see the discussion following (FA1) and (FA2) below). Besides, numerical optimisation becomes less tractable. Such an extension might be needed if we want to deal with serially correlated errors possibly present in macroeconomic panels and we can modify the theory developed in this paper to suit such a setting.

In our empirics  with a large cross-sectional panel of returns for $n$ assets over a short time span with $T$ periods, vectors $y_i$ and $\varepsilon_i$ stack the monthly returns and the idiosyncratic errors of stock $i$. Any row vector $f_t' := (f_{t,1},...,f_{t,k})$ of matrix $F$ yields the latent factor values in a given month $t$, and vector $\beta_i$ collects the factor loadings of stock $i$. In this finance application, we assume the No-Arbitrage (NA) principle to hold, so that the entries $\mu_t$ in the intercept vector  in Equation (\ref{model}) account for the (possibly time-varying) risk-free rate and (possibly non-zero) cross-sectional mean of stock betas. Specifically, under NA and assuming that the assets are embedded into a continuous economy, the intercept term in the asset return model $y_i = \mu_i + \tilde{F} \tilde \beta_i + \varepsilon_i$ is $\mu_i = r_f + 1_T \nu' \tilde \beta_i$, where $r_f$ is the $T$-dimensional vector whose entries collect the (possibly time-varying) risk-free rates, $\nu = (\nu_1, ..., \nu_k)'$ is a $k$-dimensional vector of parameters, and $1_T$ is a $T$-dimensional vector of ones (see e.g.\ the derivation in Gagliardini, Ossola and Scaillet (2016) for an Arbitrage Pricing Theory (Ross (1976)) under weak cross-sectional dependence (Chamberlain and Rothschild (1983))). We can absorb term $1_T \nu' \tilde \beta_i$ into the systematic part to get $y_i = r_f + F \tilde \beta_i + \varepsilon_i$ with $F = \tilde{F} + 1_T \nu'$. It holds irrespective of the latent factors being tradable or not. If the factors are tradable, we further have $\nu=0$ from the NA restriction. Akin to standard formulation of FA, we recenter the latent effects by subtracting their mean $\tilde \mu_{\tilde \beta}= \frac{1}{n} \sum_{i=1}^n \tilde\beta_i$, to get model (\ref{model}) with $\beta_i = \tilde \beta_i - \tilde \mu_{\tilde \beta}$ and $\mu = r_f + F\tilde \mu_{\tilde \beta}$. Instead, Chamberlain and Rothschild (1983) derive the APT with countably many assets and show that pricing errors $\zeta_i$ are square summable under NA. Then, we get $y_i = \mu + 1_T \zeta_i + F \beta_i + \varepsilon_i$ with $\sum_{i=1}^n \zeta_i^2 \leq C$, for all $n$ and a constant $C$. Our distributional results on FA estimators and LR test statistic are unaffected by the presence of $\zeta_i$, because we can show that the estimated $T \times T$ covariance matrix of excess returns  is unchanged up to terms of order $o_p(n^{-1/2})$. If the NA principle is violated, but the elements of the vector $\mu_i$ are constant, i.e., $\mu_i=\alpha_i1_T$ for some scalar $\alpha_i$, then the vector $1_T$ becomes an additional factor in the model. In any case, our starting point is the linear FA model in (\ref{model}): $y_{i,t} = \mu_t + f_t'\beta_i + \varepsilon_{i,t}$, that is the standard formulation in asset pricing. 
We cover the Capital Asset Pricing Model (CAPM) when  the single latent factor is the excess return of the market portfolio.

This paper focuses mainly on testing hypotheses on the number of latent factors $k$ when $T$ is fixed and $n\rightarrow\infty$. The fixed $T$ perspective makes FA especially well-suited for applications with short panels. Indeed, we work conditionally on the $T$ realizations of the latent factors $F$ and treat their values as $T \times k$ parameters to estimate. In comparison with the standard small $n$ and large $T$ framework in traditional asset pricing (e.g.\ Shanken (1992) with observable factors), here factors and loadings are interchanged in the sense that the $\beta_i$ and $F$ play the roles of the ``factors" and the ``factor loadings" in FA. 
We depart from classical FA since the $\beta_i$ are not considered as random effects with a Gaussian distribution but rather as fixed effects, namely non-random incidental parameters. Moreover, in Assumption \ref{Ass1}, we neither assume Gaussianity nor do we impose sphericity of the covariance matrix of the error terms. In addition, we accommodate weak cross-sectional dependence and general forms of heteroskedasticity (individual and common across assets) in idiosyncratic errors (see Section 3). Hence, the FA estimators $\hat{F}$ and $\hat{V}_{\varepsilon}$ defined below correspond to maximizers of a Gaussian pseudo likelihood. By-products of our analysis are the feasible asymptotic distributions of FA estimators of $F$ and $V_{\varepsilon}$ in more general settings than in the available literature (e.g.\  Anderson and Amemiya (1988)), which we present in Appendix B. Chamberlain (1992) studies semiparametrically efficient estimation in panel models with fixed effects and short $T$ using moment restrictions from instrumental variables. Our approach does not rely on availability of valid instruments. Another approach to deal with fixed $T$ is to increase the sampling frequency. Andersen et~al.~(2025) develop a test for determining the number of jump factors under fixed $T$, large $n$, and increasing sampling frequency asymptotics. In their framework, the idiosyncratic noise vanishes asymptotically and therefore does not need to be estimated.


Let $\hat{V}_y = \frac{1}{n} \tilde Y \tilde Y'$ 
be the sample $T \times T$ cross-sectional covariance matrix, where the $n$ columns of $\tilde Y$ are $y_i - \bar y$ and $\bar y = \frac{1}{n}\sum_{i=1}^n y_{i}$ is the vector of cross-sectional means. The covariance $\hat V_y$ is a sufficient statistic for FA estimation of the parameter $\theta = ( vec(F)',diag(V_{\varepsilon})')' $.
In the following, we use the same notation for  the matrix-to-vector diag operator and  the vector-to-matrix diag operator. Hence,  $diag(A)$ for a matrix $A$ denotes the vector in which we stack the diagonal elements of matrix $A$, and $diag(a)$ for a vector $a$ denotes a diagonal matrix with the elements of $a$ on the diagonal. Moreover, to normalize the latent factor matrix $F = [F_1:\cdots:F_k]$, we follow classical FA and set $\bar{\beta}=0$,  $\tilde{V}_{\beta}  = I_k$, and $F' \tilde{V}_{\varepsilon}^{-1}F = diag(\gamma_1,...,\gamma_k)$, where  $\bar \beta = \frac{1}{n} \sum_{i=1}^n \beta_i$ and $\tilde{V}_{\beta}= \frac{1}{n}  \beta' \beta$ are the sample mean and second moment of the betas  (we treat $\beta_i$ as fixed effects, and not random variables), and $\tilde{V}_{\varepsilon}=E[\frac{1}{n}\varepsilon\varepsilon']$ is the finite-$n$ analog of $V_{\varepsilon}$. This normalization of the factor values is sample dependent, i.e., $F = F_{(n)}$. We suppress index $n$ for the purpose of easing notation. The normalization can be made without loss of generality and serves to remove some of the fundamental indeterminacy that arises when working with latent factors (see Anderson, 2003, Chapter 14).
 Then, under our assumptions, we have $V_y := \underset{n\rightarrow \infty}{\text{plim }} \hat V_y = F F' + V_{\varepsilon}$, with $F'V_{\varepsilon}^{-1}F = diag(\gamma_1,...,\gamma_k)$. In particular, we have $V_y V_{\varepsilon}^{-1} F_j = (1+\gamma_j) F_j$, i.e., the columns of $F$ are the eigenvectors of matrix $V_y V_{\varepsilon}^{-1}$ associated with eigenvalues $1+\gamma_j$, $j=1,...,k$. The standardizations $\bar{\beta}=0$ and  $\tilde{V}_{\beta}  = I_k$ wash out the incidental parameter problem (Neyman and Scott (1948); see Lancaster (2000) for a review) since the individual loadings do not appear in $\mu_y :=\underset{n\rightarrow \infty}{\text{plim}} ~\bar y$ nor in $V_y$ and we do not need to estimate them. It explains why we are able to get consistent estimators $\hat F$ and $\hat{V}_\varepsilon$ for large $n$ and fixed $T$.  
 
As in Anderson (2003) Chapter 14 (Section 14.3.1), the FA  estimators $\hat{F}$, $\hat{V}_{\varepsilon}$ with $k$ latent factors maximize a Gaussian pseudo likelihood  written under Assumption \ref{Ass1} (Appendix B.1). Even if we write the Gaussian PML under the assumption that $\beta_i$ are i.i.d.\ $N(0,I_k)$ in Appendix B.1, we do not assume that to derive our asymptotic results. When both $n$ and $T$ are large and cross-sectional independence holds, Bai and Li (2012) show that the factor estimates given by a Gaussian PML approach and a Generalized Least Squares (GLS) approach (weighted PCA), albeit different, share the same asymptotic distribution. Our factor estimates driven by the first order conditions (FA1) and (FA2) do not share the same asymptotic distribution as the ones given by a GLS approach since the GLS estimates build on factor loadings estimates and thus suffer from a bias due to the incidental parameter problem under large $n$ and fixed $T$. 

  The FA estimators meet the first order conditions:
\medskip

\noindent
(FA1) \quad $diag( \hat{V}_y ) = diag ( \hat{F} \hat{F}' + \hat{V}_{\varepsilon})$, and \\
\noindent
(FA2) \quad  $\hat{F}$ is the $T \times k$ matrix of eigenvectors of $\hat{V}_y \hat{V}_{\varepsilon}^{-1}$ associated to the $k$ largest eigenvalues 

\quad \quad $1+\hat{\gamma}_j$, $j=1,..., k$, normalized such that $\hat{F}'\hat{V}_{\varepsilon}^{-1} \hat{F} = diag (\hat{\gamma}_1,...,\hat{\gamma}_k)$.

\medskip

\noindent 
The normalization in (FA2) applies for $\hat{\gamma}_j \geq 0$, which holds with probability approaching $1$. Otherwise, the first-order conditions of the FA estimators hold with $\hat{\gamma}_j$ replaced by its positive part. If error variance $V_{\varepsilon}$ where known, i.e., $\hat V_{\varepsilon} = V_{\varepsilon}$, condition (FA2) yields that $V_{\varepsilon}^{-1/2} \hat F$ is the matrix of the orthogonal eigenvectors of $V_{\varepsilon}^{-1/2} \hat V_y V_{\varepsilon}^{-1/2}$. It is akin to PCA applied on pre-whitened data $V_{\varepsilon}^{-1/2} Y$, up to normalization. Because $V_{\varepsilon}$ is unknown and needs to be estimated, condition (FA1) is an implicit equation that yields $\hat V_{\varepsilon}$ from the diagonal of $\hat V_y - \hat F \hat F'$. The numerical calculation of the FA estimates from (FA1) and (FA2) is described in Magnus and Neudecker (2007) Chapter 17. The number of degrees of freedom is $df =  \frac{1}{2}((T-k)^2 - T - k)$. It corresponds to the number of different elements in data matrix $\hat V_y$, i.e., $\frac{1}{2}T(T+1)$, plus the number of normalization constraints $\frac{1}{2} k(k-1)$ in equations $F'V_{\varepsilon}^{-1}F=diag$, minus the number of unknown parameters $(k+1)T$ (Anderson (2003)).  It is required that $df \geq 0$ for estimation, and we need $df >0$ to test the null hypothesis of $k$ latent factors (see Proposition \ref{thm:asy:tests} (a) below). In OA Table \ref{maxk}, we list the largest admissible number $k$ of latent factors as a function of $T$ such that $df \geq 0$. 

An alternative to maximum likelihood estimation is minimum distance estimation, also called Asymptotically Distribution-Free (ADF) estimation in covariance structure analysis (see Zheng and Bentler (2024) for a comparison of the two approaches). The two approaches are related, since we can show that the second-order expansion of the Gaussian pseudo likelihood criterion underlying the FA estimator yields (minus) a minimum distance criterion with weighting matrix $(V_y)^{-1}\otimes (V_y)^{-1}$. The asymptotic properties of minimum distance estimation under block-dependence are not available in the literature and require additional developments.

We conclude this section by introducing the classical FA likelihood ratio (LR) statistic to test the null hypothesis $H_0(k)$ of $k$ latent factors against the alternative hypothesis $H_1(k)$ of more than $k$ latent factors : 
\begin{eqnarray}\label{Definition:statistics}
	{LR}(k) := - n \sum_{j=k+1}^T \log ( 1 + \hat{\gamma}_j),
\end{eqnarray}
where the $\hat{\gamma}_j$ for $j=k+1,...,T$ are the $T-k$ smallest eigenvalues of $\hat{V}_y \hat{V}_{\varepsilon}^{-1} - I_T$. 
We have $LR(k)=\frac{n}{2} \sum_{j=k+1}^T \hat{\gamma}_j^2 +o_p(1)$, 
which results from a second-order expansion of the $\log$ function and the use of $\sum_{j=k+1}^T \hat{\gamma}_j=0$, from Lemma \ref{prop:1} in Appendix A and Anderson (2003), and $\sqrt{n} \hat \gamma_j = O_p(1)$, for $j=k+1,...,T$, from the proof of Proposition \ref{thm:asy:tests} below. Hence, the LR statistic checks if the $T-k$ smallest eigenvalues of $\hat{V}_y \hat{V}_{\varepsilon}^{-1}$ are close to $1$, which mimics the properties of the population quantities. Prewhitening $Y$ with $V_\varepsilon^{-1/2}$, gives limit variance $V_{\varepsilon}^{-1/2} V_y V_{\varepsilon}^{-1/2}  = V_{\varepsilon}^{-1/2} F F'V_{\varepsilon}^{-1/2}  + I_T$. Hence,  the $k$ largest values of  $V_y V_\varepsilon^{-1}$ are above $1$  while the subsequent ones are equal to 1. It is the scree plot (plot of eigenvalues in descending order introduced by Cattell (1966)) related to $V_y V_\varepsilon^{-1}$ that is informative on the number of factors and not the one related to $V_y$ as in standard PCA. When sphericity fails, the latter should not be used in practice  with fixed $T$ since the shape of the spectrum of $V_y V_\varepsilon^{-1}$ is not similar to the shape of the spectrum of $V_y $.  Since the check is on the $T-k$ smallest eigenvalues being close to 1, we cannot interpret the LR test as a test of the null hypothesis
that the number of factors is less than or equal to $k$ and is thus different from the proposal by Chen and Fang (2019) for inference on the rank of a matrix. In Section \ref{asymexp}, we establish feasibility of the asymptotic distribution of the LR statistic  with $n\rightarrow \infty$ and $T$ fixed under a block-dependence structure. 

\section{Regularity assumptions}
\label{regularity}

In this section, we list and comment the additional assumptions used to
derive the large sample properties of the estimators and test statistics. We often denote by $C>0$ a generic constant, and we use $\delta_j(A)$ to denote the $j$th largest eigenvalue of a symmetric matrix $A$. 
Set $\Theta$ is a compact subset of $\{ \theta = ( vec(F)',diag(V_{\varepsilon})')' \in \mathbb{R}^r ~ : ~ V_{\varepsilon}~\text{is diagonal and positive definite},~F'V_{\varepsilon}^{-1} F ~\text{is diagonal, with}$\break  $\text{diagonal elements ranked in decreasing order} \}$ with $r = (T+1)k$, and function $L_0(\theta) = \break  - \frac{1}{2} \log \vert \Sigma(\theta) \vert - \frac{1}{2} Tr\left( V_y \Sigma(\theta)^{-1} \right)$ is the population FA criterion, where $\Sigma(\theta) = F F' + V_{\varepsilon}$ and $V_y = \underset{n \rightarrow \infty}{\text{plim}}~ \hat V_ y$. Further, $\theta_0 = ( vec(F_0)',diag(V_{\varepsilon}^0)')'$ denotes the vector of true parameter values under $H_0(k)$ and is an interior point of set $\Theta$. For a $T\times T$ symmetric matrix $A = (a_{i,j})$, let us define the $\frac{1}{2}T (T+1)\times 1$  vector $vech(A)=\left(\frac{1}{\sqrt{2}} a_{11},..., \frac{1}{\sqrt{2}} a_{T,T},\{ a_{i,j} \}_{i<j} \right)'$, where the pairs of indices $(i,j)$ with $i<j$ are ranked as $(1,2),(1,3),..., (1,T), (2,3),..., (T-1,T)$. This definition of the half-vectorization operator for symmetric matrices differs from the usual one by the ordering of the elements, and the rescaling of the diagonal elements. It is more convenient for our purposes (see proof of Lemma \ref{lemma:transformation:vech}). For instance, it holds $\frac{1}{2} \Vert A \Vert^2 = vech(A)'vech(A)$. 
 Below, $\odot $ denotes the Hadamard product (i.e., element-wise matrix product), and $A^{\odot 2} := A \odot A$.

\vspace{-.5em}

\begin{assumption} \label{Ass2}
	(a) The errors are such that $\varepsilon = V_{\varepsilon}^{1/2} W \Sigma^{1/2}$, where $W=[w_1:\cdots:w_n]$ is a $T \times n$ random matrix of standardized errors terms $w_{i,t}$ that are independent across $i$ and uncorrelated across $t$, and $\Sigma = (\sigma_{i,j})$ is a positive-definite symmetric $ n \times n$ matrix, such that $\underset{n\rightarrow\infty}{\lim} \frac{1}{n}\sum_{i=1}^n \sigma_{ii}=1$. (b) Matrix $\Sigma$ is block diagonal with $J_n$ blocks of size $b_{m,n} = B_{m,n} n$, for $m=1,...,J_n$, where $J_n \rightarrow \infty$ as $n \rightarrow \infty$, and $I_m$ denotes the set of indices in block $m$. (c) There exist constants $\delta \in [0,1]$ and $C>0$ such that $\underset{i \in I_m}{\max} \sum_{j \in I_m} \vert \sigma_{i,j} \vert \leq C b_{m,n}^{\delta}$. (d) The block sizes $b_{m,n}$ and block number $J_n$ are such that $n^{2\delta} \sum_{m=1}^{J_n} B_{m,n}^{2(1+\delta)}= o(1)$.
\end{assumption}

\begin{assumption}  \label{ass:A:1} The non-zero eigenvalues of $V_y V_{\varepsilon}^{-1}-I_T$ are distinct, i.e., $\gamma_1 > ... > \gamma_k>0$.
\end{assumption}

\begin{assumption} \label{ass:A:2}
	The loadings are normalized such that $\bar \beta = \frac{1}{n} \sum_{i=1}^n \beta_i=0$ and  $\tilde V_{\beta} :=  \frac{1}{n} \sum_{i=1}^n \beta_i \beta_i ' \break =I_k$, for any $n$. Moreover, $\vert \beta_i\vert \leq C$, for all $i$.
\end{assumption}

\begin{assumption} \label{ass:A:3}
	We have $E[ w_{i,t}^8] \leq C$ and $\vert \sigma_{i,j} \vert \leq C$, for all $i,j,t$. 
\end{assumption}

\begin{assumption} \label{ass:A:4}
	Under the null hypothesis $H_0(k)$, we have: $\Sigma (\theta) = \Sigma(\theta_0)$, $\theta \in \Theta$ $\Rightarrow$ $\theta = \theta_0$, up to sign changes in the columns of $F$. 
\end{assumption}

\begin{assumption}  \label{ass:A:5}
	Matrix $M_{F_0,{V}^0_{\varepsilon}}^{\odot 2}$ is non-singular.
\end{assumption}

\begin{assumption} \label{ass:A:6}
	(a) The $\frac{T(T+1)}{2} \times \frac{T(T+1)}{2}$ symmetric matrix $D = \underset{n\rightarrow\infty}{\lim} D_n$ exists, where  $D_n = \frac{1}{n} \sum_{i=1}^n \sigma_{ii}^2 V[ vech( w_i w_i')]$. (b) We have $\delta_{T(T+1)/2}\left(V[ vech( w_i w_i')]\right) \geq \underline{c}$, for all $i \in \bar{S}$, where $\bar S \subset \{ 1,...,n \}$ with $\frac{1}{n} \sum_{i=1}^n 1_{i\in \bar S} \geq 1 - \frac{1}{2\bar C}$, for constants $\bar C , \bar c >0$, such that $\sigma_{ii} \leq \bar C$.
	(c) We have $\underset{n\rightarrow\infty}{\lim} \kappa_n= \kappa$ for a constant $\kappa \geq 0$, where $\kappa_n := \frac{1}{n}\sum_{m= 1}^{J_n}  \left(\sum_{i\neq j\in I_m} \sigma_{ij}^2 \right)$.
\end{assumption}

\begin{assumption} \label{ass:A:7}
	Under the alternative hypothesis $H_1(k)$, (a) function $L_0(\theta)$ has a unique maximizer $\theta^* = (vec(F^*)',diag(V_{\varepsilon}^*)')'$ over $\Theta$, and (b) we have $V_y \neq F^*(F^*)' + V^*_{\varepsilon}$.
\end{assumption}

\begin{assumption} \label{ass:A:8}
	Matrix $Q_{\beta}:= \underset{n\rightarrow\infty}{\lim} \frac{1}{n}  \beta' \Sigma \beta$ 
	is positive definite.
\end{assumption}


\noindent We use the block-dependence structure in Assumption \ref{Ass2}  to allow for weak cross-sectional dependence in errors and get implementable estimators of population (unknown) quantities, i.e., feasible asymptotic results under fixed $T$,  as stated in Proposition  \ref{thm:asy:tests} and discussed in Section \ref{asymexp} below. Under large $n$ and $T$, other estimation strategies for the asymptotic variance are available (see e.g.\ Bai and Ng (2006a), Maldonado and Ruiz (2021), Kim (2022)). The modeling of the errors in Assumption \ref{Ass2} (a) is similar to Onatski (2010) and Ahn and Horenstein (2013) but with more restrictive structure because of the diagonality of $V_{\varepsilon}$ and the blocks in $\Sigma$.
The block-dependence structure as in Assumption  \ref{Ass2} (b) is satisfied, for instance, when there are unobserved industry-specific factors independent among industries and over time, as in Ang, Liu, and Schwarz (2020). In empirical applications, blocks in $\Sigma$ can match industrial sectors (Fan, Furger,  and Xiu (2016), Gagliardini, Ossola, and Scaillet (2016)). As already remarked, the diagonal elements of $V_{\varepsilon}$ are the sample realizations of the common component driving the variance of the error terms at times $t=1,...,T$; see e.g.\ Barigozzi and Hallin (2016), Renault, Van Der Heijden and Werker (2023) for theory and empirical evidence pointing to variance factors.
A sphericity assumption cannot accommodate such a common time-varying component.  Assumption \ref{Ass2} (a) is coherent with Assumption \ref{Ass1}. Indeed, $ \tilde{V}_{\varepsilon}  =  V_{\varepsilon}^{1/2} \frac{1}{n} \sum_{i,j=1}^n \sigma_{i,j} E[ w_i w_j'] V_{\varepsilon}^{1/2} 
= \frac{1}{n}\sum_{i=1}^n \sigma_{ii} V_{\varepsilon}$ is diagonal. Hence, $\tilde{V}_{\varepsilon} $ is a scalar multiple of $V_{\varepsilon}$, and converges to $V_{\varepsilon}$ under the normalization $\underset{n\rightarrow\infty}{\lim}\frac{1}{n}\sum_{i=1}^n \sigma_{ii}=1$. That normalization is without loss of generality by rescaling of the parameters. Assumption \ref{Ass2} (c) builds on Bickel and Levina (2008), and $\delta < 1$ holds under sparsity, vanishing correlations or mixing dependence within blocks.  Assumption \ref{Ass2} (d) implies a minimum degree of granularity on the
blocks, i.e., largest blocks cannot be too large
and the number of blocks should grow fast enough. With blocks of equal size, Assumption \ref{Ass2} (d) holds for $J_n = n^{\bar \alpha}$ and $\bar\alpha  > \frac{2\delta}{2\delta+1}$. Having $\delta< 1$ helps relaxing this condition on block granularity, however it is not strictly necessary because we allow value $\delta=1$.

Assumption \ref{ass:A:1} removes the rotational indeterminacy (up to sign) that occurs when identifying the columns of matrix $F$ as eigenvectors of $V_yV^{-1}_\varepsilon$. Assumptions \ref{ass:A:2} and \ref{ass:A:3} require uniform bounds on factor loadings as well as on covariances and higher-order moments of the idiosyncratic errors. We can weaken Assumption \ref{ass:A:3}  to require only existence of error moments
slightly above order 4 (finite kurtosis) at the expense of further technicalities and more granular blocks in Assumption \ref{Ass2}. Assumption \ref{ass:A:4} implies global identification in the FA model
(see Lemma  \ref{App:lemma:Global:identification}). Assumptions \ref{ass:A:2}-\ref{ass:A:4} yield consistency of FA estimators (see proof of Lemma \ref{lemma:conv:espilon:beta}). Assumption  \ref{ass:A:5} is the local identification condition in the FA model (see Lemma \ref{App:lemma:Local:identification}).  
We use Assumption \ref{ass:A:6} together with Assumption \ref{ass:A:3} to invoke a CLT based on a multivariate Lyapunov condition (see proof of Lemma \ref{CLT:Zn}) to establish the asymptotic distribution of the LR statistic. To ease the verification of the Lyapunov condition, we bound a fourth-order moment of squared errors, which explains why we require finite eight-order moments in Assumption \ref{ass:A:3}. We could relax this condition at the expense of a more sophisticated proof of Lemma \ref{CLT:Zn}. The mild Assumption \ref{ass:A:6} (b)  requires that the smallest eigenvalue of $V[ vech( w_i w_i')]$ is bounded away from $0$ for all assets $i$ up to a small fraction. In Assumption \ref{ass:A:6} (c), in order to have $\kappa_n$ bounded, we need either mixing dependence in idiosyncratic errors within blocks, i.e., $\vert \sigma_{i,j} \vert \leq C \rho^{\vert i - j\vert}$ for $i,j\in I_m$ and $0 \leq \rho < 1$, or vanishing correlations, i.e., $\vert \sigma_{i,j} \vert \leq C b_{m,n}^{- \bar s}$ for all $i \neq j \in I_m$ and a constant $\bar s \geq 1/2$, with blocks of equal size.
In Assumption \ref{ass:A:7}, part (a) ensures a well-defined pseudo-true parameter value (White (1982)) under the alternative hypothesis.  Since we maximize the likelihood over a compact set, a maximizer exists. If the maximizer $\theta^{*}$ is unique w.r.t.\ $V^*_\varepsilon$, and the first $k$ eigenvalues of $V_y (V^{*}_{\varepsilon})^{-1}-I_T$ are distinct (as in Assumption \ref{ass:A:1}), then the pseudo-true values of the factors $F^*$ are unique up to column sign by (FA2). Part (b) is used to establish the consistency of the LR test under global alternative hypotheses (see proof of Proposition \ref{thm:asy:tests}). 
Finally, Assumption 
\ref{ass:A:8} is used to apply a Lyapunov CLT
(see proof of Lemma \ref{lemma:asydistr:Wn}) when deriving the asymptotic normality of the FA estimators.

\section{Feasible asymptotic distributional theory} \label{asymexp}

 In the proof of Proposition \ref{thm:asy:tests} below, we establish an asymptotic expansion for the LR test statistic under the null hypothesis of $k$ latent factors. 
For integer $p = \frac{1}{2} (T-k)(T-k+1)$, let us define the $p\times T$ matrix $\boldsymbol{X} = \left[ vech(G'E_{1,1}G) ~:~\cdots~:~ vech(G'E_{T,T}G)\right]$, where $E_{t,t}$ denotes the $T\times T$ matrix with entry 1 in position $(t,t)$ and $0$
elsewhere, and $G$ is a $T \times (T-k)$ matrix such that $F'V_{\varepsilon}^{-1} G =0$ and $G'V_{\varepsilon}^{-1} G = I_{T-k}$.  Matrix $G$ 
yields the orthogonal complement to the factor space in the scalar product induced by the inverse error variance, and is unique up to post-multiplication by an orthogonal matrix. The columns of $\boldsymbol{X}$ span the linear space $\{vech(G'DG): D \text{ diagonal} \}$. Then, the asymptotic expansion of $LR(k)$  under $H_0(k)$ is:\begin{eqnarray} \label{asy:exp:stats}
	LR(k)  =  vech(Z_n^*)'M_{\boldsymbol{X}} vech(Z_n^*) + o_p(1) ,  
\end{eqnarray} where $Z^*_n  := G' V_{\varepsilon}^{-1} Z_n V_{\varepsilon}^{-1} G$ for  $Z_n := \sqrt{n} \left( \frac{1}{n} \varepsilon \varepsilon' - \tilde{V}_{\varepsilon} \right)$, and
 matrix $M_{\boldsymbol{X}}:=I_p - \boldsymbol{X} ( \boldsymbol{X}'\boldsymbol{X})^{-1} \boldsymbol{X}'$ is idempotent of rank $p-T = df$. 
  The full-rank condition for matrix $\boldsymbol{X}$ corresponds to the  local identification condition for the PML estimator in Assumption \ref{ass:A:5} (see Lemma \ref{App:lemma:Local:identification}), analogously as in linear regression. The symmetric random matrix $Z_n$ involves average squares and cross-moments of error terms. The expansion in (\ref{asy:exp:stats}) does not depend on the diagonal elements of $Z_n$, since $vech(G'V_{\varepsilon}^{-1} diag(Z_n) V_{\varepsilon}^{-1} G)$ is spanned by the columns of $\boldsymbol{X}$, and thus is annihilated by the projection matrix $M_{\boldsymbol{X}}$. Intuitively, the diagonal elements of $Z_n$ are irrelevant because only the cross-sectional correlations of residuals for two  different dates are useful to check for omitted factors.

We now outline the distributional convergence $ Z^*_n \Rightarrow Z^*$ as $n \rightarrow \infty$ and $T$ is fixed, where $Z^*$ is a Gaussian symmetric matrix variate. By the block structure in Assumption \ref{Ass2} (b), we can write $Z^*_n$ as a sum of independent zero-mean terms:
$Z^*_n = \frac{1}{\sqrt n} \sum_{m=1}^{J_n} z^*_{m,n},
$
where the variables in the triangular array $z^*_{m,n} =\sum_{i\in I_m} G' V_{\varepsilon}^{-1}\left(\varepsilon_i \varepsilon_i'  -   E[\varepsilon_i \varepsilon_i']\right)V_{\varepsilon}^{-1}G$ are independent across $m$ and such that $E[ z^*_{m,n} ] =0$. In Appendix A, we invoke the CLT for independent heterogeneous variables to $vech( Z^*_n) = \frac{1}{\sqrt{n}} \sum_{m=1}^{J_n} vech(z^*_{m,n})$ and use Assumptions \ref{Ass2} (c) and (d) to check the Liapunov condition. 
We get $Z^*_n \Rightarrow Z^*$, where $vech(Z^*) \sim N(0,\Omega_{Z^*})$ and 
$
\Omega_{Z^*} = \underset{n \rightarrow \infty}{\lim}  \frac{1}{n} \sum_{m=1}^{J_n} V[ vech( z^*_{m,n})]
$.
Then, the  asymptotic expansion in (\ref{asy:exp:stats}) yields the asymptotic distribution for the LR test statistic defined in (\ref{Definition:statistics}) under the null hypothesis, and is given in Proposition \ref{thm:asy:tests}  (a) below. 

In Proposition \ref{thm:asy:tests} (b) hereafter, we state a feasible result for the asymptotic distribution of the LR statistic. We use  the estimate $\hat{G}$, i.e. the $T \times (T-k)$ matrix of standardized eigenvectors of $\hat V_y \hat{V}^{-1}_{\varepsilon}$ corresponding to its $T-k$ smallest eigenvalues ranked in decreasing order such that $\hat G' \hat{V}^{-1}_{\varepsilon} \hat G = I_{T-k}$, 
and set $\hat{\boldsymbol{X}}$ the matrix obtained by replacing $G$ with $\hat{G}$ in $\boldsymbol{X}$.
The GLS projection matrix orthogonal to $F$ for variance $V$ is defined by $M_{F,V} := I_T - F (F'V^{-1}F)^{-1} F'V^{-1}$. 

\begin{proposition} \label{thm:asy:tests} Let Assumptions \ref{Ass1}-\ref{ass:A:7} hold.
As $n \rightarrow \infty$ and $T$ is fixed, under the null hypothesis $H_0(k)$ of $k$ latent factors, 
(a) $LR(k) \Rightarrow \sum_{j=1}^{df} \mu_j \chi^2_j(1)$, where the $\chi^2_j(1)$ are independent chi-square variables with one degree of freedom, and the $\mu_j$ are the $df$ non-zero eigenvalues of matrix $M_{\boldsymbol{X}}\Omega_{Z^*}M_{\boldsymbol{X}} $.
(b) $\hat{\mu}_j \overset{p}{\rightarrow} \mu_j$, where $\hat{\mu}_j$, $j=1,...,df$, are the non-zero eigenvalues of matrix $M_{\hat{\boldsymbol{X}}}\hat{\Omega}_{Z^*}M_{\hat{\boldsymbol{X}}} $,  and $\hat{\Omega}_{Z^*} = \frac{1}{n} \sum_{m=1}^{J_n} vech( \hat{z}^*_{m,n}) vech( \hat{z}^*_{m,n} )'
$ with $\hat{z}^*_{m,n} = \sum_{i\in I_m}  \hat{G}'\hat{V}_{\varepsilon}^{-1}\hat{\varepsilon}_i \hat{\varepsilon}_i'\hat{V}_{\varepsilon}^{-1}\hat{G} $ and $\hat\varepsilon_i = M_{\hat F, \hat V_{\varepsilon}} (y_i - \bar y)$. Under the alternative hypothesis $H_1(k)$ of more than $k$ latent factors,  (c)  $LR(k)\geq C n$, w.p.a.\ $1$ for a constant $C>0$, and $\hat{\mu}_j = O_p( n \sum_{m=1}^{J_n} B_{m,n}^2)= o_p(n)$.
\end{proposition}

Proposition \ref{thm:asy:tests} (a)  shows that we depart from classical FA theory since we have convergence to a weighted sum of chi-square variates (see Section 4 of Robin and Smith (2000) for discussion of such random variable)  instead of a single chi-square variate (see the discussion in Section 5). Specification testing in the GMM framework ($J$-test of Hansen (1982)) without an optimal weighting matrix also gives this type of asymptotic distribution. It differs from the distribution of a mixture of chi-squared distributions found in testing when a parameter is on the boundary of the maintained hypothesis (see e.g.\ Andrews (2001)).  In Proposition \ref{thm:asy:tests} (b), matrix $\hat{G}$ is consistent for $G$ up to a right-rotation. The eigenvalues $\hat{\mu}_j$ are unaffected by such rotation because eigenvalues are invariant under pre and post multiplication by an orthogonal matrix and its transpose. In Proposition  \ref{prop:transformation:invariance} in Appendix B.6, we study how $\boldsymbol{X}$ and 
$M_{\boldsymbol{X}}\Omega_{Z^*}M_{\boldsymbol{X}} $ are transformed under different choices for the rotation of $G$. The eigenvalues $\mu_j$ are invariant to such rotation as expected. With fixed $T$, the GLS residuals $\hat{\varepsilon}_i$ are  asymptotically close to $M_{F,V_{\varepsilon}} \varepsilon_i$ and not to the true errors $\varepsilon_i$. However, it does not impede the consistency of the eigenvalues $\hat{\mu}_j$ underlying our feasible inference, since $G' V_{\varepsilon}^{-1} M_{F,V_{\varepsilon}} = G' V_{\varepsilon}^{-1}$. 
When we apply the CLT, centering is done implicitly since $M_{\boldsymbol{X}}vech(G'V_{\varepsilon}^{-1}\varepsilon_i \varepsilon_i'V_{\varepsilon}^{-1}G)=M_{\boldsymbol{X}}vech(G'V_{\varepsilon}^{-1}(\varepsilon_i \varepsilon_i'-E[\varepsilon_i \varepsilon_i'])V_{\varepsilon}^{-1}G)$. Indeed  $vech(G'V_{\varepsilon}^{-1}E[\varepsilon_i \varepsilon_i']V_{\varepsilon}^{-1}G)$ is spanned by the columns of $\boldsymbol{X}$ since $E[\varepsilon_i \varepsilon_i']=\sigma_{ii}V_{\varepsilon}$ is diagonal. We can consistently estimate the critical values of the statistics by simulating a large number of draws of $\sum_{j=1}^{df} \hat{\mu}_j \chi^2_j(1)$.   Hence, even if FA residuals are not consistent estimates of the true errors in short panels, we are still able to supply a feasible asymptotic distributional theory for our empirical applications under a block-dependence structure. Proposition \ref{thm:asy:tests} (c) gives test consistency against global alternatives (see Section \ref{local:sec} for local alternatives and AUMPI properties).

\section{Impact of assuming Gaussian errors and sphericity} \label{discu}
In this section, we particularize the general distributional results of Proposition \ref{thm:asy:tests} to three important cases, namely (i) Gaussian errors, (ii) settings where the asymptotic distribution under Gaussian errors still holds for the LR test statistics (up to scaling),  and (iii) spherical errors. We further discuss the impact of wrongly assuming Gaussian errors or sphericity (PCA) when conducting inference on the number of latent factors. This section should help guard against misuse of standard software routines in applied work.

\subsection{Gaussian errors} \label{Gaussian}

Let us consider the case where the errors $\varepsilon_i \overset{ind}{\sim} N(0,\sigma_{ii} V_{\varepsilon})$ are independent Gaussian vectors. From classical FA theory, we expect that the statistic $LR(k)$ admits asymptotically a chi-square distribution with $df$ degrees of freedom in the cross-sectionally homoskedastic case, i.e., $\sigma_{ii}=1$ for all assets $i$. We cannot expect that this distributional result applies to the Gaussian framework in full generality, since - even in such a case - our setting corresponds to a pseudo model (because the  $\sigma_{ii}$ may be heterogeneous across $i$, and the $\beta_i$ are treated as fixed effects, namely non-random incidental parameters, instead of Gaussian random effects).  Under the normality assumption for the error terms, we have ${\varepsilon}^*_i := G' V_{\varepsilon}^{-1} \varepsilon_i \overset{ind}{\sim} N(0,\sigma_{ii} I_{T-k})$. Thus, by the Liapunov CLT, the distributional limit of  $\frac{1}{\sqrt{q}} Z_n^* =  \sqrt{n/q} \left( \frac{1}{n} {\varepsilon^*} (\varepsilon^*)' - \frac{1}{n}E\left[ {\varepsilon^*} (\varepsilon^*)'\right]\right)$ is in the Gaussian Orthogonal Ensemble (GOE) for dimension $T-k$ (see e.g.\ Tao (2012)), i.e., $\frac{1}{\sqrt q} vech(Z^*) \sim N (0, I_p)$, where $q:= \underset{n\rightarrow\infty}{\lim} \frac{1}{n} \sum_{i=1}^n \sigma_{ii}^2$. Then, from (\ref{asy:exp:stats}) we get $LR(k) \Rightarrow q\chi^2(df)$, i.e., we get convergence to a scaled chi-square variate $q\chi^2(df)$. In the cross-sectionally homoskedastic case, we have $q=1$ yielding the classical $\chi^2 (df)$ result. On the contrary, cross-sectional heterogeneity in the unconditional idiosyncratic variances yields $q>1$ and a deviation from classical FA theory even in the Gaussian case. In the Gaussian case, unobserved heterogeneity across asset idiosyncratic variances leads to an oversized LR  test if we use critical values from the chi-square table without proper scaling. In our Monte Carlo experiments with non-Gaussian errors, we get size distortion by around 80 percentage points when $T = 24$ for a nominal size of 5\%, and thus an average estimated number of factors over 3 instead of 2. 

\subsection{Validity of the scaled asymptotic chi-square test} \label{scaled}

In this subsection, we investigate sufficient conditions for the validity of the convergence of the LR statistic to a scaled chi-square variate $\chi^2(df)$, but in special cases beyond Gaussianity of errors. For this purpose,  let us recall that the asymptotic expansion of $LR(k)$ in (\ref{asy:exp:stats}) only involves the out-of-diagonal elements of $Z_n$. Under independent Gaussian errors (Section \ref{Gaussian}), by the Liapunov CLT, we have $Z_n\Rightarrow Z$, where $Z_{t,s} \sim N (0, q V_{\varepsilon,tt} V_{\varepsilon,ss})$, for $t > s$, mutually independent, where the $V_{\varepsilon,tt}$ are the diagonal elements of matrix $V_{\varepsilon}$. We deduce that any setting featuring the same joint asymptotic distribution for the out-of-diagonal elements of random matrix $Z_n$ leads to an asymptotic distribution for the LR statistic similar to the Gaussian case.  

\begin{proposition}  \label{prop:gen:Gaussian}
Let Assumptions \ref{Ass1}-\ref{ass:A:6} hold with (a) $\underset{n \rightarrow \infty}{\lim} \frac{1}{n} \sum_{i=1}^{n}  E[ \varepsilon_{i,t} \varepsilon_{i,s} \varepsilon_{i,r}\varepsilon_{i,p}] =  q V_{\varepsilon,tt} V_{\varepsilon,ss}$, when $t = r > s = p$, for a constant $q>0$, and $=0$ in all other cases with $t>s$ and $r>p$, and (b) let $\kappa = \underset{n \rightarrow \infty}{\lim} \frac{1}{n} \sum_{m=1}^{J_n} \sum_{i\neq j \in I_m} \sigma_{ij}^2$ as in Assumption \ref{ass:A:3} (b). Then, $LR(k) \Rightarrow \bar q  \chi^2(df)$ under $H_0(k)$ for $\bar q := q + \kappa$.
\end{proposition}

Conditions (a) and (b) in Proposition \ref{prop:gen:Gaussian} generalize the correctness of the scaled chi-square test beyond Gaussianity and error independence across time and assets. Under Assumption \ref{Ass2}, Condition (a) is satisfied if the standardized error terms $w_{i,t}$ are conditionally homoskedastic martingale difference sequences. However, Condition (a) excludes empirically relevant cases such as  ARCH processes for $w_{i,t}$, because, in that case, $\frac{1}{V_{\varepsilon,tt} V_{\varepsilon,ss}} E[ \varepsilon_{i,t}^2 \varepsilon_{i,s}^2]$  depends on lag $t-s$.  Hence, serial correlation in squared idiosyncratic errors is responsible for the deviation of the LR test from the scaled chi-square asymptotic distribution. This setting is covered by the general results in Proposition \ref{thm:asy:tests}.
Anderson and Amemiya (1988) establish the asymptotic distribution of FA estimates assuming that the error terms are i.i.d.\ across sample units and deploy an assumption that is analogue to Condition (a) above in their Corollary 2. The i.i.d.\ assumption in our case implies $\sigma_{ii}=1$ for all $i$, which results in a cross-sectionally homoskedastic setting.  That setting is irrealistic in our application, as it would imply that the idiosyncratic variance is the same for all assets. Our results show that establishing the asymptotic distribution of the test statistics, especially the AUMPI property of LR test (see Section \ref{local:sec}), in a general setting with non-Gaussian errors, heterogeneous idiosyncratic variances and ARCH effects (individual and common across assets), is challenging, but still possible. If the $\sigma_{ii}$ are treated as i.i.d.\ random effects independent of the errors, and we exclude cross-sectional correlation of errors to simplify, we recover the i.i.d.\ condition of the data. However, the random $\sigma_{ii}$ yields a stochastic common factor across time that breaks the condition in Corollary 2 of Anderson and Amemiya (1988).

When the assumptions of Proposition \ref{prop:gen:Gaussian} are satisfied, scaling the LR test statistic by a consistent estimate of $1/\bar{q}$ is sufficient to recover an asymptotic $\chi^2(df)$ distribution, as in the Gaussian case. The idea of scaling a test statistic to make it more robust to deviations from Gaussianity is well known in the psychometrics literature, for example, in the Satorra-Bentler scaled difference chi-square test (Satorra and Bentler (2001, 2010)). In our context, the Satorra-Bentler scaling factor corresponds to a consistent estimate of $df / \sum_{j=1}^{df} \mu_j$, and ensures that the  weighted sum of chi-square variates described in Proposition \ref{thm:asy:tests} has the same mean as a $\chi^2(df)$ variate. Under the conditions of Proposition \ref{prop:gen:Gaussian}, we have $\frac{1}{df}\sum_{j=1}^{df}\mu_j = \bar{q}$, so that $1/\bar{q}$ coincides  with the Satorra-Bentler scaling factor. Unreported Monte Carlo experiments under the design of Section \ref{Montecarlo} show that this scaling approach performs well in finite samples. 

\subsection{Spherical errors} \label{spheri_imp}

If errors are spherical, i.e., matrix $V_{\varepsilon} =  \bar \sigma^2 I_T$ is a multiple of the identity with unknown parameter $\bar \sigma^2 >0$, then the asymptotic distribution of $LR(k)$ corresponds to a special case of Proposition \ref{thm:asy:tests} (a). If sphericity is imposed in the estimation procedure, i.e., $\hat{V}_{\varepsilon}$ becomes $\hat{V}_{\varepsilon,c}=\hat{\sigma}^2I_T$, the constrained FA estimator boils down to the Principal Component Analysis (PCA) estimator; see Anderson and Rubin (1956) Section 7.3. If we have independent individual ARCH processes across assets and no common ARCH effects, PCA is still valid since one inherits sphericity from cross-sectional averaging. Let $\hat{F}_c$ denote the FA estimator under sphericity constraint. Then, $\hat{F}_c$ is the matrix of eigenvectors of matrix $\hat{V}_y$ standardized such that $\hat{F}_c'\hat{F}_c =  diag(\hat {\delta}_1 - \hat{\sigma}^2,...,\hat{\delta}_k - \hat{\sigma}^2)$, and $\hat{\sigma}^2 = \frac{1}{T-k} \sum_{j=k+1}^{T} \hat{\delta}_j$, where $\hat{\delta}_j$ denotes the $j$th largest eigenvalue of matrix $\hat{V}_y$. 
Then, the constrained LR statistic becomes $LR_c(k) := -n\sum_{j=k+1}^T \log ( 1  + \frac{\hat \delta_j  - \hat \sigma^2}{\hat\sigma^2} )$ and reduces to the LR statistic invoqued by Onatski (2023) in his discussion of FGS. We have $\hat{\gamma}_j = \hat{\delta}_j / \hat{\sigma}^2 - 1$, and $LR_c(k) = \frac{n}{2 \hat \sigma^4} \sum_{j=k+1}^T ( \hat \delta_j^2 - \hat\sigma^2)^2 + o_p(1)$, i.e., the constrained LR statistic is asymptotically equivalent to the sum of squared deviations of the $T-k$ smallest eigenvalues from their mean. 
 Under spherical errors, Onatski (2023) shows $LR_c(k) \Rightarrow \frac{1}{2} ( Tr [(Z^*)^2] - \frac{1}{T-k} [ Tr(Z^*)]^2) = vech(Z^*)'M_{\boldsymbol{x}}vech(Z^*)$, where $M_{\boldsymbol{x}}=I_p-\boldsymbol{x}(\boldsymbol{x}'\boldsymbol{x})^{-1}\boldsymbol{x}'$ with $\boldsymbol{x}=vech(I_{T-k})$,  $Z^*=\frac{1}{\bar{\sigma}^4}G'ZG$, $Z$ is the distributional limit of $Z_n$, and $G$ is
a $T \times (T-k)$ matrix such that $F'G = 0$ and $G'G = \bar \sigma^2I_{T-k}$. Hence the asymptotic distribution of the constrained LR statistic under sphericity is driven by $\sum_{j=1}^{p-1}\mu_j\chi^2_j(1)$, where the $\chi^2_j(1)$ are independent and the  $\mu_j$ are the non-zero eigenvalues of matrix $M_{\boldsymbol{x}}\Omega_{Z^*}M_{\boldsymbol{x}}$, with $\Omega_{Z^*}=V[vech(Z^*)]$. Under Gaussian errors, it simplifies to $LR_c(k)\Rightarrow q\chi^2(p-1)$ (see Section \ref{Gaussian}). In a PCA setting when sphericity fails to hold, a massive over-rejection by over 80 percentage points for a nominal size of 5\% is again observed in our Monte Carlo experiments. It is true both for the constrained LR test and for the eigenvalue spacing test of FGS. We get an average estimated number of factors often above 10 instead of 2 when $T = 24$.

\section{Local asymptotic power} \label{local:sec}

In this section, we study the asymptotic power properties of the test statistics against local alternatives in which we have $k$ (strong) factors plus a weak factor.\footnote{Recently, Barigozzi and Hallin (2025) show that exchangeability of assets in the whole cross-section rules out weak factors as defined in this section. In equity markets, ordering matters because of industry sectors and other firm characteristics such as size and book-to-market.} Weak factors can result from strong factors that load only on a subsample of assets of size $O(\sqrt n)$ or from factors with vanishing loadings of size $O(n^{-1/4})$ diluted in the cross-section.  Specifically, under $H_{1,loc}(k)$, we have $\sqrt n \gamma_{k+1} \rightarrow c_{k+1}$ as $n \rightarrow \infty$, with $c_{k+1}>0$. The (drifting) DGP is $Y = \mu 1_n' + F \beta' + F_{k+1} \beta_{loc}' + \varepsilon$, where $\beta_{loc}$ is the loading vector for the $(k+1)$th factor, and the factor vector is normalized such that $F_{k+1} = \sqrt{\gamma_{k+1}} \rho_{k+1}$ with $\rho_{k+1}' V_{\varepsilon}^{-1} \rho_{k+1} = 1$ and $F'V_{\varepsilon}^{-1} \rho_{k+1} = 0$. Thus, we can write $\rho_{k+1} = G \xi_{k+1}$ for a $T-k$ dimensional vector $\xi_{k+1}$ with unit norm. Scalar $c_{k+1}$ and vector $\xi_{k+1}$ yield the (normalized) strength and the direction of the local alternative.  The root-$n$ condition defining $H_{1,loc}(k)$ represents the weakest level of factor strength that our test can detect. This detection threshold matches the convergence rate of the FA estimators. Intuitively, vanishing factors satisfying $\sqrt{n}\gamma_{k+1}\to 0$ cannot be detected because their signal is dominated by estimation noise. In contrast, the test achieves asymptotic power of 1 against semi-strong factors for which $\sqrt{n}\gamma_{k+1} \to \infty$, because their signal dominates estimation noise. The hedge case $H_{1,loc}(k)$ corresponds to the signal being of the same order as the estimation noise, for which asymptotic power is between 0 and 1.

\subsection{Asymptotic distributions under local alternatives} \label{locpow}

We derive an asymptotic expansion of  the LR statistics under $H_{1,loc}(k)$ using similar arguments as in the proof of Proposition \ref{thm:asy:tests} (a) (see the proof of Proposition \ref{thm:local:alternative} in Appendix A for the derivation):
\begin{eqnarray} 
	LR(k) =  vech(Z_{n,loc}^*)'M_{\boldsymbol{X}} vech(Z_{n,loc}^*) + o_p(1) ,  \label{asy:exp:stats:alternative}
\end{eqnarray}
where $Z_{n,loc}^*=Z^*_n+c_{k+1}\xi_{k+1} \xi_{k+1}'$. From the CLT, we have $Z_{n,loc}^*\Rightarrow Z_{loc}^*$ where $Z_{loc}^*=Z^*+c_{k+1}\xi_{k+1} \xi_{k+1}'$. Matrix variate $Z^*_{loc}$ is a non-central symmetric Gaussian matrix. The non-zero mean depends in general on both $c_{k+1}$ and $\xi_{k+1}$, while the variances and covariances of the elements of $Z^*_{loc}$ are the same as those of $Z^*$. Then, we deduce from (\ref{asy:exp:stats:alternative})  that the asymptotic distribution of the $LR(k)$ statistic under the local alternative hypothesis  is a weighted sum
of $df$ mutually independent non-central chi-square variables, with non-centrality parameters depending on $vech(\Delta):=M_{\boldsymbol{X}} vech(c_{k+1}\xi_{k+1}\xi'_{k+1})$.
\begin{proposition}  \label{thm:local:alternative} Let Assumptions  \ref{Ass1}-\ref{ass:A:6} hold.
Under the local alternative hypothesis $H_{1,loc}(k)$, we have as $n \rightarrow \infty$ and $T$ is fixed, $LR(k) \Rightarrow \sum_{j=1}^{df} \mu_j \chi^2 ( 1, \lambda_j^2)$, \label{chiweight}
where $\lambda_j^2 = \mu_j^{-1} [v_j' vech(\Delta)]^2$, and the $\mu_j$ and $v_j$ are the non-zero eigenvalues and the associated standardized eigenvectors of matrix $M_{\boldsymbol{X}}\Omega_{ Z^*}M_{\boldsymbol{X}}$.
\end{proposition}
 The non-centrality term $vech(\Delta)$ is in charge of  the asymptotic local power of the statistics. When this vector is null, the asymptotic local power is zero. Indeed, for some local alternatives the 
$(k+1)$th weak factor can be absorbed in the diagonal covariance matrix $V_{\varepsilon}$ of the error terms. More precisely, in Appendix B.4 ii) we show that $V_y + \frac{c_{k+1}}{\sqrt{n}} \rho_{k+1} \rho_{k+1}' = F^* (F^*)' + V_{\varepsilon}^* + \frac{1}{\sqrt{n}} G \Delta G' + o(1/\sqrt{n})$ for some $T \times k$ matrix $F^*$ and diagonal matrix $V_{\varepsilon}^*$, which yields asymptotically a $k$-factor model when $\Delta=0$. 
We have $\lambda^2 := \sum_{j=1}^{df} \mu_j \lambda_j^2 = \frac{1}{2} \Vert \Delta \Vert^2$, i.e., the half squared Frobenius norm of the matrix measuring local distance from the $k$-factor specification. It follows that the asymptotic local power of the LR statistic is non null as long as $\lambda^2>0$, i.e., it has non-trivial asymptotic power against any proper local alternative. 

 Under the normality of errors,  or more generally the conditions of Proposition \ref{prop:gen:Gaussian}, using that matrix $\frac{1}{\sqrt{\bar q}} Z^*$ is in the GOE for dimension $T-k$, i.e., $vech(Z^*) \sim N(0,\bar q I_{p})$, we have $LR(k) \Rightarrow \bar q \chi^2( df, \lambda^2/\bar q)$ from (\ref{asy:exp:stats:alternative}). The local power is a function solely of the squared Euclidean norm of the vector $vech(\Delta)$ measuring local distance from the $k$-factor specification, divided by $\bar q$. 
 
 
\subsection{AUMPI tests}
\label{sectionAUMPI}

In this subsection, we investigate asymptotic local optimality  of the LR statistic for testing hypotheses on the number of latent factors. In our framework with composite null and alternative hypotheses and multi-dimensional parameter, we cannot expect in general  to establish Uniformly Most Powerful (UMP) tests. Instead, we can establish an optimality property by restricting the class of tests to invariant tests (e.g.\ Choi et al.\ (1996), Lehmann and Romano (2005)). We focus on statistics with test functions $\phi$ written on the elements of symmetric matrix $\hat{S} = \hat{V}_{\varepsilon}^{-1/2}M_{\hat{F},\hat{V}_{\varepsilon}} ( \hat{V}_y  - \hat{V}_{\varepsilon} ) M_{\hat{F},\hat{V}_{\varepsilon}}'  \hat{V}_{\varepsilon}^{-1/2}$. This matrix measures to which extent we can approximate the difference between the sample variance-covariance $\hat V_y$ and diagonal matrix $\hat V_{\varepsilon}$ by a symmetric matrix of reduced rank $k$, with range spanned by the range of $\hat{F}$. The non-zero eigenvalues of $\hat S$ are $\hat \gamma_{j}$ for $j=k+1,...,T$ (see Lemma \ref{prop:1} in Appendix A) which implies that  the LR  test statistic  is function of the elements of  $\hat{S}$.
To eliminate the redundancy implied by the deterministic relations holding for the elements of $\hat{S}$, we actually consider the test class $\mathscr{C} = \left\{ \phi ~: ~  \phi = \phi( \hat W) \right\}$ with $\hat W: = \sqrt{n} \hat{\boldsymbol{D}}'  vech(\hat{S}^*)$, where $\hat{S}^* = \hat{G}' \hat{V}_{\varepsilon}^{-1/2} \hat{S} \hat{V}_{\varepsilon}^{-1/2} \hat{G}$ and $\hat{\boldsymbol{D}}$ is a $p \times df$ full-rank matrix, such that $M_{\hat{\boldsymbol{X}}} = \hat{\boldsymbol{D}} \hat{\boldsymbol{D}}'$ and $\hat{\boldsymbol{D}}' \hat{\boldsymbol{D}} = I_{df}$. We have that $\hat{S}^*=diag(\hat{\gamma}_{k+1},\dots,\hat{\gamma}_{T})$, since $\hat{S}$ has the eigendecomposition $\hat{V}_{\varepsilon}^{-1/2}\hat{G}diag(\hat{\gamma}_{k+1},\dots,\hat{\gamma}_{T})\hat{G}'\hat{V}_{\varepsilon}^{-1/2}$ from Lemma \ref{prop:1} (a).  Diagonal matrix $\hat{S}^*$ contains the information in $\hat{S}$ beyond mechanical orthogonality of its rows and columns to $\hat{V}_{\varepsilon}^{-1/2} \hat{F}$. Vector $\hat W$ contains the information in $\sqrt{n} vech(\hat{S}^*)$ beyond mechanical orthogonality to $\hat{\boldsymbol{X}}$. From Lemma \ref{prop:1} (c), we have $0=diag (\hat{S}) =2\hat{V}^{-1}_\varepsilon\hat{\boldsymbol X}'vech(\hat{S}^*)$. Therefore, $vech(\hat{S}^*)$ lies in the orthogonal complement of the range of $\hat{\boldsymbol X}$ in sample.


Matrices $\hat{G}$ and $\hat{\boldsymbol{D}}$ are both consistent up to post-multiplication by an orthogonal matrix. This point yields a group of orthogonal transformations under which we require the test statistics to be invariant, because the choices for the normalization of $\hat{G}$ and $\hat{\boldsymbol{D}}$ should be immaterial for the testing outcome. Here, we do not deal with invariance to data transformations but rather with invariance to parameterization of $\hat{G}$ and $\hat{\boldsymbol{D}}$. However, if we consider tests based on the elements of vector $\hat{W}$, this difference is immaterial. In Appendix B.6, we show that the maximal invariant under this group is provided by $\hat W' \hat W = n vech(\hat{S}^*)' M_{\hat{\boldsymbol{X}}} vech(\hat{S}^*)$. Since $\sqrt{n} vech(\hat{S}^*) $ belongs to the range of matrix $M_{\hat{\boldsymbol{X}}}$, we have
$\hat W' \hat W = \frac{n}{2} \Vert \hat{S}^* \Vert^2  = \frac{n}{2} \Vert \hat{S} \Vert^2$.
Therefore, the invariant tests are functions of the squared norm of $\hat{S}$, which is asymptotically equivalent to the LR statistic (up to the factor $1/2$).

In the Gaussian case, or more generally under the conditions of Proposition \ref{prop:gen:Gaussian}, the LR statistic follows asymptotically a scaled non-central chi-square distribution with $df$ degrees of freedom and non-centrality parameter $\lambda^2/\bar{q} = \sum_{j=1}^{df}\lambda_j ^2$ as shown in the previous subsection.
Thus, we can simplify the null and alternative hypotheses of our testing problem asymptotically and locally to a one-sided test with
null hypothesis $H_0(k) : \lambda^2=0$ vs.\ alternative hypothesis $H_{1,loc}(k) : \lambda^2 > 0$. The scaling constant $q>0$ plays no role in the power analysis. It means that the LR test is an AUMPI test (Lehmann and Romano (2005) Chapters 3 and 13). Indeed, the density $g(z;df,\lambda^2)$ of the $\chi^2(df,\lambda^2)$ distribution is Totally Positive of order 2 ($TP2$) in $z$ and $\lambda^2$ (Eaton (1987) Example A.1 p.\ 468); see Miravete (2011) for a review of applications of TP2 in economics. A density, which is $TP2$ in $z$ and $\lambda^2$, has the Monotone Likelihood Ratio (MLR) property (Eaton (1987) p.\ 467). Since $g(z;df, \lambda^2)/g(z;df,0)$ is an increasing function in $z$, it gives the AUMPI property. Bollerslev et al.\ (2025) also achieve to show the MLR property (with technical arguments markedly dfferent from the techniques commonly employed in high-frequency financial econometrics), and thus the AUMPI property for their Marubozu test.

In the general case with $df > 1$, when neither Gaussianity nor the conditions of Proposition \ref{prop:gen:Gaussian} apply, we cannot use the same reasoning, since the density $f(z;\lambda_1, ..., \lambda_{df})$ of $\sum_{j=1}^{df} \mu_j \chi^2(1,\lambda_j^2)$, with $\mu_j>0$, $j=1, ...,df$, is not a function of $\lambda^2 = \sum_{j=1}^{df}\mu_j \lambda_j^2 $ only, and thus cannot be $TP2$ in $z$ and $\lambda^2$. Instead, our strategy to get a new AUMPI result uses a power series representation of the density of $\sum_{j=1}^{df} \mu_j \chi^2(1,\lambda_j^2)$ in terms of central chi-square densities from  Kotz, Johnson, and Boyd (1967). Under the sufficient condition (\ref{cond:MLR}) in Proposition \ref{thm:AUMPI}, the density ratio $\frac{f(z;\lambda_1, ..., \lambda_{df})}{f(z;0, ..., 0)}$ is monotone increasing in $z$.

\begin{proposition} \label{thm:AUMPI} Let Assumptions  \ref{Ass1}-\ref{ass:A:6} hold.
(a) Let us assume that, for any DGP in the subset $\bar H_{1,loc}(k) \subset H_{1,loc}(k)$ of the local alternative hypothesis, we have  for any integer $m \geq 3$:
\begin{eqnarray}
\sum_{j > l \geq 0, j+l = m} \frac{(j-l)  \Gamma(\frac{df}{2})^2}{\Gamma( \frac{df}{2} + j)\Gamma( \frac{df}{2} + l)} [c_j(\lambda_1,...,\lambda_{df}) c_l(0,...,0) - c_l(\lambda_1,...,\lambda_{df}) c_j(0,...,0)] \geq 0,  \quad \label{cond:MLR}
\end{eqnarray}
where $\Gamma(\cdot)$ is the Gamma function, $c_j(\lambda_1,...,\lambda_{df}) := E[Q(\lambda_1,...,\lambda_{df})^j]/j!$ for $Q(\lambda_1,...,\lambda_{df}) = \frac{1}{2} \sum_{j=1}^{df}(\sqrt{\nu_j} X_j + \sqrt{1-\nu_j} \lambda_j)^2$, $\nu_j = 1- \frac{1}{\mu_j}\mu_{1}$ with the $\mu_j$ ranked in increasing order,  and $X_j \sim N(0,1)$ are mutually independent. Then, the statistic $LR(k)$ yields an AUMPI test against $\bar H_{1,loc}(k)$.
(b) Suppose that either $\lambda_1^2 + (1-\nu_2)\lambda_2^2 \geq \nu_2$ and $(1-\nu_2)\lambda_2^2 \geq \frac{1}{2}\nu_2$ when $df=2$, or
\begin{equation} \label{condition:lambda}
1\{i=0\} \lambda_1^2 + \sum_{j=2}^{df-1} \rho_j^i  (1-\nu_j) \lambda_j^2 + ( 1 - \nu_{df}) \lambda_{df}^2 \geq \frac{\nu_{df}}{i+1} \left( df - 2 - \sum_{j=2}^{df-1} \rho_j^{i+1}\right),
\end{equation}
 for all $i\geq 0$, where $\rho_j := \frac{\nu_j}{\nu_{df}}$, when $df\geq 3$. Then, Inequalities (\ref{cond:MLR}) hold for any $m\geq 3$.
\end{proposition}

The proof of Proposition \ref{thm:AUMPI}  relies on using complete exponential Bell's polynomial (Bell (1934)) in the proof of Lemma \ref{prop:recursion}.
Conditions (\ref{cond:MLR}) involve polynomial inequalities in the parameters $\lambda_j$ of the alternative hypothesis, and parameters $\nu_j$ of the weights of the non-central chi-square distributions, $j=1,...,df$. It is challenging to establish an explicit characterization of the $\lambda_j$ and $\nu_j$ equivalent to Inequalities (\ref{cond:MLR}), unless $df=1$.\footnote{\label{footnote19}Inequalities (\ref{cond:MLR}) with $df=1$ are easily proved to hold. In such a case, we can use the asymptotic distribution of a scaled chi-square variable and its MLR property.} By deploying a novel characterization of the $c_j(\lambda_1,...,\lambda_{df})$ in terms of a recurrence relation (Lemma \ref{prop:recursion}),  we establish explicit sufficient conditions in part b) of Proposition \ref{thm:AUMPI}.  Inequalities  (\ref{condition:lambda}) are linear in the $\lambda_j^2$, and define a non-empty convex domain in the $(\lambda_1^2,..,\lambda_{df}^2)$ space, that does not contain the origin $\lambda_1=...=\lambda_{df}=0$  (unless the DGP is such that $\nu_2=...=\nu_{df}$, in which case the RHS of (\ref{condition:lambda}) is nil for all $i$ and thus any $\lambda_j^2$ meet the inequalities). Proposition \ref{thm:AUMPI} (b) implies that, for a given set of values of $df$, the  MLR property holds if $\lambda_j \geq \underline{\lambda}$ for all $j$, uniformly for $\nu_j \leq \bar \nu$, where $\underline{\lambda}>0$ is a constant that depends on $\bar \nu < 1$. Vanishing values of the $\nu_j$ correspond to homogenous weights $\mu_j$, i.e., the scaled non-central chi-square distribution with $df$ degrees of freedom. Hence, the AUMPI property in Proposition \ref{thm:AUMPI} holds in neighborhoods of DGPs that match the conditions of Proposition \ref{prop:gen:Gaussian} (e.g.\ Gaussian errors) for alternative hypotheses that are sufficiently separated from the null hypothesis. Besides, Proposition \ref{thm:AUMPI} shows that the Gaussian case is not the only design delivering an AUMPI test. Further, in the SMC, we establish a new analytical representation of the coefficients  $c_k(\lambda_1,...,\lambda_{df})$ in terms of matrix product iterations. That analytical representation allows us to check numerically the validity of Inequalities (\ref{cond:MLR}) for given $df$, $\lambda_j$, $\nu_j$, and $m=1,....,M$, for a large bound $M$ (see Appendix C). 
In Appendix C, when Inequalities (\ref{condition:lambda}) are met, we always conclude to the MLR property  in the numerical checks as predicted by the theory of Proposition \ref{thm:AUMPI}. There, we also provide numerical evidence that the domain of validity of the MLR property is relevant for our empirical application. Thus, the LR statistic has optimal power properties for testing the number of latent factors in short subperiods of our stock return dataset. 
The sufficient conditions (\ref{cond:MLR}) and (\ref{condition:lambda}) in Proposition \ref{thm:AUMPI} yielding the monotone property of density ratios have potentially broad application outside the current setting to show AUMPI properties since other test statistics share an asymptotic distribution characterized by a positive definite quadratic form in normal vectors (Khatri (1980)).

\section{Monte Carlo experiments}
\label{Montecarlo}

This section gives a Monte Carlo assessment of size and power and selection procedure for the number of factors for the LR test under non-Gaussian errors. Let us start with a description of the DGP we use in our simulations.
In the DGP, the betas are $\beta_i ~ \overset{i.i.d.}{\sim} ~ N(0,I_k)$, with $k=3$, and the matrix of factor values is $F = V_{\varepsilon}^{1/2} U \Gamma^{1/2}$, where $U = \tilde F  ( \tilde F' \tilde F)^{-1/2}$ and $vec(\tilde F) \sim N(0,I_{Tk})$. We generate the diagonal elements of $V_{\varepsilon} =diag ( h_{1},...,h_{T})$ through a common time-varying component in idiosyncratic volatilities (Renault, Van Der Heijden and Werker (2023)) via the ARCH $h_{t} = 0.6 + 0.5 h_{t-1} z_{t-1}^2$, with $z_t \sim IIN(0,1)$. This common component induces a deviation from spherical errors. The diagonal matrix $\Gamma = T diag (3,2,n^{-\bar{\kappa}})$ yields $\frac{1}{T} F' V_{\varepsilon}^{-1} F = diag (3,2,n^{-\bar{\kappa}})$, i.e., the "signal-to-noise" ratios equal $3$, $2$ and $n^{-\bar{\kappa}}$ for the three factors. We take $\bar{\kappa}=\infty$ to study the size of $LR(2)$. To study the power of $LR(2)$, we take $\bar{\kappa}=0$ to get a global alternative and $\bar{\kappa}=1/2$ to get a local alternative (weak factor). We generate the idiosyncratic errors by $\varepsilon_{i,t} = h_t^{1/2} h_{i,t}^{1/2} z_{i,t}$, where $h_{i,t} = c_i + \alpha_i h_{i,t-1} z_{i,t-1}^2$, with $z_{i,t} \sim IIN(0,1)$ mutually independent of $z_t$. We use the constraint $c_i= \sigma_{ii}(1-\alpha_i)$ with uniform draws for the idiosyncratic variances $V[\varepsilon_{i,t}]=\sigma_{ii} ~ \overset{i.i.d.}{\sim} ~ U[1,4]$, so that $V[\varepsilon_{i,t}/h_t^{1/2}] = \frac{c_i}{1-\alpha_i} = \sigma_{ii} $. Such a setting allows for cross-sectional heterogeneity in the variances of the scaled $\varepsilon_{i,t}/h_t^{1/2}$. The ARCH parameters are uniform draws $\alpha_i ~ \overset{i.i.d.}{\sim} ~ U[0.2,0.5]$ with an upper boundary of the interval ensuring existence of fourth-order moments. 
We generate $5,000$ panels of returns of size $n \times T$ for each of the $100$ draws of the $T\times k$ factor matrix $F$ and common ARCH process $h_t$, $t=1,...,T$, in order to keep the factor values constant within repetitions, but also to study the potential heterogeneity of size and power results across different factor paths.  The factor betas $\beta_i$, idiosyncratic variances $\sigma_{ii}$, and individual ARCH parameters $\alpha_i$ are the same across all repetitions in all designs of the section. We use three different cross-sectional sizes $n=500,1000, 5000$, and three  values of time-series dimension $T=6,12,24$. The variance matrix $\hat{\Omega}_{\bar{Z}^*}$ is computed using the parametric structure of Lemma \ref{lemma:parametric:Omega:Z:bar:star}. We get the $T-1$ estimated parameters by least squares, as detailed in OA Section E.5.3 i). The $p$-values are computed over $5,000$ draws.

We provide the size and power results in \% in Table \ref{MC:Table1}. Size of $LR(2)$ is close to its nominal level $5\%$, with size distortions smaller than $1\%$, except for the case $T=24$ and $n=500$. The impact of the factor values on size is small for $T$ above $6$. The labels global power and local power refer to $\bar{\kappa}=0$ and $\bar{\kappa}=1/2$, and power computation is not size adjusted. The global power is equal to 100\%, while the local power  ranges from 80\% to  85\% for $T=6$, and 
is equal to 100\% for $T=12$ and $T=24$. The approximate constancy of local power w.r.t.\ $n$, for large $n$, is coherent with theory implying convergence to asymptotic local power. 
In the last panel of Table \ref{MC:Table1}, we provide the average of the estimated number $\hat{k}_{LR}$ of factors, obtained by sequential testing with $LR(k)$, for $k=0,\dots,k_{max}$, with $k_{max}=2,7,17$ for $T=6,12,24$ (see Table \ref{maxk} of OA). We follow the procedure adopted in our empirical application and described in Section 8.2, with size $\alpha_n = 10/n$. If we reject for all $k=0,\dots,k_{max}$, then the estimated number of factors is set to $\hat k_{LR} = k_{max} + 1$. For all sample sizes $T=6,12,24$, the average estimated number of factors is very close to the true number 2 with estimated proportions close or equal to 100\%. We can conclude that our selection procedure for the number of factors works well in our simulations. 
We have also run some similar simulations with a true number $k=2$ for $T=36$ and $k=10$ for $T=24,36$, and our approach performs equally well in terms of size, power, and selection when $n=5000$. When $T$  becomes larger and $n=500,1000$, the test  becomes  oversized (up to 8 percentage points above 5\%), hence more liberal, but the selection procedure still selects  the correct $k$ most of the time. We get estimated proportions around 95\% (with a range (2,3), (10,12) for $k=2, 10$) for $n=500$, and 100\% for the other cases. We still get 100\% for $n=5000$ and $T=60$ but the estimated proportions drop at 85\% (range 2,6)) and 74\% (range (10,13)) when $n= 500$, and 93\% (range (2,3)) and 95\% (range (10,12)) when $n= 1000$. When $T$ becomes large,  the number of degrees of freedom becomes large, and the performance deteriorates under small $n$. 

\begin{table}  
	
	
	\hspace{-0.3cm}
	\begin{tabular}{|c|ccc|ccc|ccc|ccc|}
		\hline
		& \multicolumn{3}{c|}{Size (\%)} & \multicolumn{3}{c|}{Global Power (\%)}  & \multicolumn{3}{c|}{Local Power (\%)}
		& \multicolumn{3}{c|}{ $1_{\{\hat{k}_{LR}=2\}}$} \\
		$T$ &    6   &   12   &  24    &     6   &    12  & 24  &    6    &    12   & 24  &  6    &    12   & 24      \\ \hline
		$n=500$  & 6.0   & 5.2  & 6.7  &   100 & 100     & 100  & 80 & 100 & 100 & 0.98 & 0.99 & 0.95 \\
		& (2.8)  & (0.3) & (0.4) & (0.1)  & (0.0)   & (0.0) & (20.5)& (0.0) & (0.0) & (2,3)& (2,3) & (2,3) \\
		$n=1000$ &   5.6 &  4.9 & 5.5  &  100  & 100    & 100  & 81 &  100 & 100 & 0.99 &  0.98 & 1.00  \\
		& (2.3)  & (0.3) & (0.3) & (0.0)  & (0.0)   & (0.0) & (21.1) & (0.0) & (0.0) & (2,3) & (2,3) & (2,2) \\
		$n=5000$ &  5.3  & 5.0  & 4.9  &   100  & 100    & 100  & 85 & 100 & 100 & 1.00 & 1.00 & 1.00 \\
		& (0.9)  & (0.3) & (0.3) & (0.0)   & (0.0)   & (0.0)  & (20.4) &(0.0) & (0.0)  & (2,2) &(2,2) & (2,2) \\ \hline
	\end{tabular}
	\caption{For each sample size combination $(n,T)$, we provide the average size and power in \%  for the statistic $LR(2)$ (first three panels), and the average of the indicator function $1_{\{\hat{k}_{LR}=2\}}$ with the estimated number $\hat{k}_{LR}$  of factors  obtained by sequential testing (last panel). Nominal size is $5\%$ for the first three panels, and $\alpha_n = 10/n$ for the last panel. Global power refers to the global alternative $\bar{\kappa}=0$, and local power refers to the local alternative $\bar{\kappa}=0.5$. In parentheses, we report the standard deviations for size and power (first three panels), and the range of $\hat{k}_{LR}$ (last panel) across $100$ different draws of the factor path.}
	\label{MC:Table1}
\end{table}

\section{Empirical application to short panels of stock returns} \label{appli}

We showcase relevance of the LR statistic for testing hypotheses about the number of latent factors driving stock returns in short subperiods of the Center for Research in Securities Prices (CRSP) panel. The FA likelihood ratio (LR) statistic compares the log-likelihood of a $k$-factor model, $V_y = FF' + V_\varepsilon$ (where $F$ is $T\times k$ and $V_\varepsilon$ is diagonal and positive definite), with that of the unrestricted model $V_y$. Hence, it can be interpreted as a goodness-of-fit test (Anderson, 2003, Chapter 14). Our empirical analysis starts by identifying the smallest value of $k$ for which this representation holds for the covariance matrix of stock returns. We implicitly assume that there exists an admissible $k$ for which such representation holds. If not, our test should reject for all values of $k$. Then, we decompose the cross-sectional variance of returns into systematic and idiosyncratic components. We also check whether there is spanning between the estimated latent factors and standard observed factors.

\subsection{Testing for the number of latent factors} \label{nbfact}

\begin{figure}[H]
	\footnotesize
	\begin{center}
		\vspace{0cm}  \caption{\small We display the p-values for the statistic $LR(k)$ for the subperiods from January 1963 to December 2021,  stopping at the smallest $k$ such that $H_0(k)$ is not rejected at level $\alpha_n=10/n_{max}$. If no such $k$ is found then p-values are displayed up to $k_{max}$. We use rolling windows of $T=20$ months moving forward by $12$ months each time. The first bar of p-values covers the whole 20 months. Other bars cover the last 12 months of the 20 months subperiod. We flag bear market phases with grey shaded vertical bars.  
		}
		\label{figure:pval}
		 \includegraphics[scale=1,width=.9\textwidth]{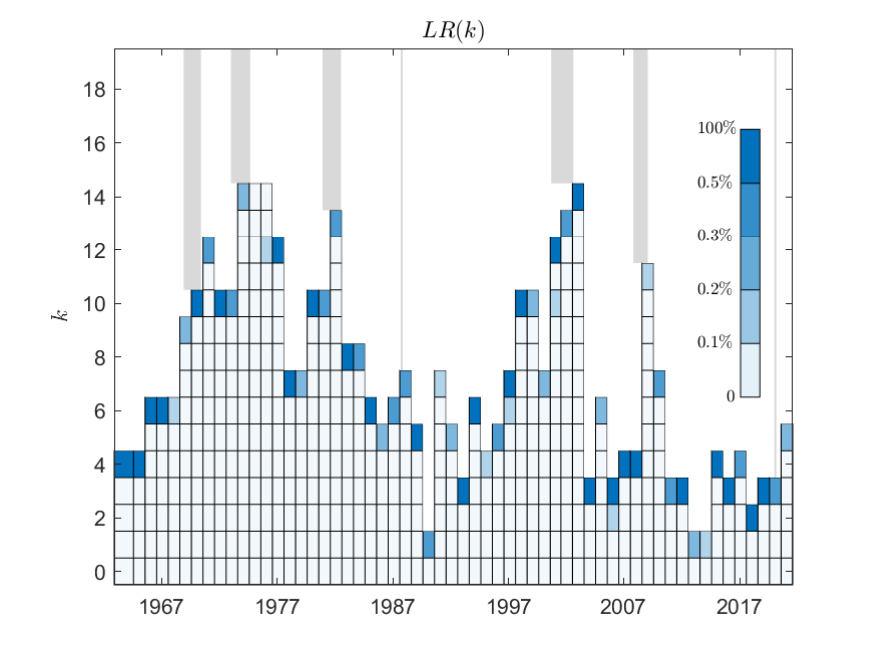}
		
	\end{center}
\end{figure}

We consider monthly returns of U.S. common stocks trading on the NYSE, AMEX or NASDAQ between January 1963 and December 2021, and having a non-missing Standard Industrial Classification (SIC) code. We partition subperiods into bull and bear market phases according to the classification methodology of Lunde and Timmermann (2004). We fix their parameter values $\lambda_1=\lambda_2=0.2$ for the classification based on the nominal S\&P500 index. Bear periods are close to NBER recessions. We implement the tests using a rolling window of $T=20$ months, moving forward 12 months each time (adjacent windows overlap by 8 months), thereby ensuring that we can test up to $14$ latent factors in each subperiod. The size of the cross-section $n$ ranges from $1768$ to $6142$, and the median is $3680$. We only consider stocks with available returns over the whole subperiod, so that our panels are balanced. 
In each subperiod, we sequentially test $H_0(k)$ vs\  $H_1(k)$, for $k=0,\dots,k_{max}$, where $k_{max}=14$ is the largest nonnegative integer such that $df>0$ (see Table \ref{maxk} in OA). We compute the variance-covariance estimator $\hat{\Omega}_{{Z}^*}$ in Proposition  \ref{thm:asy:tests} using a block structure implied by the partitioning of stocks by the first two digits of their SIC code. The number of blocks ranges from 61 to 87 over the sample, and the number of stocks per block ranges from 1 to  641. The median number of blocks is 76 and the median number of stocks per block is 21. We display the p-values of the statistic $LR(k)$ over time for each subperiod in Figure \ref{figure:pval}, stopping at the smallest $k$ such that $H_0(k)$ is not rejected at level $\alpha_n=10/n_{max}$, where  $n_{max}$ is the largest cross-sectional sample size over all subperiods, so that  $\alpha_n=0.16\%$ in our data. If no such $k$ is found then p-values are displayed up to $k_{max}$. The $n$-dependent size adjustment controls for the over-rejection problem induced by sequential testing (see Section  \ref{decomposition} below). We use the same $\alpha_n$ across all windows to prevent variations in the estimated number of factors that arise solely from changes in test size. Besides, the use of $n_{max}$ ensures a conservative procedure. Overall, the results point to a higher number of latent factors during bear market phases compared to bull market phases and a decrease of the number of factors over time.\footnote{We also investigate stability of the factor structure by dividing each window of 20 months into two overlapping subperiods of 16 months (overlap of 12 months) and by estimating canonical correlations between the betas in each subperiod (see SMC). We find that the fraction of the number of latent factors which are common factors is 1 in 70\% of the windows. The fraction is between 0.8 and 1 in 25\% of the windows. It is between 0.5 and 0.8 in the remaining periods.}  It remains true for the three-month recession periods 1987/09-1987/11 and 2020/01-2020/03, which represent only a fraction of their respective subperiods, although there are "bull" market periods finding a similar number of latent factors. In particular, our results based on a fixed $T$ and large $n$ approach contradict the comprehension of a single factor model during market downturns due to estimated correlations between equities approaching $1$. It is consistent with the presence of risk factors, such as tail risk or liquidity risk, only showing in distress periods and yielding a more heterogeneous behaviour among stocks and industries (fire sales during the 2008 global financial crisis for example). Factors can include scaled factors induced by time-varying factor loadings
(see Cochrane (2005) for a discussion) and weak factors since the LR test has local asymptotic power against them (see Section  \ref{local:sec}).
A rise in the estimated $k$ often happens towards the end of the recession periods. It is consistent with the methodology  of Lunde and Timmermann (2004) being early in detecting bear periods (early warning system). The average estimated number of factors is around $7$, close to the 4 to 6 factors found by PCA in Bai and Ng (2006b) on large time spans of individual stocks. With fixed $T$, the selection procedure of  Zaffaroni (2025), being  by construction more conservative  than a (multiple) testing procedure (see the discussion on p.\ 508 of Gagliardini, Scaillet, and Ossola (2019)), yields a smaller number of factors. As a robustness check, we have also repeated the testing procedure using different significance levels $\alpha$. Setting $\alpha = 0.5\%$ yields an average of 8 factors, while $\alpha = 0.05\%$ yields an average of 6 factors. In all cases, the pattern of a higher number of factors during recessions remains unchanged. As discussed in Section \ref{discu}, inference can differ substantially if we wrongly assume Gaussian errors or sphericity (PCA). We find  that imposing cross-sectional independence (resp., Gaussianity and cross-sectional independence) for the LR test gives most of the time an increase by 1 or 2 (1 or 3). We have an average increase of 7 under sphericity (PCA) as expected from the massive over-rejection reported in Section \ref{spheri_imp} for Monte Carlo experiments
when errors are not spherical.

\subsection{Decomposing the cross-sectional variance} \label{decomposition}

\begin{figure}
	\footnotesize
	\begin{center}
		\vspace{-2cm}  \caption{\small  We display $\left(\overline{\hat{V}}_{y}\right)^{1/2}$  for total cross-sectional volatility, $\left(\overline{\hat F' \hat F}\right)^{1/2}$ for systematic volatility,  $\left(\overline{\hat V}_{\varepsilon}\right)^{1/2}$ for idiosyncratic volatility, as well as $\hat{R}^2$ and $\hat{R}^2$ under a single-factor model for the subperiods from January 1963 to December 2021. We flag bear market phases with grey shaded vertical bars and use the same rolling windows as in Figure \ref{figure:pval}. 
		}
		\label{sys_idio}
		\includegraphics[scale=1,width=1\textwidth]{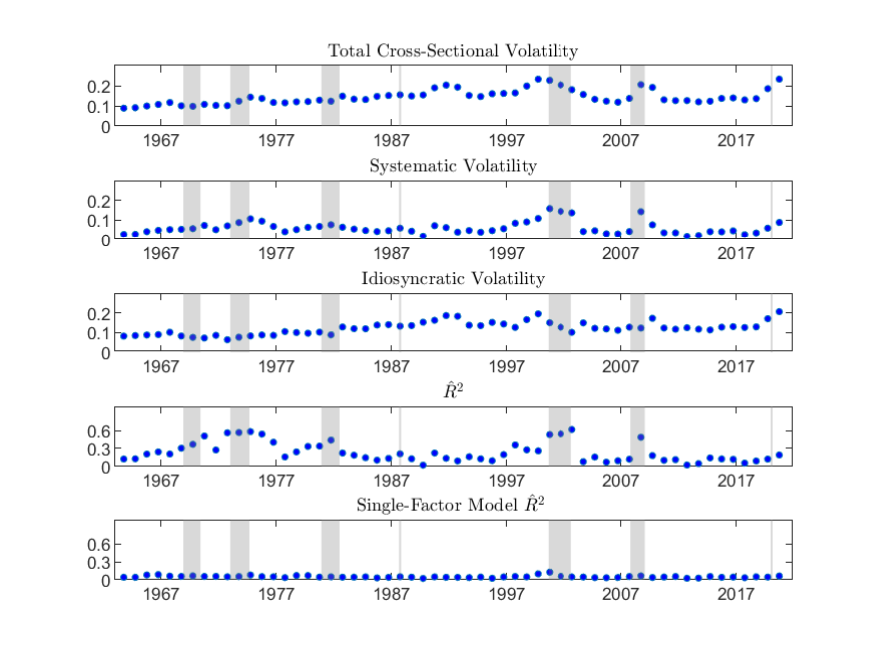}
		
	\end{center}
\end{figure}

We can use the $p$-values displayed in Figure \ref{figure:pval} to obtain a consistent estimator of the number of latent factors  in each subperiod by allowing the asymptotic size $\alpha$ go to zero as $n\to \infty$ in a sequential testing procedure (e.g., P{\"o}tscher (1983) or Robin and Smith (2000) Section 5). Indeed, let $\hat{k}$ be defined as the smallest nonnegative integer $k$ satisfying $\text{pval}(k)>\alpha_n$, where $\text{pval}(k)$ is the p-value from testing $H_0(k)$, and $\alpha_n$ is a sequence in $[0,1]$ with $\alpha_n\to 0$. In the previous section, we set $\alpha_n = 10/n_{max}$. This choice satisfies the theoretical rule $\log \alpha_n/n \to 0$ given in P{\"o}tscher (1983). It also aims at minimizing Type I errors induced by multiple testing across subperiods.
If no such $k$ is found after sequentially testing $H_0(k)$, for $k=0,\dots, k_{max}$ at level $\alpha_n$, then we take $\hat{k}=k_{max}+1$. The Monte Carlo results  show that such a selection procedure works well. We use the estimate $\hat{k}$ at each subperiod to decompose the path of the cross-sectional variance of stock returns into its systematic and idiosyncratic parts: $\hat{V}_{y, tt} = \hat f_t' \hat f_t + \hat V_{\varepsilon,tt}$, where $\hat{F}$ and $\hat{V}_\varepsilon$ are the FA estimates obtained by extracting $\hat{k}$ latent factors. The condition (FA1) ensures that the decomposition holds for any $t$. Such a decomposition is invariant to the choice of normalization for the latent factors. 
If we look at time averages on a subperiod, we get the decomposition $\overline{\hat{V}}_{y} = \overline{\hat f' \hat f} + \overline{\hat V}_{\varepsilon}$, where the overline indicates averaging $\hat{V}_{y, tt} = \hat f_t' \hat f_t + \hat V_{\varepsilon,tt}$ on $t$. In Figure  \ref{sys_idio}, the blue dots  correspond to  the square root of those quantities for the volatilities, while  the ratios $\hat R^2 = \overline{\hat f' \hat f}/\overline{\hat{V}}_{y}$ and $\hat R^2$  under a single-factor model in the two last panels give measures of goodness-of-fit. We do not plot the whole paths date $t$ by date $t$, but only averages,  for readability.  If we sum over time instead of averaging the estimated variances, we get a quantity similar to an integrated volatility (see e.g.\ Barndorff-Nielsen and Shephard (2002), Andersen et al.\ (2003), and references in A{\"i}t-Sahalia and Jacod (2014)), and $\hat R^2$ is the ratio of such quantities. We can observe an uptrend in total and idiosyncratic volatilities, while the systematic volatility appears to remain stable over time even if  the number of factors has overall decreased over time.\footnote{Cross-sectional independence increases estimated systematic risk in average by 0.6\% and decreases estimated idiosyncratic volatility in average by 0.5\%, so that estimated $R^2$ is inflated in average by 4\% per month.  We get the same magnitude under cross-sectional independence and Gaussianity. Under sphericity, PCA estimates give an increase of 2\%, a decrease  of 6\%, and an increase of 13\% for the same quantities.
}
As a result, $\hat{R}^2$ is lower on average after the year 2000, indicating a noisier environment.
During the 2007-2008 financial crisis, we can observe a rise in systematic volatility, causing  $\hat{R}^2$ to reach 59\% during that period.   In bear markets, $\hat R^2$ is often higher. It means that over a bear subperiod, the systematic risk explains a large part of the cross-sectional total variance even if  it is not driven by a single factor as reported in Section \ref{nbfact}. Figure  \ref{sys_idio} also signals that $\hat{R}^2$ under a single-factor model can be way below the one given by the multifactor model. It also means that the idiosyncratic volatility is overestimated if we use a single latent factor only.
The plots of the equal-weighted market and firm volatilities used as measures of total and idiosyncratic volatility from a single observed factor (CAPM) decomposition in Campbell et al.\ (2023) show similar patterns as our panels in Figure  \ref{sys_idio}. As in Campbell et al.\ (2023), we have also made the estimation on value-weighted returns and we confirm that the results are qualitatively  similar.
Section 4 of Campbell et al.\ (2023) discusses economic forces (firm fundamentals and investor sentiments)  driving the observed time-series variation in average idiosyncratic volatility. 

\begin{figure}[H]
	\footnotesize
	\begin{center}
		\caption{\small We display the decomposition of the $\hat k$ factor contributions to   $\hat{R}^2$  for the subperiods from January 1963 to December 2021. We flag bear market phases with grey shaded vertical bars and use the same rolling windows as in Figure \ref{figure:pval}.}
		\label{figure:R2decomp}
		\includegraphics[scale=1,width=1\textwidth]{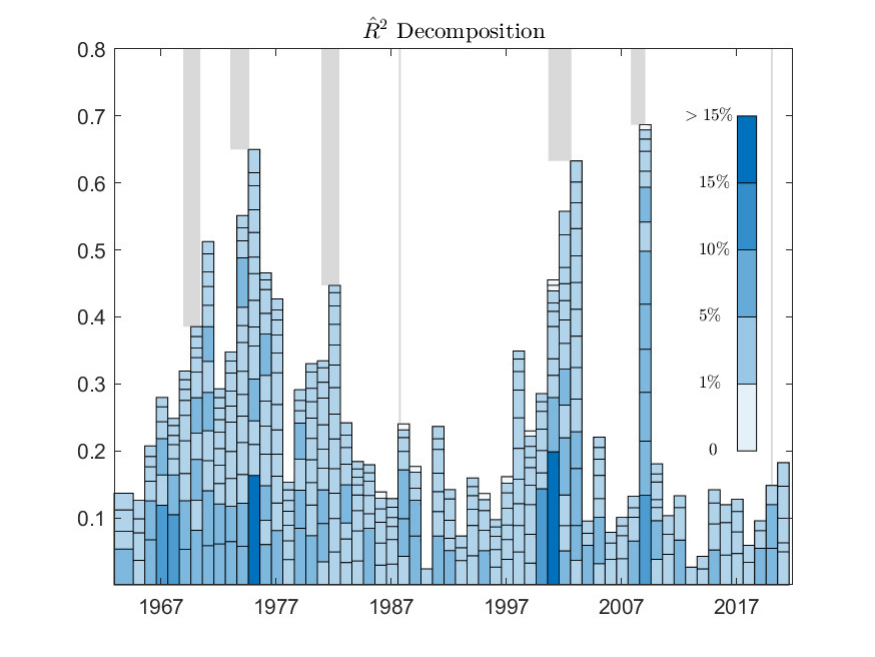}
		
	\end{center}
\end{figure}

The variance decomposition displayed in Figure  \ref{sys_idio} is closely linked to model performance measures routinely computed in empirical asset pricing. Indeed, our FA procedure avoids the estimation of the factor loadings. If we had estimated factor loadings, we would have computed the ``total $R^2$'' measure defined as $\hat R^2_{tot,n} :=  1  - \frac{\sum_i\sum_t (\tilde y_{i,t} - \hat\beta_i'\hat f_t)^2}{\sum_i\sum_t \tilde y_{i,t}^2}$ to gauge the latent model performance (see e.g.\ Kelly, Pruitt and Su (2019)), where we use the cross-sectionally demeaned returns $\tilde y_i$ to get rid of
the intercept terms. One can show then for large $n$ and small $T$ that: \begin{equation}
	\hat R^2_{tot,n} = \frac{1}{\sum_t  V_{y,tt}}\left( \sum_t  f_t' (I_K - AMSE_{\beta})  f_t   \right) + o_p(1), \label{totalR2}
\end{equation}
where $AMSE_{\beta} := \underset{n\rightarrow\infty}{\lim} \frac{1}{n} \sum_{i=1}^n E[ (\hat \beta_i - \beta_i) (\hat \beta_i - \beta_i)']$ is the Asymptotic Mean Squared Error for beta estimation. The estimation risk for betas, which does not necessarily vanish with fixed $T$, decreases the total $R^2$. The upper bound for the total $R^2$ in (\ref{totalR2}) is $\frac{\sum_t  f_t'f_t}{\sum_t  V_{y,tt}}$. It is exactly the quantity that we consistently estimate by our reported measure $\hat R^2$ in Figure  \ref{sys_idio}.
 Besides, we can show that a similar upper bound $\frac{\sum_{j=1}^r \sum_{t}  f_{t,j}^2}{\sum_{t} V_{y,tt}}$ holds if we restrict the total $R^2$ to be computed on the first $r$ factors with $r=1, ..., k$. We report the estimated increases of them in Figure \ref{figure:R2decomp}.  That plot gives a decomposition of the factor contributions to $\hat R^2$. It complements the two lowest  panels of Figure  \ref{sys_idio}  since the value of the first bucket corresponds to $\hat R^2$ under a single factor and
the peak (cumulated increases) corresponds to $\hat R^2$ under $\hat k$ factors.
The quantities 
$\frac{\sum_{j=1}^r \sum_{t}  \hat f_{t,j}^2}{\sum_{t}\hat V_{y,tt}}$  increase gradually with $r$. The marginal increase for $r=\hat k$ ($r=\hat k-1$) w.r.t.\  $r=\hat k -1$ ($r=\hat k -2$)  is similar
irrespective of $\hat k$ being small or large and ranges between 1\% and 5\%.  We find similar values for standard models with observed factors (such as the one used in Section \ref{spann}) on large time spans of individual stocks. Hence, we do not think that the large $\hat k$ observed in market downturns is exclusively due to picking  weak factors since our test has a local asymptotic power against weak factors as shown in Section \ref{locpow}.  Overall, the message of Figure \ref{figure:R2decomp} is that we need several factors to exhaust $\hat R^2$  under $\hat k$ factors and  cannot limit ourselves to the two or three first ones when $\hat k$ is larger. Instead of using $\hat R_{tot,n}^2$, we can use a weighted version defined by $\hat R_{\hat V_{\varepsilon},n}^2 = 1  - \frac{\sum_i (\tilde y_i - \hat F \hat \beta_i)' \hat V_{\varepsilon}^{-1} (\tilde y_i - \hat F \hat \beta_i)}{\sum_i \tilde y_i' \hat V_{\varepsilon}^{-1} \tilde y_i}$, where $\hat \beta_{i} = ( \hat F' \hat V_{\varepsilon}^{-1} \hat F)^{-1} \hat F' \hat V_{\varepsilon}^{-1} \tilde y_i$. Then, one can show that $\hat R_{\hat V_{\varepsilon},n}^2 = \frac{k+ \sum_{j=1}^k \hat \gamma_j}{ T  + \sum_{j=1}^k \hat \gamma_j}$, which extends the usual result linking the unweighted $R^2$ to eigenvalues in standard PCA. Besides, we get a weighted estimated $R^2$ equal to $\frac{r+ \sum_{j=1}^r \hat \gamma_j}{ T  + \sum_{j=1}^r \hat \gamma_j}$ if we compute it on the first $r$ factors with $r=1, ..., k$. A plot of them results in a picture similar to Figure \ref{figure:R2decomp}, with again a marginal increase for $r=\hat k$ ($r=\hat k-1$) w.r.t.\  $r=\hat k -1$ ($r=\hat k -2$) irrespective of $\hat k$ being small or large, but here with decreasing increments when we move to larger $r$. It also means the scree plot of $\hat \gamma_1$, ..,  $\hat \gamma_{\hat k}$ continues to decrease beyond the first few eigenvalues even when $\hat k$ is large.


\subsection{Spanning with observed factors} \label{spann}

As discussed in Bai and Ng (2006b), we get economic interpretation of latent factors with observed factors when we have spanning between the latent factors and the observed factors to be used as proxies in asset pricing (Shanken (1992)). When $n$ and $T$ are large, Bai and Ng (2006b)  (see also Andreou et al.\ (2019, 2025)) exploit the asymptotic normality of the empirical canonical correlations between the two sets of factors to investigate spanning under a symmetric role of the two sets. When $T$ is fixed, we cannot follow that path and we suggest the following strategy based on testing for the rank of a matrix in order to get the information about the number of potentially redundant observed factors. Let us consider $k^O \geq k$ empirical factors that are excess returns of portfolios. If $k^O < k$, empirical factors cannot span the latent space by construction. The condition $k^O \geq k$ eases discussion but is not needed for the rank tests. Let $\hat F^O$ denote the $T \times k^O$ matrix of their values with  row $t$ given by the transpose of  $\hat f^O_t  = \frac{1}{n} \sum_{i=1}^n (y_{i,t} - r_{f,t}) z_{i,t}$, where $\frac{1}{n} z_{i,t}$ is a $k^O \times 1$ vector of time-varying portfolio weights (long or short positions) based on stocks characteristics. Let matrix $F^O$ with rows $f_t^O = \underset{n\rightarrow\infty}{\lim} \frac{1}{n} \sum_{i=1}^n E[(y_{i,t} - r_{f,t}) z_{i,t}]$ be the corresponding large-$n$ population limit. The notation $\hat F^O$ makes clear that the sample average of weighted excess returns is an estimate of the population values $F^O$. We need to take this nontrivial feature  into account in the asymptotic analysis of the rank test statistics when $n \rightarrow \infty$ (see SMC for details).  From the factor model under NA, $y_{i,t} = r_{f,t} + f_t'\tilde \beta_i+ \varepsilon_{i,t}$ (see Section  \ref{sectest}), and assuming cross-sectional non-correlation of idiosyncratic errors and portfolio weights, we get $F^O = F \Phi'$, where $\Phi = \underset{n\rightarrow\infty}{\lim} \frac{1}{n} \sum_{i=1}^n E[z_{i,t}] \tilde \beta_i'$ is assumed independent of $t$,  $t=1, ...,T$. Hence, the range of $F^O$ is a subset of the range of $F$, namely the latent factors span the observed factors (in the population limit sense) by construction. Moreover, $\text{Rank}(F^O) \leq k$. We can test the null hypothesis that $F$ and $F^O$ span the same linear spaces, namely matrices $F$ and $F^0$ have the same range. Such a null hypothesis is equivalent to the rank condition: $\text{Rank}(F^O) = k$. 

We build on the rank testing literature; see e.g.\ Cragg and Donald (1996), Robin and Smith (RS, 2000), Kleibergen and Paap (KP, 2006), Al-Sadoon (2017).  We use in particular the RS and KP statistics. For those tests, the null hypothesis is that a given matrix has a reduced rank $r$ against the alternative hypothesis that the rank is greater than $r$. Hence, to test for spanning by the empirical factors, we  consider the null hypothesis $H_{0,sp}(r): \text{Rank}(F^O) = r$ against the alternative hypothesis $H_{1,sp}(r): \text{Rank}(F^O) > r$, for any integer $r<k$. Spanning holds if we can reject $H_{0,sp}(r)$ for any $r < k$.  Ahn, Horenstein and Wang (2018) use that technology in a fixed-$n$ large-$T$ setting, and find that ranks of beta matrices estimated from either portfolios, or individual stocks, excess returns are often substantially smaller than the (potentially large) number $k^O$ of observed factors. The explanation in large economies  is that the portfolio beta matrices coincide with $\Phi$, and thus they cannot have a rank above the (potentially small) number $k$ of latent factors.

\begin{figure}[H]
	\footnotesize
	\begin{center}
		\caption{\small The upper panel displays the p-values for the  KP statistic for the subperiods from January 1963 to December 2021, for the rank test of the null hypothesis $H_{0,sp}(r)$ that $F^O$ has rank $r$ against the alternative hypothesis of rank larger than $r$, for any integer $r \leq k-1$. The empirical matrix $\hat F^O$ is computed with the time-varying portfolio weights of the FF5 model plus momentum. We stop at the smallest $r$ such that $H_{0,sp}(r)$ is not rejected at level $\alpha_n=10/n_{max}$. If no such $r$ is found then p-values are displayed up to $k-1$. The red horizontal segments give $\hat k-1$, i.e., the estimated number of latent factors obtained from Figure \ref{figure:pval} minus $1$. We flag bear market phases with grey shaded vertical bars, and use the same rolling windows as in Figure \ref{figure:pval}. The lower panel displays the ratio $\hat R^2_O$ to $\hat R^2$ for the FF3, 6 and 14 factor models.}
		\label{figure:spanning}
		\includegraphics[scale=1,width=1\textwidth]{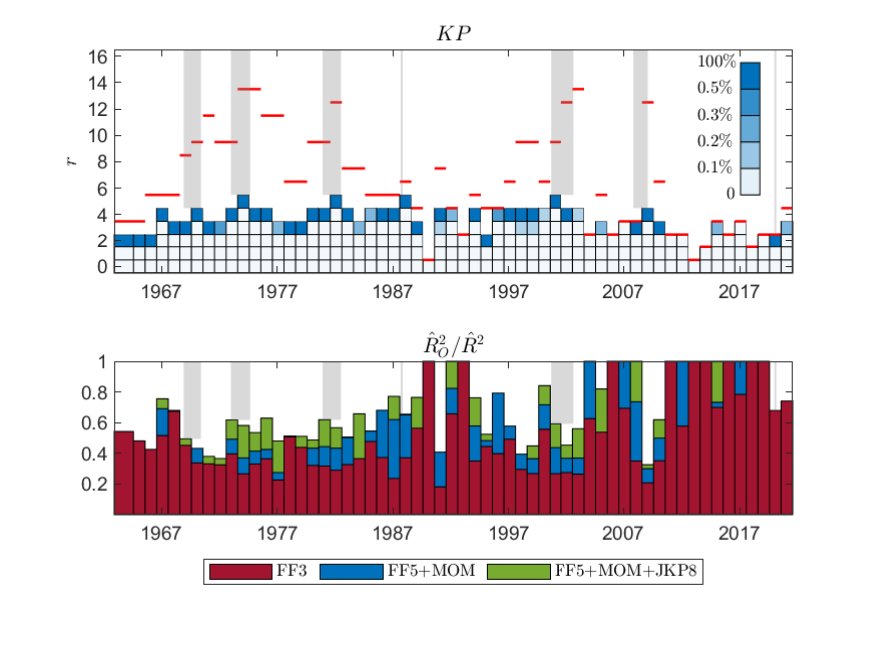}
		
	\end{center}
\end{figure}

We first construct the empirical matrix $\hat F^O$ with the time-varying portfolio weights of the Fama-French five-factor model (Fama and French (2015), FF5) plus the momentum (MOM) factor (Jegadeesh and Titman (1993), Carhart (1997)), i.e., $k^O= 6$, based on value-weighting and June portfolio updates. In the upper panel of Figure \ref{figure:spanning} for the KP statistic, we can observe that the rank tests point most of the time at a low reduced rank $r$ either 2 , 3, or 4, with only occasionally 5, for the matrix $\hat F^O$. The average  (median) rank is around 3 (3). The picture  is similar for the RS statistic. Observed factors struggle spanning latent factors since their associated linear space is of a dimension $r$ smaller than the one of the latent factor space. Given the increase of the estimated number of latent factors documented in Section \ref{nbfact}, we observe a larger gap between the dimensions of the two factor spaces under cross-sectional independence (resp., Gaussianity and cross-sectional independence) and under sphericity. The discrepancy between the dimensions of the two factor spaces has decreased over time. 

In order to better understand how observed factors contribute to the latent factor model fit, we choose the following procedure in our fixed $T$ setting. Recall that the FA normalization of the latent factors and loadings is such that $V_y = F F' + V_{\varepsilon}$ and $F' V_{\varepsilon}^{-1} F$ is diagonal. Then, the population $R^2$ measure is $R^2 = \frac{Tr(F F')}{Tr(V_y)}= \frac{\sum_{t=1}^T f_t'f_t}{\sum_{t=1}^T V_{y,tt}} = \sum_{j=1}^k R_j^2$, where $R_j^2 = \frac{\sum_{t=1}^T f_{t,j}2}{\sum_{t=1}^T V_{y,tt}}$ is the $R^2$ contribution of the $j$th latent factor (see Section \ref{decomposition}).  Any rotation of the latent factors, i.e., $\tilde F = F R$ with $R$ an orthogonal $k \times k$ matrix, keeps the normalization of the loadings to have unit variance, so that $V_y = \tilde F \tilde F' + V_{\varepsilon}$ (albeit $\tilde F' V_{\varepsilon}^{-1} \tilde F$ is not diagonal anymore, i.e., the rotated factors are not orthogonal to each other w.r.t.\ the scalar product induced by $V_{\varepsilon}^{-1}$). The $R^2$ measure is invariant and $R^2 = \frac{\sum_{t=1}^T \tilde f_t' \tilde f_t}{\sum_{t=1}^T V_{y,tt}}$. Among the possible rotations of the latent factors, let us consider one such that the first $r$ latent factors span the space of observed factors of dimension $r$, namely $F^O = F \Phi'$, with $\Phi$ having rank $r \leq k$ and SVD given by $\Phi = U S V'$. Then, the columns of $\tilde F^O := F^O U S^{-1} = F V$ are linearly independent and span the $r$-dimensional space of the observed factors. Now, let $V_{\perp}$ be a $k \times (k-r)$matrix with orthonormal columns that span the orthogonal complement of the range of $V$, so that $R = [V : V_{\perp}]$ is an orthogonal matrix. Then, the first $r$ columns of rotated factors $\tilde F = F R$ coincide with $\tilde F^O$. We define the $R^2$ of the observed factors as the $R^2$ of the first $r$  rotated latent factors, namely $R^2_O :=  \frac{Tr(\tilde F^O \tilde F^{O \prime})}{Tr(V_y)}= \frac{\sum_{t=1}^T (\tilde f_t^O)'\tilde f_t^O}{\sum_{t=1}^T V_{y,tt}} =  \frac{Tr(F V V' F')}{Tr(V_y)}$. It involves the projection matrix $V V'$ onto the orthogonal complement of the kernel of $\Phi$.  Here, we rotate the latent factors, such that the first $r$ rotated factors span exactly the space of the observed factors, and then we define $R^2_O$ as the $R^2$ of those $r$ rotated factors. The decomposition of $R_O^2$ into the contributions of each observed factor is not uniquely defined because the observed factors are redundant when $r < k_O$. In other words, since we can write any column of $F_O$ as linear combination of other $r$ columns, singling-out the independent contribution of any given column is not possible. In the lower panel of Figure \ref{figure:spanning}, we can observe that the estimated contribution $\hat R_O^2$ of the 6 observed factors to $\hat R^2$ is often low in the early periods. If we use a 3 factor model with market, size and value factors (Fama and French (1993), FF3), it often provides good explanatory power in recent years.
The additional 3 factors do not bring much contribution in recent periods but more in early periods when the FF3 model is not fully explanatory. Both observed factor models struggle when $\hat k$ is large around market downturns. The picture is similar if we use a 4 factor model (FF3 + MOM).

Let us now delve into less traditional factors than the 6 factors considered in the above to check whether they can provide additional explanatory power if needed. We investigate the database assembled by Jensen et al.\  (2023). The database targets factors based on 153 stock characteristics gathered into 13 clusters made of ($x$) characteristics: Accruals (6), Debt Issuance (5), Investment (19), Low Leverage (4), Low Risk (13), Momentum (8), Profit Growth (7), Profitability (10), Quality (12), Seasonality (10), Short-Term Reversal (4), Size (5), Value (16).  
Jensen et al.\  (2023) find that the average within-cluster pairwise correlation is above 0.5 for nine out of 13 clusters. The detected redundancy (see also Feng et al.\ (2020)) signifies that the zoo of factors (Cochrane (2011)) is limited to a few representative animals. 
In order to construct an expanded set of factors, we pick one factor in each cluster not represented by the original set of factors so that we reach 14 factors in total. The 8 additional factors correspond to change in current operating working capital, growth in book debt (3 years), book leverage, market beta, change in sales minus change in Selling, General \& Administrative expenses, operating leverage, years 2-5 lagged returns (annual), and short-term reversal (see Jensen et al.\  (2023) and references therein for a detailed description). Because of the redundancy, the empirical results in the below are similar for other choices of factors within each cluster.
This procedure allows to  get  the rank of  $F^O$ closer to $\hat k$ when needed, but only by 1 or 2 additional factors w.r.t.\ FF5 + MOM. It means that less traditional factors slightly help
to reconcile the large number of latent  factors found in market downturns and distressed times.
The estimated contribution $\hat R_O^2$ of the observed factors to $\hat R^2$ is larger around those times, but only by an increase of about 10 percentage points in $\hat R_O^2/\hat R^2$ as seen in the lower panel of Figure \ref{figure:spanning} for FF5 + MOM + JKP8. Hence, the empirical message with 14 observed factors is mostly unchanged w.r.t.\ FF5 + MOM. Building a non-linear single-factor asset pricing model instead of relying on standard  construction with long-short portfolios might be a promising avenue (Borri et al.\ (2024)).


\section{Concluding remarks\label{section:Concluding remarks}}
In this paper, we develop a new theory of latent Factor Analysis in short panels beyond the Gaussian and i.i.d.\ cases. We establish the AUMPI property of the LR statistic for testing hypotheses on the number of latent factors. Our results for short subperiods of the CRSP panel of US stock returns contradict the comprehension of a single factor during market downturns. In bear markets, systematic risk driven by a  latent multifactor structure explains a large part of the
cross-sectional variance, and is not spanned by traditional empirical factors with a discrepancy between the dimensions of the two factor spaces decreasing over time. Less traditional factors help in providing a better fit, but only marginally.
The estimated paths of total and idiosyncratic volatilities month after month feature an uptrend through time. 

\section*{References}

\noindent Ahn, S., and Horenstein, A., 2013. Eigenvalue ratio test for the number of factors. Econometrica 81 (3), 1203-1227.

\noindent Ahn, S., Horenstein, A., and Wang, N., 2018. Beta matrix and common factors in stock returns. Journal of Financial and Quantitative Analysis 53 (3), 1417-1440.

\noindent  Ahn, S., Lee, H.,  and Schmidt, P., 2001. GMM estimation of linear panel data models with time-varying individual effects. Journal of
Econometrics 101 (2), 219-255.

\noindent   Ahn, S., Lee, H., and Schmidt, P., 2013. Panel data models with multiple time-varying individual effects. Journal of Econometrics 174 (1), 1-14.

\noindent Aigner, D., Hsiao, C., Kapteyn A., and Wansbeek, T., 1984. Latent variable models  in econometrics, in Handbook of Econometrics, Volume II, Z.\ Griliches and M.D.\ Intriligator Eds., 1321-1393. 

\noindent  A{\"i}t-Sahalia, Y., and Jacod, J., 2014. High-frequency financial econometrics. Princeton University Press.

\noindent Al-Sadoon, M., 2017. A unifying theory of tests of rank. Journal of Econometrics 199 (1), 49-62.

\noindent  Ang, A.,  Liu, J., and Schwarz, K., 2020. Using stocks or portfolios in tests of factor models. Journal of Financial and Quantitative Analysis 55 (3), 709-750.

\noindent Andersen, T., Bollerslev, T., Diebold, F., and Labys, P., 2003. Modeling and forecasting realized volatility. Econometrica 71 (2), 579-625.

\noindent Andersen, T., Ding, Y., Todorov, V., and Yiu, S., 2025. The factor structure for jump risk.  Northwestern University working paper.

\noindent Anderson, T. W., 1963. Asymptotic theory for Principal Components Analysis. Annals of Mathematical Statistics 34, 122-148.

\noindent Anderson, T. W., 2003. An introduction to multivariate statistical analysis. Wiley.

\noindent Anderson, T. W., and Rubin, H., 1956. Statistical inference in factor analysis. Proceedings of the Third Berkeley Symposium in Mathematical Statistics and Probability 5, 11-150.

\noindent Anderson, T. W. and Amemiya, Y., 1988. The asymptotic normal distribution of estimators in factor analysis under general conditions. Annals of Statistics 16 (2), 759-771. 

\noindent Ando, T., and  Bai, J., 2015. Asset pricing with a general multifactor structure. Journal of Financial Econometrics 13 (3), 556-604.

\noindent Andreou, E., Gagliardini, P., Ghysels, E. and Rubin, M., 2019. Inference in group factor models with an application to mixed frequency data. Econometrica 87 (4), 1267-1305.

\noindent Andreou, E., Gagliardini, P., Ghysels, E. and Rubin, M., 2025. Spanning latent and observable factors. Journal of Econometrics 248, 105743.

\noindent Andrews, D., 2001. Testing when a parameter is on the boundary of the maintained hypothesis. Econometrica 69 (3), 683-734.

\noindent  Bai, J., 2003. Inferential theory for factor models of large dimensions. Econometrica 71 (1), 135-171.

\noindent Bai, J., 2009. Panel data models with interactive effects. Econometrica 77 (4), 1229-1279.

\noindent Bai, J., and Li, K., 2012. Statistical analysis of factor models of high dimension. Annals of Statistics 40 (1),
436-465.

\noindent Bai, J., and Li, K., 2016. Maximum likelihood estimation and inference for approximate factor models of high dimension. Review of Economics and Statistics 98 (2),
298-309.

\noindent Bai, J., and Ng, S., 2002. Determining the number of factors in approximate factor models. Econometrica 70 (1),
191-221.

\noindent Bai, J. and Ng, S., 2006a. Confidence intervals for diffusion index forecasts and inference for factor-augmented regressions. Econometrica 74 (4), 1133-1150.

\noindent Bai, J., and Ng, S., 2006b. Evaluating latent and observed factors in macroeconomics and finance. Journal of Econometrics 131 (1),
507-537.

\noindent Barigozzi, M., and Hallin, M., 2016. Generalized dynamic factor models and volatilities: recovering the market volatility shocks. Econometrics Journal 19 (1), 33-60.

\noindent Barigozzi, M., and Hallin, M., 2025. The dynamic, the static, and the weak.
Factor models and the analysis of high-dimensional time series. Forthcoming in Journal of Time Series Analysis.

\noindent  Barndorff-Nielsen, O., and Shephard, N., 2002. Econometric analysis of realized volatility and its use in estimating stochastic volatility models. Journal of the Royal Statistical Society, Series B, 64 (2), 253-280.

\noindent Bell, E. T., 1934. Exponential polynomials. Annals of Mathematics 35 (2), 258-277.

\noindent Bernstein, D., 2009. Matrix mathematics: Theory, facts and formulas. Princeton University Press. 

\noindent Bickel, P. J., and Levina, E., 2008. Covariance regularization by thresholding. Annals of Statistics 36 (6), 2577-2604.

\noindent Bollerslev, T.,  Li, J., Li, Y., and Zhang, Q., 2025. Illuminating important economic news by candlesticks:
Optimal testing meets technical analysis. Duke University  working paper .

\noindent Borri, N.,  Chetverikov, D., Liu, Y., and Tsyvinski, A., 2024. One factor to bind the cross-section of returns.  Yale University working paper .

\noindent Campbell,  J., Lettau, M., Malkiel, B., and Xu, Y., 2023. Idiosyncratic equity risk two decades later. Critical Finance Review 12, 203-223.

\noindent Caner, M., and Han, X., 2014. Selecting the correct number of factors in approximate factor models: the large panel case with group bridge estimator. Journal of
Business and Economic Statistics 32 (3), 359-374.

\noindent Carhart, M., 1997. On persistence in mutual fund performance. Journal of Finance 52 (1), 57-82. 

\noindent  Cattell, R., 1966. The scree test for the number of factors. Multivariate Behavioral Research 1 (2), 245-276.

\noindent Chamberlain, G., 1992. Efficiency bounds for semi-parametric regression. Econometrica 60 (3), 567-596.

\noindent  Chamberlain, G., and Rothschild, M., 1983. Arbitrage, factor structure, and mean-variance analysis on large asset markets. Econometrica 51 (5), 1281-1304.

\noindent  Chen, Q., and Fang, Z., 2019. Improved inference on the rank of a matrix. Quantitative Economics 10, 1787-1824.

\noindent  Choi, S., Hall, W., and Schick, A., 1996. Asymptotically uniformly most powerful tests in parametric and semiparametric models. Annals of Statistics 24 (2), 841-861.

\noindent  Cochrane, J., 2005. Asset pricing. Princeton University Press.

\noindent  Cochrane, J., 2011. Presidential address: Discount rates.  Journal of Finance 66 (4), 1047-1108.

\noindent  Connor, G., and Korajczyk, R., 1986. Performance measurement with the arbitrage pricing theory: A new framework for analysis. Journal of  Financial Economics 15 (3), 373-394.

\noindent Connor, G., and Korajczyk, R., 1993. A test for the number of factors in an approximate factor model. Journal of Finance 48 (4), 1263-1291.

\noindent Cragg, J., and Donald, S., 1996. On the asymptotic properties of LDU-based tests of the rank of a matrix. Journal of the American Statistical Association 91 (435), 1301-1309.

\noindent Eaton, M., 1987. Multivariate statistics. A vector space approach. Institute of Mathematical Statistics Lecture notes-Monograph series, Vol.\ 53. 

\noindent Engle, R., 1984. Wald, likelihood ratio, and Lagrange multiplier tests in econometrics, in Handbook of Econometrics, Volume II, Z.\ Griliches and M.D.\ Intriligator Eds., 775-826. 

\noindent Fama, E., and French, K., 1993. Common risk factors in the returns on stocks and
bonds. Journal of Financial Economics 33 (1), 3-56.

\noindent Fama, E., and French, K., 2015. A five-factor asset pricing model. Journal of Financial Economics 116 (1), 1-22.

\noindent  Fan, J., Furger, A., and Xiu, D., 2016. Incorporating global industrial classification standard into portfolio allocation: A simple factor-based large covariance matrix estimator with high-frequency data. Journal of Business and Economic Statistics 34 (4), 489-503.

\noindent Feng, G., Giglio, S., and Xiu, A., 2020. Taming the factor zoo: a test of new factors. Journal of Finance 75 (3), 1327-1370.

\noindent Fortin, A.-P., Gagliardini, P., and Scaillet, O., 2023. Eigenvalue tests for the number of latent factors in short panels. Journal of Financial Econometrics, https://doi.org/10.1093/jjfinec/nbad024.

\noindent Freyberger, J., 2018. Non-parametric panel data models with interactive fixed effects. Review of Economic Studies 85 (3), 1824-1851.

\noindent  Gabaix, X., Koijen, R., Richmond, R., and Yogo, M., 2023. Asset embeddings.  University of Chicago working paper.

\noindent Gagliardini, P., Ossola, E., and Scaillet, O., 2016. Time-varying risk premium in large cross-sectional equity datasets. Econometrica 84 (3), 985-1046.

\noindent  Gagliardini, P., Ossola, E., and Scaillet, O., 2019. A diagnostic criterion for approximate factor structure. Journal of Econometrics 21 (2), 503-521.

\noindent  Gagliardini, P., Ossola, E., and Scaillet, O., 2020. Estimation of large dimensional conditional factor models in finance,  in Handbook of Econometrics, Volume 7A, S.\ Durlauf, L. Hansen, J.\ Heckman, and R.\ Matzkin Eds., 219-282.

\noindent Giglio, S., and Xiu, D., 2021. Asset pricing with omitted factors. Journal of Political Economy, 129, 1947-1990.

\noindent   Gobillon, L., and Magnac, T., 2016. Regional policy evaluation: Interactive fixed effects and synthetic controls. Review of Economics and Statistics 98 (3), 535-551.




\noindent  Hansen, L., 1982. Large sample properties of Generalized Method of Moments estimators.
Econometrica 50, 1029-1054.

\noindent  Hayakawa, K., Pesaran, H., and Smith, V.,  2023. Short $T$ dynamic panel data models with individual, time and interactive effects.
Journal of Applied Econometrics 38, 940-967.



\noindent Jegadeesh, N., and Titman, S., 1993. Returns to buying winners and selling losers: Implications for stock market efficiency. Journal of Finance 48 (1), 65-91. 

\noindent  Jensen, T., Kelly, B., and Pedersen, L., 2023. Is there a replication crisis in finance? Journal of Finance 78 (5), 2465-2518.

\noindent   J{\"o}reskog,  K., 1970. A general method for analysis of covariance structures.  Biometrika 57 (2), 239-251.

\noindent  Kapetanios, G., 2010. A testing procedure for determining the number of factors in approximate factor models with large datasets. Journal of Business and Economic
Statistics 28 (3), 397-409.

\noindent  Kelly, B.,  Pruitt, S., and Su, Y., 2019. Characteristics are covariances: A unified model of risk and return. Journal of Financial Economics 134 (3), 501-524.

\noindent Khatri, C.,1980. Quadratic forms in normal variables, in Handbook of Statistics, Volume 1, P.\ Krishnaiah Ed., 443-469. 

\noindent Kim, M., 2022. Robust inference for diffusion-index forecasts with cross-sectionally dependent data. Journal of Business and Economic Statistics 40 (3), 1153-1167,

\noindent Kim, S., and Skoulakis, G., 2018. Ex-post risk premia estimation and asset pricing tests using large cross-sections: the regression-calibration approach. Journal of Econometrics 204 (2), 159-188.

\noindent  Kleibergen, F., and Paap, R., 2006. Generalized reduced rank tests using the singular
value decomposition. Journal of Econometrics 133 (1), 97-126.

\noindent  Kleibergen, F., and Zhan, Z., 2025. Risk premia from the cross-section of individual assets. Journal of Econometrics 252, 106108.

\noindent Kotz, S., Johnson, N., and Boyd, D., 1967. Series representations of distributions of quadratic forms in Normal variables II. Non-central case. Annals of Mathematical Statistics 38 (3), 838-848. 

\noindent Lancaster, T., 2000. The incidental parameter problem since 1948. Journal of Econometrics 95 (2), 391-413.

\noindent Lehmann, E.,  and Romano, D., 2005. Testing statistical hypotheses, Springer Texts in Statistics.

\noindent Lunde, A., and Timmermann, A., 2004. Duration dependence in stock prices: An analysis of bull and bear markets. Journal of Business and Economic Statistics 22 (3), 253-273.

\noindent Magnus, J., and Neudecker, H., 2007. Matrix differential calculus, with applications in statistics and econometrics. Wiley.

\noindent Maldonado, J. and Ruiz, E., 2021. Accurate confidence regions for principal components factors. Oxford Bulletin of Economics and Statistics 83 (6), 1432-1453.

\noindent Miravete, E., 2011. Convolution and composition of totally positive random variables in economics. Journal of Mathematical Economics 47 (4), 479-490.

\noindent Moon, H.R., and Weidner, M., 2015. Linear regression for panel with unknown number of factors as interactive fixed effects. Econometrica 83 (4), 1543-1579.

\noindent Neyman, J., and Scott, E., 1948. Consistent estimation from partially consistent observations. Econometrica 16(1), 1-32.

\noindent Onatski, A.,  2009. Testing hypotheses about the number of factors in large factor models. Econometrica  77 (5), 1447-1479.

\noindent Onatski, A., 2010. Determining the number of factors from empirical distribution of eigenvalues. Review of Economics and Statistics 92 (4), 1004-1016.

\noindent Onatski, A. 2023. Comment on ``Eigenvalue tests for the number of latent factors in short panels" by A.-P.\ Fortin, P.\ Gagliardini and O.\ Scaillet, Journal of Financial Econometrics, \\ https://doi.org/10.1093/jjfinec/nbad028.

\noindent Pesaran, M.H., 2006. Estimation and inference in large heterogeneous panels with a multifactor error structure. Econometrica 74 (4), 967-1012. 

\noindent P{\"o}tscher, B., 1983. Order estimation in ARMA-models by Lagrangian multiplier tests, Annals of
Statistics 11 (3), 872-885.

\noindent Raponi, V., Robotti, C., and Zaffaroni, P., 2020. Testing beta-pricing models using large cross-sections. Review of Financial Studies 33 (6), 2796-2842.

\noindent Renault, E., Van Der Heijden, T., and Werker, B., 2023. Arbitrage pricing theory for idiosyncratic variance factors. Journal of Financial Econometrics 21 (5), 1403-1442.

\noindent  Robin, J.-M., and Smith, R., 2000. Tests of rank. Econometric Theory 16 (2), 151-175.

\noindent Romano, J., Shaikh, A., and Wolf, M., 2010. Hypothesis testing in econometrics.  Annual Review of Economics 2, 75-104.

\noindent  Ross, S., 1976. The arbitrage theory of capital asset pricing, Journal of Economic Theory
13 (3), 341-360. 

\noindent  Satorra, A., and Bentler, P.,  2001. A scaled difference chi-square test statistic for
moment structure analysis. Psychometrika 66, 507-514.

\noindent Satorra, A., and Bentler, P., 2010. Ensuring positiveness of the scaled difference chi-square test statistic. Psychometrika 75, 243-248.

\noindent Shanken, J. 1992. On the estimation of beta pricing models. Review of Financial Studies 5 (1), 1-33.

\noindent Stock, J., and Watson, M., 2002. Forecasting using principal components from a large number of predictors.
Journal of the American Statistical Association 97 (460), 1167-1179.

\noindent Tao, T. 2012. Topics in random matrix theory. Graduate Studies in Mathematics, Volume 132, American Mathematical Society.

\noindent White, H., 1982. Maximum likelihood estimation of misspecified models. Econometrica 50 (1), 1-25.

\noindent Zaffaroni, P., 2025. Factor models for conditional asset pricing. Journal of Political Economy 133 (8), 2615-2642. 

\noindent Zheng, B., and  Bentler, P., 2024. Enhancing model fit evaluation in SEM: Practical
tips for optimizing chi-square tests. Structural Equation Modeling: A Multidisciplinary
Journal 32, 136-141.

\bigskip

\newpage


\newpage

\newpage

 $ $
\newpage

\begin{center}
\Large \textbf{ONLINE APPENDIX}\\
Latent Factor Analysis in Short Panels\\
Alain-Philippe Fortin, Patrick Gagliardini, and Olivier Scaillet
\end{center}

We give proofs of Propositions  \ref{thm:asy:tests}-\ref{thm:AUMPI} of the paper in Section A.  We also prove Lemmas \ref{prop:1}-\ref{prop:sequence:lambda:general} supporting them. We provide additional theory in Appendix B, namely the characterization of the pseudo likelihood and the PML estimator (B.1), the conditions for global identification and consistency (B.2), the asymptotic expansions for the FA estimators (B.3), the local analysis of the first-order conditions of FA estimators (B.4), the asymptotic normality of FA estimators (B.5), the definition of invariant tests (B.6), 
and proofs of additional lemmas (B.7). We give numerical checks of Inequalities (\ref{cond:MLR}) of Proposition   \ref{thm:AUMPI} in Appendix C.  
Finally,  we collect   the maximum value of $k$ as a function of $T$ in Appendix D.

\section*{A Proofs}
\subsection*{Proofs of Propositions  \ref{thm:asy:tests}-\ref{thm:AUMPI}}

\setcounter{equation}{0}\def\theequation{A.\arabic{equation}}

\textbf{Proof of Proposition \ref{thm:asy:tests}:} (a) The proof of this part is made in three steps. (i) We first establish the link between the LR statistic and the norm of matrix $\hat{S} = \hat{V}_{\varepsilon}^{-1/2}M_{\hat{F},\hat{V}_{\varepsilon}} ( \hat{V}_y  - \hat{V}_{\varepsilon} ) M_{\hat{F},\hat{V}_{\varepsilon}}'  \hat{V}_{\varepsilon}^{-1/2}$, namely we prove $LR(k) = \frac{n}{2} \Vert \hat S \Vert^2  + o_p(1)$. The next lemma is instrumental to this step.
\begin{lemma} \label{prop:1} Under Assumption \ref{Ass1},
(a) the eigenvalues of matrix $\hat{S}$ are: $\hat{\gamma}_j$, for $j=k+1,...,T$, and $0$, with multiplicity $k$, where $1 + \hat{\gamma}_j$ for $j=k+1,...,T$ are the $T-k$ smallest eigenvalues of $\hat{V}_y \hat{V}_{\varepsilon}^{-1}$, (b) the squared Frobenius norm is $\Vert \hat{S} \Vert^2 = \sum_{j=k+1}^T  \hat{\gamma}_j^2$, and (c)  $diag( \hat{S} )=0$. 
\end{lemma}

\noindent Then, we apply a second-order expansion of the log function in the RHS of (\ref{Definition:statistics}). The first-order term vanishes because $\sum_{j=k+1}^T \hat \gamma_j = tr( \hat S) = 0$ by Lemma \ref{prop:1} a) and c). The second-order term equals $\frac{n}{2} \Vert \hat S \Vert^2$ by Lemma \ref{prop:1} b). The remainder (third-order) term  is $o_p(1)$ because we have $\sqrt n \hat \gamma_j = O_p(1)$ for $j=k+1,...,T$. This bound results from the expansion of the sample covariance:
\begin{equation} \label{proof:asy:exp:Vhat}
\hat{V}_y = \tilde V_y +\frac{1}{\sqrt{n}}\Psi_y+o_p(\frac{1}{\sqrt{n}}) = \tilde V_y +O_p(\frac{1}{\sqrt{n}}), 
\end{equation}
where $\tilde V_y := F F' + \tilde{V}_\varepsilon$ and $\Psi_y := \frac{1}{\sqrt{n}} ( \varepsilon \beta F' +  F \beta' \varepsilon') + \sqrt{n} \left( \frac{1}{n} \varepsilon \varepsilon' - \tilde{V}_{\varepsilon} \right)$, see Equation  (\ref{formula:Vhaty}) and Lemma \ref{lemma:conv:espilon:beta} in Appendix B.2, and $\hat V_{\varepsilon} = \tilde V_{\varepsilon} + O_p(\frac{1}{\sqrt{n}})$, see Equation (\ref{asy:FA:estimators}). Then, $\hat V_y \hat V_{\varepsilon}^{-1} = \tilde V_y \tilde V_{\varepsilon}^{-1} + O_p(\frac{1}{\sqrt{n}})$, matrix $\tilde V_y \tilde V_{\varepsilon}^{-1} $ has unit eigenvalues for order $j=k+1,...,T$, and the eigenvalues of matrices $\hat V_y \hat V_{\varepsilon}^{-1}$ and $\tilde V_y \tilde V_{\varepsilon}^{-1}$ differ by quantities of order $O_p(\frac{1}{\sqrt{n}})$ by Weyl's inequalities.

(ii) Next, let us establish the asymptotic expansion of $n \Vert \hat S \Vert^2$ in order to show equation (\ref{asy:exp:stats}). Since $\hat{G}\hat{G}'\hat{V}^{-1}_\varepsilon=M_{\hat{F},\hat{V}_\varepsilon}$, we have $\hat{S}=\hat{V}^{-1/2}_\varepsilon\hat{G}\hat{S}^*\hat{G}'\hat{V}^{-1/2}_\varepsilon$, where $\hat{S}^*=\hat{G}'\hat{V}^{-1}_\varepsilon(\hat{V}_y-\hat{V}_\varepsilon)\hat{V}^{-1}_\varepsilon\hat{G}$. Besides, we have $0=diag(\hat{S})$ (see Lemma \ref{prop:1} (c)). Therefore, 
$0=diag(\hat{S})=\hat{V}^{-1}_\varepsilon diag(\hat{G}\hat{S}^*\hat{G}') =2\hat{V}^{-1}_\varepsilon\hat{\boldsymbol{X}}'vech(\hat{S}^*)$, i.e., $vech(\hat{S}^*)$ is in the orthogonal complement of the range of $\hat{\boldsymbol{X}}$. \footnote{To see this step, write $\hat{G}=(\hat{g}_{t,i})=[\hat{g}_1:\cdots : \hat{g}_{T-k}]$.  By definition of the $vech$ operator,  $vech(\hat{G}'E_{t,t}\hat{G})= \left[ \frac{1}{\sqrt{2}}\hat{g}^2_{t,1}  ~:~\cdots~:~\frac{1}{\sqrt{2}}\hat{g}^2_{t,T-k} ~:~\{\hat{g}_{t,i}\hat{g}_{t,j}\}_{i<j} \right]'$. Therefore,  $\hat{\boldsymbol{X}}' = [ \frac{1}{\sqrt{2}}\hat{g}_1 \odot \hat{g}_1 ~:~\cdots~:~\frac{1}{\sqrt{2}}\hat{g}_{T-k}\odot \hat{g}_{T-k} ~:~\{\hat{g}_i \odot \hat{g}_j \}_{i<j} ]$. Thus, for any $(T-k)\times (T-k)$ symmetric matrix $A=(a_{i,j})$, $diag(\hat{G}A\hat{G}')=\sum_{i=1}^{T-k} a_{i,i}diag(\hat{g}_i\hat{g}'_i)+2\sum_{i<j}a_{i,j}diag(\hat{g}_i\hat{g}'_j)=\sum_{i=1}^{T-k} a_{i,i}(\hat{g}_i\odot\hat{g}_i)+2\sum_{i<j}a_{i,j}(\hat{g}_i\odot\hat{g}_j)=2\hat{\boldsymbol{X}}'vech(A)$.} It follows from the local identification Assumption \ref{ass:A:5} that $M_{\hat{\boldsymbol{X}}}$ is well-defined and thus $vech(\hat{S}^*)=M_{\hat{\boldsymbol{X}}}vech(\hat{S}^*)$.\footnote{Assumption \ref{ass:A:5} is equivalent to $\boldsymbol{X}$ having full column rank by Lemma \ref{App:lemma:Local:identification} in Appendix B.4. Besides, from Proposition  \ref{prop:transformation:invariance} in Appendix B.6 and the fact that $\hat{G}\hat{O}=G+o_p(1)$ for some rotation matrix $\hat{O}$ (see below),  we have $\hat{\mathscr{R}}\hat{\boldsymbol{X}}=\boldsymbol{X}+o_p(1)$ from some orthogonal matrix $\hat{\mathscr{R}}$ . Hence, $\hat{\boldsymbol{X}}$ is invertible with probability approaching $1$. } Next, we have  \begin{equation} \label{B1} M_{\hat{\boldsymbol{X}}}vech(\hat{S}^*)=M_{\hat{\boldsymbol{X}}}vech(\hat{G}'\hat{V}^{-1}_\varepsilon(\hat{V}_y-\hat{V}_\varepsilon)\hat{V}^{-1}_\varepsilon\hat{G})=M_{\hat{\boldsymbol{X}}}vech(\hat{G}'\hat{V}^{-1}_\varepsilon(\hat{V}_y-\tilde{V}_\varepsilon)\hat{V}^{-1}_\varepsilon\hat{G}),\end{equation} because the kernel of $M_{\hat{\boldsymbol{X}}}$ is $\{vech(\hat{G}'D\hat{G}): D \text{ diagonal} \}$. Besides, we have the expansion $\sqrt{n}vech(\hat{G}'\hat{V}^{-1}_\varepsilon(\hat{V}_y-\tilde{V}_\varepsilon)\hat{V}^{-1}_\varepsilon\hat{G})=vech(\hat{Z}^*_n)+o_p(1)$, where $\hat{Z}^*_n=\hat{G}'\hat{V}^{-1}_\varepsilon Z_n \hat{V}^{-1}_\varepsilon\hat{G}$. It is because expansion (\ref{proof:asy:exp:Vhat}) and $\hat{G}'\hat{V}^{-1}_\varepsilon F=\hat{G}'\hat{V}^{-1}_\varepsilon M_{\hat{F},\hat{V}_\varepsilon} F=O_p(\frac{1}{\sqrt{n}})$ by the root-$n$ consistency of FA estimators (see Appendix B.5.1). Using $\|\hat{S}\|^2=\|\hat{S}^*\|^2$, it follows that \begin{equation} \label{B2}\frac{n}{2}\|\hat{S}\|^2=nvech(\hat{S}^*)'vech(\hat{S}^*)=nvech(\hat{S}^*)'M_{\hat{\boldsymbol{X}}}vech(\hat{S}^*)=vech(\hat{Z}^*_n)'M_{\hat{\boldsymbol{X}}}vech(\hat{Z}^*_n)+o_p(1).\end{equation}
From $M_{\hat{F},\hat{V}_\varepsilon}=M_{F,V_\varepsilon}+o_p(1)$, we have $\hat{G}\hat{O}=G+o_p(1)$ for some (possibly data-dependent) $(T-k)\times (T-k)$ orthogonal matrix $\hat{O}$. Since $vech(\hat{Z}^*_n)'M_{\hat{\boldsymbol{X}}}vech(\hat{Z}^*_n)$ is invariant to post-multiplication of $\hat{G}$ by an orthogonal matrix (see Proposition \ref{prop:transformation:invariance} in Appendix B.6), from (\ref{B2}) we get $\frac{n}{2}\|\hat{S}\|=vech(Z^*_n)'M_{\boldsymbol X}vech(Z^*_n)+o_p(1)$, which - together with step (i) - yields asymptotic expansion (\ref{asy:exp:stats}).

(iii) Let us now establish the asymptotic normality of $vech(Z^*_n)$. For any integer $m$, we let $A_m$ denote the unique $m^2\times \frac{m(m+1)}{2}$ matrix satisfying $vec(S)=A_mvech(S)$ for any $m\times m$ symmetric matrix $S$.\footnote{The explicit form for $A_m$ is $A_m =\left[ \sqrt{2}( e_1 \otimes e_1):\cdots : \sqrt{2}(e_m\otimes e_m) : \{ e_i \otimes e_j + e_j \otimes e_i \}_{i<j} \right]$, with $e_i$ being the $i$th unit vector of dimension $m$.} Duplication matrix $A_m$  satisfies $A_m'A_m=2I_{\frac{m(m+1)}{2}}$, $A_mA_m'=I_{m^2}+K_{m,m}$, and $K_{m,m}A_m=A_m$, where $K_{m,m}$ is the commutation matrix (see also Magnus, Neudecker (2007) Theorem 12 in Chapter 2.8). Then, we have $vech(Z^*_n)=\boldsymbol{R}'vech(\mathscr{Z}_n)$,
where $\mathscr{Z}_n =  V_{\varepsilon}^{-1/2} Z_n V_{\varepsilon}^{-1/2}$, $\boldsymbol{R}=\frac{1}{2}A'_T(Q\otimes Q)A_{T-k}$, and $Q=V_{\varepsilon}^{-1/2}G$. Matrix $\boldsymbol{R}$ satisfies $\boldsymbol{R}'\boldsymbol{R}=I_p$.  The next lemma establishes the asymptotic normality of $vech(\mathscr{Z}_n)$.

\begin{lemma} \label{CLT:Zn}
(a) Under Assumptions  \ref{Ass1}-\ref{Ass2},  \ref{ass:A:2}, \ref{ass:A:6} (a)-(b), we have $\Omega_n^{-1/2} vech( \mathscr{Z}_n )  \Rightarrow N(0, I_{\frac{T(T+1)}{2}})$ as $n \rightarrow \infty$ and $T$ is fixed, where $\Omega_n = D_n + \kappa_n I_{\frac{T(T+1)}{2}}$, and  $\kappa_n = \frac{1}{n}\sum_{m= 1}^{J_n}  \left(\sum_{i\neq j\in I_m} \sigma_{ij}^2 \right)$. If additionally Assumption \ref{ass:A:6} (c) holds, then $vech(\mathscr{Z}_n)  \Rightarrow N(0, \Omega)$, with $\Omega := D + \kappa I_{\frac{T(T+1)}{2}}$.
\end{lemma} 
Lemma \ref{CLT:Zn} yields the asymptotic normality of $vech(Z^*_n)$, namely $vech(Z^*_n)\Rightarrow N(0,\Omega_{Z^*})$, with $\Omega_{Z^*}=\boldsymbol{R}'\Omega\boldsymbol{R}$. Part (a) then follows from expansion (\ref{asy:exp:stats}) and the standard result on the distribution of idempotent quadratic forms of Gaussian vectors. 

%
%

(b) We have
$
\hat {z}^*_{m,n} = \sum_{i\in I_m} \hat{G}' \hat V_{\varepsilon}^{-1} \left( \tilde y_i \tilde y_i ' \right) \hat V_{\varepsilon}^{-1} \hat G $ with $\tilde y_i = y_i - \bar y$, since $\hat\varepsilon_i = M_{\hat F, \hat V_{\varepsilon}} \tilde y_i$ and $\hat{G}'  \hat V_{\varepsilon}^{-1} M_{\hat F, \hat V_{\varepsilon}}  =\hat{G}'  \hat V_{\varepsilon}^{-1}$.  We get $ \hat{z}^*_{m,n} =  \sum_{i\in I_m} \hat{G}' \hat V_{\varepsilon}^{-1}  \left( \tilde\varepsilon_i \tilde\varepsilon_i' \right)  \hat V_{\varepsilon}^{-1} \hat G   + \sum_{i\in I_m} \hat{G}' \hat V_{\varepsilon}^{-1}  \left(  F \beta_i \beta_i' F'  \right)  \hat V_{\varepsilon}^{-1} \hat G \break + \sum_{i\in I_m} \hat{G}' \hat V_{\varepsilon}^{-1}  \left( F \beta_i \tilde\varepsilon_i' + \tilde\varepsilon_i \beta_i' F'  \right)  \hat V_{\varepsilon}^{-1} \hat G   =: \tilde{z}^*_{m,n} + z^*_{m,n,1} + z^*_{m,n,2}$, where $\tilde \varepsilon_i = \varepsilon_i - \bar \varepsilon$ by using $\tilde y_i \tilde y_i' = \tilde \varepsilon_i \tilde\varepsilon_i' + F \beta_i \beta_i' F' + F \beta_i \tilde \varepsilon_i' + \tilde \varepsilon_i \beta_i' F' $.
Then, we can decompose $\hat{\Omega}_{Z^*} $ into a sum of a leading term and other terms, which are asymptotically negligible, so that 
$\hat{\Omega}_{Z^*} = \tilde{\Omega}_{Z^*} + o_p(1)$, with $\tilde{\Omega}_{Z^*} = \frac{1}{n} \sum_{m=1}^{J_n} vech( \bar{ z}^*_{m,n}) vech( \bar{z}^*_{m,n})'  $, with $\bar{z}^*_{m,n}$ defined as $\tilde{z}^*_{m,n}$ after replacing $\tilde\varepsilon_i$ with $\varepsilon_i$. 
Let us now show that $M_{\hat{\boldsymbol{X}}}\tilde{\Omega}_{Z^*}M_{\hat{\boldsymbol{X}}}=M_{\boldsymbol{X}}\Omega_{Z^*}M_{\boldsymbol{X}}+o_p(1)$ up to pre- and post-multiplication by a rotation matrix and its inverse.
We have $M_{\hat{\boldsymbol{X}}}vech \left( \hat{G}' \hat V_{\varepsilon}^{-1}  \left( \varepsilon_i \varepsilon_i' \right)  \hat V_{\varepsilon}^{-1} \hat G \right) = M_{\hat{\boldsymbol{X}}}vech \left( \hat{G}' \hat V_{\varepsilon}^{-1}  \left( \varepsilon_i \varepsilon_i'  - \sigma_{ii} V_{\varepsilon}\right)\hat V_{\varepsilon}^{-1} \hat G \right)$, because of the kernel of $M_{\hat{\boldsymbol{X}}}$ . Moreover, from the properties of matrix $A_m$ introduced in part (a), we have:
$
vech \left( \hat{G}' \hat V_{\varepsilon}^{-1}   ( \varepsilon_i \varepsilon_i'  - \sigma_{ii} V_{\varepsilon})   \hat V_{\varepsilon}^{-1} \hat G \right) \nonumber
\break = vech \left( \tilde Q'   ( e_i e_i'  - \sigma_{ii} I_T) \tilde Q \right)= \hat{\boldsymbol{R}}'  vech( e_i e_i'  - \sigma_{ii} I_T ), 
$
where $e_i = V_{\varepsilon}^{-1/2} \varepsilon_i$, and $\hat{\boldsymbol{R}}  := \frac{1}{2} A_{T} '  ( \tilde Q \otimes \tilde Q) A_{T-k}$ with $\tilde Q = V_{\varepsilon}^{1/2}   \hat V_{\varepsilon}^{-1} \hat G$. We get $M_{\hat{\boldsymbol{X}}} vech(\bar{z}^*_{m,n}) = {M}_{\hat{\boldsymbol{X}}} \hat{\boldsymbol{R}}'   vech(  \zeta_{m,n})$,  where $\zeta_{m,n} := \sum_{i\in I_m} ( e_i e_i'  - \sigma_{ii} I_T )$. Besides, $vech(\mathscr{Z}_n) = \frac{1}{\sqrt{n}} \sum_{m=1}^{J_n}vech( \zeta_{m,n})$. Then, $M_{\hat{\boldsymbol{X}}}\tilde{\Omega}_{Z^*}M_{\hat{\boldsymbol{X}}} \break =  {M}_{\hat{\boldsymbol{X}}} \hat{\boldsymbol{R}}'\tilde{\Omega}_n \hat{\boldsymbol{R}}{M}_{\hat{\boldsymbol{X}}}$ for  $\tilde{\Omega}_n := \frac{1}{n} \sum_{m=1}^{J_n}vech(\zeta_{m,n})vech( \zeta_{m,n})'$. Further, $E[\tilde \Omega_n]=  V[ vech(\mathscr{Z}_n)] \break = \Omega_n$. Moreover, $\tilde \Omega_n - E[\tilde \Omega_n] = o_p(1)$, by using $vec(\tilde{\Omega}_n) = \frac{1}{n}  \sum_{m=1}^{J_n}  vech(\zeta_{m,n}) \otimes vech( \zeta_{m,n})$ and $\Vert V[vec(\tilde{\Omega}_n)] \Vert \leq C \frac{1}{n^2}  \sum_{m=1}^{J_n} E\left[ \Vert vech(\zeta_{m,n}) \Vert^4 \right] = o(1)$, where the latter bound is shown in the proof of Lemma \ref{CLT:Zn} using Assumption \ref{Ass2} (d). Additionally, by Assumption \ref{ass:A:6}, we have $\Omega_n = \Omega + o(1)$. Thus, $\tilde{\Omega}_n = \Omega+ o_p(1)$. Now, from the proof of part (a) we have $\hat{G} \hat{O} = G + o_p(1)$ for some $(T-k)\times (T-k)$ orthogonal matrix $\hat O$. Then, by Proposition \ref{prop:transformation:invariance} (e) in Appendix B.6, we have $\hat{\boldsymbol{R}} {M}_{\hat{\boldsymbol{X}}} \hat{\mathscr{R}}^{-1} = \boldsymbol{R} M_{\boldsymbol{X}} + o_p(1)$, for a $p$ dimensional orthogonal matrix 
$\hat{\mathscr{R}} \equiv \mathscr{R} (\hat O)$. We conclude that $\hat{\mathscr{R}} M_{\hat{\boldsymbol{X}}}\tilde{\Omega}_{Z^*}M_{\hat{\boldsymbol{X}}} \hat{\mathscr{R}}^{-1}$ is a consistent estimator of $M_{\boldsymbol{X}}{\Omega}_{Z^*}M_{\boldsymbol{X}}$ as $n\rightarrow\infty$ and $T$ is fixed. Part (b) then follows from the continuity of eigenvalues for symmetric matrices, and their invariance under pre- and post-multiplication by an orthogonal matrix and its transpose.

(c) Under $H_1(k)$ and Assumption \ref{ass:A:7} (a), we have $\hat F \overset{p}{\rightarrow} F^*$ and $\hat V_{\varepsilon} \overset{p}{\rightarrow} V_{\varepsilon}^*$. Then,  $\hat{S} \overset{p}{\rightarrow} S^*$ with $S^* = (V_{\varepsilon}^*)^{-1/2} M_{F^{*},V_{\varepsilon}^*} ( V_y - V_{\varepsilon}^*) M_{F^{*},V_{\varepsilon}^*}'  (V_{\varepsilon}^*)^{-1/2} \neq 0$. Indeed, if $S^*$ were the null matrix, then we would have $M_{F^{*},V_{\varepsilon}^*} ( V_y - V_{\varepsilon}^*) M_{F^{*},V_{\varepsilon}^*}' =0$, which implies $V_y - V_{\varepsilon}^* = P_{F^*,V_{\varepsilon}^*}( V_y - V_{\varepsilon}^*) + ( V_y - V_{\varepsilon}^*)P'_{F^*,V_{\varepsilon}^*} - P_{F^*,V_{\varepsilon}^*}( V_y - V_{\varepsilon}^*)P'_{F^*,V_{\varepsilon}^*}$, with $P_{F^*,V_{\varepsilon}^*}= I_T - M_{F^*,V_{\varepsilon}^*}$. From the probability limits of Equation (FA2) for pseudo values, we have  $P_{F^*,V_{\varepsilon}^*}( V_y - V_{\varepsilon}^*) = ( V_y - V_{\varepsilon}^*)P'_{F^*,V_{\varepsilon}^*} = P_{F^*,V_{\varepsilon}^*}( V_y - V_{\varepsilon}^*)P'_{F^*,V_{\varepsilon}^*}=F^*(F^*)'$ (see proof of Lemma \ref{prop:1} (c)). Thus $V_y=F^*(F^*)'+V_{\varepsilon}^*$, in contradiction with Assumption \ref{ass:A:7} (b). Thus, $ n \Vert \hat S \Vert^2 \geq C n$, w.p.a.\ 1, for a constant $C>0$. Moreover, using $vech( \hat{z}^*_{m,n}) = vech( \hat G' \hat{V}_{\varepsilon}^{-1} ( \sum_{i \in I_m} \tilde y_i \tilde y_i') \hat{V}_{\varepsilon}^{-1} \hat G )$ and the conditions on $\Theta$, we get $\Vert vech( \hat{z}^*_{m,n})  \Vert \leq C \sum_{i \in I_m} \Vert \tilde y_i \Vert^2$. Then, from Assumptions \ref{ass:A:2} and \ref{ass:A:3},  $E [\Vert M_{\hat{\boldsymbol{X}}} \hat{\Omega}_{Z^*}M_{\hat{\boldsymbol{X}}} \Vert ] \leq C \frac{1}{n} \sum_{m=1}^{J_n} b_{m,n}^2 = O( n  \sum_{m=1}^{J_n} B_{m,n}^2 )$. Moreover, $\sum_{m=1}^{J_n} B_{m,n}^2 =o(1)$. Indeed, Assumption \ref{Ass2} (d) implies $B_{m,n} \leq c n^{-\frac{\delta}{\delta+1}}$ uniformly in $m$, for any $c>0$ and $n$ large enough, and hence $\sum_{m=1}^{J_n} B_{m,n}^2= c n^{-\frac{\delta}{\delta+1}}\sum_{m=1}^{J_n} B_{m,n} \leq c$, for any $c>0$ and $n$ large. Part (c) follows from the Lipschitz continuity of eigenvalues for symmetric matrices.

\textbf{Proof of Proposition \ref{prop:gen:Gaussian}:}  We have $LR(k) = \frac{n}{2} \Vert \hat{S} \Vert^2 + o_p(1) = vech( Z_n^* )'M_{\boldsymbol{X}}vech( Z_n^* ) + o_p(1)$ from expansion $(\ref{asy:exp:stats})$. Moreover, the kernel of matrix $M_{\boldsymbol{X}}$ implies that  $M_{\boldsymbol{X}}vech( Z_n^* ) = A ( F, V_{\varepsilon} ) z^{AD}_n$, where vector $z^{AD}_n$ stacks the $T(T-1)/2$ above-diagonal elements of matrix $Z_n$  and $A ( F, V_{\varepsilon} )$ is a deterministic matrix whose elements only depend on $F, V_{\varepsilon}$. From Conditions (a) and (b) of Proposition \ref{prop:gen:Gaussian}, and Lemma \ref{CLT:Zn}, we have $z^{AD}_n \Rightarrow N(0,\Omega_{z})$, where the diagonal matrix $\Omega_{z}$ is the same as if the errors were independent normally distributed - up to replacing $q$ with ${q}+\kappa$.

\textbf{Proof of Proposition \ref{thm:local:alternative}:} Let us first get the asymptotic expansion of $\hat{V}_y-\tilde{V}_\varepsilon = \frac{1}{n} \tilde Y \tilde Y' -\tilde{V}_\varepsilon$. With the drifting DGP $Y= \mu 1_n' + F \beta' + F_{k+1} \beta_{loc}' + \varepsilon$, and using $\bar \beta=0$, $\bar \beta_{loc} =0$, $\frac{1}{n} [\beta : \beta_{loc}]'[\beta : \beta_{loc}]  = I_{k+1}$ and Lemma \ref{lemma:conv:espilon:beta} (a) in Appendix B, we get $\hat{V}_y-\tilde{V}_\varepsilon = FF' + \frac{1}{\sqrt{n}} \Psi_{y,loc} +  R_y$, where
\begin{equation} \label{app:Psi_y:local}
\Psi_{y,loc} = c_{k+1} \rho_{k+1} \rho_{k+1}'  
+ \frac{1}{\sqrt{n}} ( \varepsilon \beta F' +  F \beta' \varepsilon') + \sqrt{n} \left( \frac{1}{n} \varepsilon \varepsilon' - \tilde{V}_{\varepsilon} \right),
\end{equation}
and $R_y = \frac{1}{n}( \varepsilon \beta_{loc} F_{k+1}' + F_{k+1} \beta_{loc}' \varepsilon') + [ F_{k+1} F_{k+1}' - n^{-1/2} c_{k+1} \rho_{k+1} \rho_{k+1}'] + o_p(\frac{1}{\sqrt n})$. Using $F_{k+1} = \sqrt{\gamma_{k+1}} \rho_{k+1}$ and $\sqrt{n} \gamma_{k+1} = c_{k+1} + o(1)$, we get $R_y = o_p( 1/\sqrt{n})$. Subsituting the  expansion for $\hat{V}_y-\tilde{V}_\varepsilon$ into (\ref{B1}), and repeating the arguments leading to expansion $(\ref{asy:exp:stats})$ yields expansion $(\ref{asy:exp:stats:alternative})$. From Lemma \ref{CLT:Zn}, we get $vech(Z_{n,loc}^*)\Rightarrow N(c_{k+1}vech(\xi_{k+1} \xi_{k+1}'),\Omega_{Z^*})$ as $n \rightarrow \infty$. The result then follows from the standard result on the distribution of idempotent quadratic forms of non-central Gaussian vectors.

\textbf{Proof of Proposition \ref{thm:AUMPI}:} The proof of part (a) is in three steps. (i) The testing problem asymptotically simplifies to the null hypothesis $H_0 : \lambda_1=...=\lambda_{df}=0$ vs.\ the alternative hypothesis $H_1 : \exists \lambda_j > 0, j=1, ...,df$. Let us define $\boldsymbol{\lambda}_0 = (0, ...,0)'$ for the null hypothesis and pick a given vector $\boldsymbol{\lambda}_1 = (\lambda_1, ...,\lambda_{df})'$ in the alternative hypothesis, and consider the test of $\boldsymbol{\lambda}_0$ versus $\boldsymbol{\lambda}_1$ (simple hypothesis). By Neyman-Pearson Lemma, the most powerful test for $\boldsymbol{\lambda}_0$ versus $\boldsymbol{\lambda}_1$ rejects the null hypothesis when $f(z;\lambda_1, ..., \lambda_{df})/f(z; 0, ...,0)$ is large, i.e., the test function is $\phi(z) = \boldsymbol{1}\left\{ \frac{f(z;\lambda_1, ..., \lambda_{df})}{f(z; 0, ...,0)} \geq C \right\}$ for a constant $C>0$ set to ensure the correct asymptotic size. 

(ii) Let us now show that the density ratio $\frac{f(z;\lambda_1, ..., \lambda_{df})}{f(z; 0, ...,0)}$ is an increasing function of $z$. To show this, we can rely on an expansion of the density of $\sum_{j=1}^{df} \mu_j \chi^2 (1,\lambda_j^2)$ in terms of central chi-square densities
(Kotz, Johnson, and Boyd (1967) Equations (144) and (151)):
\begin{eqnarray} f(z;\lambda_1, ..., \lambda_{df}) = \sum_{k=0}^\infty \bar{c}_k(\lambda_1, ..., \lambda_{df}) g(z; df + 2k, 0),\label{fnoncentral} \end{eqnarray}
where  the coefficients
$\displaystyle \bar{c}_k(\lambda_1, ..., \lambda_{df})  = A e^{-\sum_{j=1}^{df} \lambda_j^2/2} E[Q(\lambda_1, ..., \lambda_{df})^k]/k!$ involve moments of the quadratic form $\displaystyle Q(\lambda_1, ..., \lambda_{df})=(1/2)\sum_{j=1}^{df}\left(\nu_j^{1/2} X_j + \lambda_j(1-\nu_j)^{1/2} \right)^2$ of the mutually independent variables $X_j \sim N(0,1)$, $A = \prod_{j=1}^{df} \mu_j^{-1/2}$, and $\nu_j = 1-  \frac{1}{\mu_j} \min_{\ell} \mu_{\ell}$. Without loss of generality for checking the monotonicity, we have rescaled the density so that $\min_j \mu_j =1$. Then, from (\ref{fnoncentral}), we get the ratio:
$ \frac{f(z;\lambda_1, ..., \lambda_{df})}{
f(z;0, ..., 0)} = \frac{\sum_{k=0}^\infty \bar{c}_k(\lambda_1, ..., \lambda_{df}) g(z; df + 2k, 0)}{\sum_{k=0}^\infty \bar{c}_k(0, ..., 0)g(z;df+2k,0)}.$
By dividing both the numerator and the denominator by the central chi-square density $g(z;df,0)$, we get
$
\frac{f(z;\lambda_1,...,\lambda_{df})}{f(z;0,...,0)} = e^{-\sum_{j=1}^{df} \lambda_j^2/2} \frac{\sum_{k=0}^{\infty} c_k(\lambda_1,...,\lambda_{df}) \psi_k(z)}{\sum_{k=0}^{\infty} c_k(0,...,0) \psi_k(z)} =: e^{-\sum_{j=1}^{df} \lambda_j^2/2} \Psi (z; \lambda_1,...,\lambda_{df}),
$
where $\psi_k(z) := $ $g(z;df+2k,0)/g(z;df,0) = \frac{\Gamma(\frac{df}{2})}{2^k \Gamma( \frac{df}{2} + k)} z^k$ is the ratio of central chi-square distributions with $df+2k$ and $df$ degrees of freedom, and $c_k(\lambda_1,...,\lambda_{df}) = E[ Q(\lambda_1,...,\lambda_{df})^k]/k!$. We use the short notation $c_k(\lambda):= c_k(\lambda_1,...,\lambda_{df})$ and $c_k(0) := c_k(0,...,0)$. The factor $e^{-\sum_{j=1}^{df} \lambda_j^2/2}$ does not impact on the monotonicity of the density ratio. 
We take the derivative of $\Psi (z; \lambda_1,...,\lambda_{df})$ with respect to argument $z$ and get
$\partial_z \Psi (z; \lambda_1,...,\lambda_{df}) =
\frac{\left(  \sum_{k=1}^{\infty} c_k(\lambda) \psi_k'(z) \right) \left( 1+ \sum_{k=1}^{\infty} c_k(0) \psi_k(z) \right)
}{\left( \sum_{k=0}^{\infty} c_k(0) \psi_k(z) \right)^2}
 -
\frac{
\left(  1+ \sum_{k=1}^{\infty} c_k(\lambda) \psi_k(z) \right) \left( \sum_{k=1}^{\infty} c_k(0) \psi_k'(z) \right)}{\left( \sum_{k=0}^{\infty} c_k(0) \psi_k(z) \right)^2}.
$
The sign is given by the difference of the numerators, which is
$
 \sum_{k=1}^{\infty} [ c_k(\lambda)-c_k(0)] \psi_k' (z) 
+ \sum_{k,l=1, k \neq l}^{\infty} c_k(\lambda) c_l(0) [ \psi_k'(z) \psi_l(z) - \psi_k (z) \psi_l'(z)] 
= \sum_{k=1}^{\infty} [ c_k(\lambda)-c_k(0)] \psi_k' (z) 
+ \sum_{k,l=1, k > l}^{\infty} [c_k(\lambda) c_l(0) - c_l(\lambda) c_k(0)]  [ \psi_k'(z) \psi_l(z) - \psi_k (z) \psi_l'(z)].
$
We use $\psi_k'(z) = \frac{\Gamma(\frac{d}{2}) k}{2^{k} \Gamma( \frac{d}{2} + k)} z^{k-1}$ and $\psi_k'(z) \psi_l(z) - \psi_k (z) \psi_l'(z) = (k-l) \frac{\Gamma(\frac{d}{2})^2}{2^{k+l} \Gamma( \frac{d}{2} + k)\Gamma(\frac{d}{2} + l)}  z^{k+l-1} $ for $k >l$ and $z \geq 0$. 
The difference of the numerators in the derivative of the density ratio becomes: \break
$
  \frac{1}{2}\frac{\Gamma(\frac{d}{2})}{\Gamma( \frac{d}{2} + 1)}[c_1(\lambda)-c_1(0)] +  \frac{1}{2^2} \frac{2 \Gamma(\frac{d}{2})}{\Gamma( \frac{d}{2} + 2)}[ c_2(\lambda)-c_2(0)]  z  
+ \sum_{m=3}^{\infty} \frac{1}{2^m}\left(  m \frac{\Gamma(\frac{d}{2})}{\Gamma( \frac{d}{2} + m)}[ c_m(\lambda)-c_m(0)]  \right.$ \break $\left.   + \sum_{k > l \geq 1, k+l = m}  \frac{ (k-l) \Gamma(\frac{d}{2})^2}{\Gamma( \frac{d}{2} + k)\Gamma( \frac{d}{2} + l)}[c_k(\lambda) c_l(0) - c_l(\lambda) c_k(0)]
\right) z^{m-1} = \sum_{m=1}^{\infty} \frac{1}{2^m}\kappa_m z^{m-1},$
with $\kappa_m := \sum_{k > l \geq 0, k+l = m} (k-l) \frac{\Gamma(\frac{d}{2})^2}{\Gamma( \frac{d}{2} + k)\Gamma( \frac{d}{2} + l)} $ $ [c_k(\lambda) c_l(0) - c_l(\lambda) c_k(0)]$. A direct calculation shows that $\kappa_1 ,\kappa_2 \geq 0$. Hence, a sufficient condition for monotonicity of the density ratio is $\kappa_m \geq 0$, for all $m \geq 3$, i.e., Inequalities (\ref{cond:MLR}). 
Thus, the test rejects for large values of the argument, i.e., $\phi(z) = \boldsymbol{1}\{ z  \geq \bar{C}\}$, where the constant $\bar{C}$ is determined by fixing the asymptotic size under the null hypothesis. 

(iii) Since the test function $\phi$ does not depend on $\boldsymbol{\lambda}_1$, it is AUMPI in the class of hypothesis tests based on the LR statistic (or the squared norm statistic). It yields part (a).

Let us now turn to the proof of part (b). From the definition of the $\kappa_m$ coefficients written as $\kappa_m = \sum_{j>l \geq 0, j+l=m} \frac{(j-l)\Gamma(\frac{df}{2})^2}{\Gamma(\frac{df}{2}+j)\Gamma(\frac{df}{2}+l)} c_j(0) c_l(0) [ \frac{c_j(\lambda)}{c_j(0)} - \frac{c_l(\lambda)}{c_l(0)}]$, it is sufficient to get $\kappa_m \geq 0$, for all $m$, that sequence $\frac{c_j(\lambda)}{c_j(0)}$, for $j=0,1,...$, is increasing. To prove that, we link the coefficients $c_j(\lambda)$ to the complete exponential Bell's polynomials (Bell (1934)) and establish the following  recurrence.

\begin{lemma}  \label{prop:recursion}
We have $c_{l+1}(\lambda) = \frac{1}{l+1} \sum_{i=0}^l \left( \frac{1}{2} \sum_{j=1}^{df} \nu_j^{i} \left[  \nu_j   + (i+1)(1-\nu_j) \lambda_j^2 \right]  \right) c_{l-i}(\lambda)$, for $l \geq 0$.
\end{lemma}

We use $\frac{c_l(\lambda)}{c_l(0)}= \frac{\tilde c_l(\lambda)}{\bar\gamma_l}$, where we obtain the sequences $\bar\gamma_l := c_l(0) \nu_{df}^{-l}$ and $\tilde c_l(\lambda) := c_l(\lambda) \nu_{df}^{-l}$  by standardization with $\nu_{df}^{-l}$. From Lemma \ref{prop:recursion}, we have $\bar\gamma_{l+1} = \frac{1}{l+1} \sum_{i=0}^l \frac{1}{2} \left( 1 + \sum_{j=2}^{df-1} \rho_j^{i+1} \right) \bar\gamma_{l-i}$ with $\bar\gamma_0=1$, and $\tilde c_{l+1}(\lambda) = \frac{1}{l+1} \sum_{i=0}^l \left( \frac{1}{2} \sum_{j=1}^{df} \rho_j^i \left[  \rho_j + \frac{i+1}{\nu_{df}}  (1-\nu_j) \lambda_j^2 \right]\right) \tilde c_{l-i}(\lambda)$  
with $\tilde c_0(\lambda) =1$ (note that $\rho_1 = 0$ and $\rho_{df} = 1$). To prove that sequence$\frac{\tilde c_l(\lambda)}{\bar\gamma_l}$ is increasing,  the next lemma provides a sufficient condition from  "separation" of the coefficients that define the recursive relations.

\begin{lemma} \label{prop:sequence:lambda:general}
Let $(a_i)$ be a real sequence, and let $b_i = \frac{1}{2}\left(1 + \sum_{j=2}^{df-1}\rho_j^i \right)$, for $i \geq 1$, where $0 \leq \rho_j \leq 1$.  Let sequences $(g_l)$ and $(c_l)$ be defined recursively by $g_{l+1} = \frac{1}{l} (b_1 g_l + b_2 g_{l-1} + ... + b_l)$ and $c_{l+1} = \frac{1}{l} (a_1 c_l + a_2 c_{l-1} + ... + a_l)$, with $g_1 = c_1= 1$. Suppose that $a_i \geq \max\{\frac{df-1}{2},1\}$, for all $i$ (separation condition). Then, sequence $(\frac{c_l}{g_l})$ is increasing. 
\end{lemma}

We apply Lemma \ref{prop:sequence:lambda:general} to sequences $\tilde c_l(\lambda)$ and $\bar\gamma_l$. We detail the case $df\geq 3$ (for $df=2$ the analysis is simpler). The separation condition $ \frac{1}{2} \sum_{j=1}^{df} \rho_j^i \left[  \rho_j + \frac{i+1}{\nu_{df}}  (1-\nu_j) \lambda_j^2 \right] \geq \frac{df-1}{2}$, for $i=0$, yields $\lambda_1^2 + \sum_{j=2}^{df}  (1-\nu_j) \lambda_j^2 \geq \nu_{df} \left( df - 2 - \sum_{j=2}^{df-1} \rho_j \right)$, and, for $i \geq 1$, it yields $\sum_{j=2}^{df-1} \rho_j^i  (1-\nu_j) \lambda_j^2 + ( 1 - \nu_{df}) \lambda_{df}^2 \geq \frac{\nu_{df}}{i+1} \left( df - 2 - \sum_{j=2}^{df-1} \rho_j^{i+1}\right)$. Inequalities (\ref{condition:lambda}) follow.


\subsection*{Proofs of Lemmas \ref{prop:1}-\ref{prop:sequence:lambda:general}}

\textbf{Proof of Lemma \ref{prop:1}:} Let $\hat{U}$ be the $T\times k$ matrix whose orthonormal columns are the eigenvectors for the $k$ largest eigenvalues of matrix $\hat{V}_{\varepsilon}^{-1/2} \hat{V}_y \hat{V}_{\varepsilon}^{-1/2}$. Those eigenvalues are $1+\hat{\gamma}_j$, $j=1,...,k$, while it holds $\hat{F} = \hat{V}_{\varepsilon}^{1/2} \hat{U} \hat{\Gamma}^{1/2}$, where $\hat{\Gamma} = diag(\hat{\gamma}_1,...,\hat{\gamma}_k)$. We have
$
I_T - \hat{U} \hat{U}' = I_T - \hat{V}_{\varepsilon}^{-1/2} \hat{F} \hat{\Gamma}^{-1} \hat{F}' \hat{V}_{\varepsilon}^{-1/2} $ $
=  I_T - \hat{V}_{\varepsilon}^{-1/2} \hat{F} (\hat{F}'\hat{V}_{\varepsilon}^{-1} \hat{F})^{-1} \hat{F}' \hat{V}_{\varepsilon}^{-1/2} $ $
= \hat{V}_{\varepsilon}^{-1/2} M_{\hat{F},\hat{V}_{\varepsilon}} \hat{V}_{\varepsilon}^{1/2} 
= \hat{V}_{\varepsilon}^{1/2} M_{\hat{F},\hat{V}_{\varepsilon}}' \hat{V}_{\varepsilon}^{-1/2}.
$
Thus,
$
\hat{S} 
=  ( I_T - \hat{U} \hat{U}' ) \left (   \hat{V}_{\varepsilon}^{-1/2} \hat{V}_y \hat{V}_{\varepsilon}^{-1/2}  - I_T \right) ( I_T - \hat{U} \hat{U}' ) .
$
By the spectral decomposition of  $\hat{V}_{\varepsilon}^{-1/2} \hat{V}_y \hat{V}_{\varepsilon}^{-1/2}$, we get $( I_T - \hat{U} \hat{U}' ) \left (   \hat{V}_{\varepsilon}^{-1/2} \hat{V}_y \hat{V}_{\varepsilon}^{-1/2}  - I_T \right) $  $( I_T - \hat{U} \hat{U}' ) = \sum_{j=k+1}^T \hat{\gamma}_j \hat{P}_j$, where the $\hat{P}_j$ are the orthogonal projection matrices onto the eigenspaces for the $T-k$ smallest eigenvalues. Then, Part (a) follows. 
Part (b) is a consequence of the squared Frobenius norm of a  symmetric matrix being equal to the sum of its squared eigenvalues. For part (c), let $P_{\hat{F},{\hat{V}}_{\varepsilon}}=I_T-M_{\hat{F},\hat{V}_{\varepsilon}}$ and note that
$
 \hat{F}\hat{F}'= P_{\hat{F},{\hat{V}}_{\varepsilon}}(\hat{V}_y-\hat{V}_\varepsilon) + (\hat{V}_y-\hat{V}_\varepsilon)P_{\hat{F},{\hat{V}}_{\varepsilon}}'-P_{\hat{F},{\hat{V}}_{\varepsilon}}(\hat{V}_y-\hat{V}_\varepsilon)P_{\hat{F},{\hat{V}}_{\varepsilon}}' $ $=\hat{V}_y-\hat{V}_\varepsilon -M_{\hat{F},\hat{V}_{\varepsilon}}(\hat{V}_y-\hat{V}_\varepsilon)M_{\hat{F},\hat{V}_{\varepsilon}}' , $
where the first equality is because the three terms on the RHS are all equal to $ \hat{F}\hat{F}'$ by (FA2). The conclusion follows from (FA1) and $\hat{V}_{\varepsilon}$ being diagonal. 

\textbf{Proof of Lemma \ref{CLT:Zn}:} We have $\mathscr{Z}_n=\frac{1}{\sqrt{n}} (W\Sigma W' -Tr(\Sigma)I_T)$. Hence, $(\mathscr{Z}_n)_{tt}=\frac{1}{\sqrt{n}} \sum_{i,j}(w_{i,t} w_{j,t}-1_{\{i=j\}})\sigma_{ij}=\frac{1}{\sqrt{n}} \sum_{m=1}^{J_n} \zeta^{tt}_{m,n},$ with
$\zeta^{tt}_{m,n}=\sum_{i\in I_m} [w^2_{i,t}- 1]\sigma_{ii}+2\sum_{\underset{i<j}{i,j\in I_m}}w_{i,t} w_{j,t}\sigma_{ij},$ together with 
$(\mathscr{Z}_n)_{ts}=\frac{1}{\sqrt{n}} \sum_{i,j}w_{i,t} w_{j,s}\sigma_{ij}=\frac{1}{\sqrt{n}} \sum_{m=1}^{J_n} \zeta^{ts}_{m,n},  t\neq s,$ with
$\zeta^{ts}_{m,n}=\sum_{i\in I_m} w_{i,t}w_{i,s}\sigma_{ii}+\sum_{\underset{i<j}{i,j\in I_m}}w_{i,t} w_{j,s}\sigma_{ij}+\sum_{\underset{i>j}{i,j\in I_m}}w_{i,t} w_{j,s}\sigma_{ij}, t\neq s, $ so that
$vech(\mathscr{Z}_n)=\frac{1}{\sqrt{n}}\sum_{m=1}^{J_n} vech(\zeta_{m,n}),$
where $\zeta_{m,n}$ is the $T\times T$ matrix having element $\zeta^{ts}_{m,n}$ in position $(t,s)$. Hence, $vech(\mathscr{Z}_n)$ is the row sum of a triangular array $\{vech(\zeta_{m,n})\}_{1\leq m\leq n}$ of independent centered random vectors. Let $\Omega_{m,n}:=V[vech(\zeta_{m,n})]$. Using Assumption \ref{Ass2} (a), we compute
(i) $E[(\zeta^{tt}_{m,n})^2]=\sum_{i\in I_m} (E[w^4_{i,t}]- 1)\sigma^2_{ii}+2\sum_{\underset{i\neq j}{i,j\in I_m}}\sigma_{ij}^2$; (ii)
$E[(\zeta^{ts}_{m,n})^2]=\sum_{i\in I_m} E[w^2_{i,t}w^2_{i,s}]\sigma_{ii}^2+\sum_{\underset{i\neq j}{i,j\in I_m}}\sigma_{ij}^2, t\neq s $;  (iii)
$E[\zeta^{tt}_{m,n}\zeta^{ss}_{m,n}]=\sum_{i\in I_m} E[w^2_{i,t}w^2_{i,s}- 1]\sigma^2_{ii}, t\neq s  $; (iv)
$E[\zeta^{tt}_{m,n}\zeta^{rp}_{m,n}]=\sum_{i\in I_m} E[w^2_{i,t}w_{i,r}w_{ip}] \sigma^2_{ii}, r\neq p$; (v)
$E[\zeta^{ts}_{m,n}\zeta^{rp}_{m,n}]$ $=\sum_{i\in I_m} E[w_{i,t}w_{i,s}w_{i,r}w_{i,p}]\sigma^2_{ii},  t\neq s, r\neq p.  $
It follows that $V[vech(\mathscr{Z}_n)]=\frac{1}{n}\sum_{m=1}^{J_n} \Omega_{m,n} =D_n + \kappa_n I_{\frac{T(T+1)}{2}}=\Omega_n $. The eigenvalues of $D_n$ are bounded away from $0$ under Assumption \ref{ass:A:6} (b), because for any unit vector $\xi \in \mathbb{R}^{T(T+1)/2}$, we have $
\xi' D_n \xi \geq \frac{1}{n} \sum_{i=1}^n 1_{i\in \bar S} \sigma_{ii}^2 \xi' V[ vech( w_i w_i')] \xi \geq \underline{c} \frac{1}{n} \sum_{i=1}^n 1_{i\in \bar S} \sigma_{ii}^2 \geq \underline{c} \left( 1 - \frac{1}{n} \sum_{i=1}^n (1-1_{i\in \bar S}) \sigma_{ii} \right)^{2}  \geq 
\underline{c} \left( 1 - \bar C \frac{1}{n} \sum_{i=1}^n (1-1_{i\in \bar S})  \right)^{2}
\geq \frac{\underline{c}}{4}$, for all $n$.
We use the multivariate Lyapunov condition 
$\|\Omega^{-1/2}_n\|^4 \frac{1}{n^2} \sum_{m=1}^{J_n}E[\|vech(\zeta_{m,n})\|^4] \to 0$ to invoke a CLT.
Since $\|A^{-1/2}\|^4\leq \frac{k^2}{\delta^2_k(A)}$ and $\|x\|^4\leq k \sum_{j=1}^{k} x^4_j$, for any $k\times k$ positive semi-definite matrix $A$ and $k\times 1$ vector $x$, it suffices to check that 
$\frac{1}{n^2} \sum_{m=1}^{J_n}E[(\zeta^{ts}_{m,n})^4]\to 0 , $
for all $t,s$. Besides, we can show that there exists a constant $M>0$, such that $E[(\zeta^{ts}_{m,n})^4]\leq M b_{m,n}^{2(1+\delta)}, $
for all $m,n,t,s$. We get 
$\frac{1}{n^2} \sum_{m=1}^{J_n}E[(\zeta^{ts}_{m,n})^4]\leq M \frac{1}{n^2} \sum_{m=1}^{J_n} b_{m,n}^{2(1+\delta)}=Mn^{2\delta}\sum_{m=1}^{J_n}B_{m,n}^{2(1+\delta)} = o(1)$, under Assumption \ref{Ass2} (d). Then, $\Omega_n^{-1/2}vech(\mathscr{Z}_n)\Rightarrow N(0,I_{\frac{T(T+1)}{2}})$ by the multivariate Lyapunov CLT. Under Assumptions \ref{ass:A:6} (a)-(c), $\Omega_n \rightarrow \Omega$ follows from the Slutsky theorem, and $\Omega$ is positive definite.

\textbf{Proof of Lemma \ref{prop:recursion}:} We have $c_j(\lambda) = \frac{1}{j!} E[Q^j] =  \frac{1}{j!} \frac{d^j \Psi (0)}{du^j}$ where $\Psi(u) := E[\exp(uQ)] = \exp[ \psi (u)]$ is the Moment Generating Function (MGF) of $Q= \frac{1}{2} \sum_{j=1}^{df}  (\sqrt \nu_j X_j + \sqrt{1-\nu_j} \lambda_j)^2 $ with $X_j \sim i.i.d. N(0,1)$. By the independence of variables  $X_j$, we get $\Psi(u) = \prod_{j=1}^{df} E[ \exp( \frac{u}{2} (\sqrt \nu_j X_j + \sqrt{1-\nu_j} \lambda_j)^2 ]$ where $
E[ \exp( \frac{u}{2} (\sqrt \nu_j X_j + \sqrt{1-\nu_j} \lambda_j)^2 ]
=  (1 - \nu_j u)^{-1/2} e^{ \frac{1}{2} \frac{(1-\nu_j)u}{1-\nu_j u} \lambda_j^2 }$, 
for $u< 1/\nu_j$. Thus we get the log MGF $\psi(u) = \frac{1}{2} \sum_{j=1}^{df} \left[ -  \log( 1 - \nu_j u) + \frac{(1-\nu_j)u}{1-\nu_j u} \lambda_j^2 \right]$, 
for $u < 1/\nu_{df}$. 
Its $l$th order derivative evaluated at $u=0$ is
\begin{equation} \label{derivative:psi}
\psi^{(l)}(0) = \frac{(l-1)!}{2} \sum_{j=1}^{df} \nu_j^{l-1} \left[  \nu_j   + l(1-\nu_j) \lambda_j^2 \right], \quad l \geq 0.
\end{equation}

By using the Faa di Bruno formula for the derivatives of a composite function, we have $\frac{d^l}{d u^l} e^{\psi(u)} =  e^{\psi(u)} B_l ( \psi'(u), \psi^{''}(u),...,\psi^{(l)}(u))$, where $B_l$ is the $l$th complete exponential Bell's polynomial (Bell (1934)). Hence, $\Psi^{(l)}(0) =  B_l ( \psi'(0), \psi^{''}(0),...,\psi^{(l)}(0))$. The complete Bell's polynomials satisfy the recurrence relation $B_{l+1} ( x_1,x_2,...,x_{l+1}) =  \sum_{i=0}^l \binom{l}{i} B_{l-i}(x_1,...,x_{l-i}) x_{i+1}$. Thus, $\Psi^{(l+1)}(0) = \sum_{i=0}^l \binom{l}{i} \Psi^{(l-i)}(0) \psi^{(i+1)}(0)$. After standardization with the factorial term, and using equation (\ref{derivative:psi}), the conclusion follows. 

\textbf{Proof of Lemma \ref{prop:sequence:lambda:general}:} The proof is in four steps. (i) We first show that $(c_i)$ is increasing, i.e., $G_i^c := c_{i+1} - c_i \geq 0$ for all $i$. For this purpose, from the recursive relation defining $c_{i+1}$ we have:
\begin{eqnarray*}
c_{i+1} &=& \frac{1}{i} \left( a_1 (c_{i-1}+G_{i-1}^c)  + a_2 (c_{i-2}+G_{i-2}^c) +  \cdots + a_{i-1} (c_1+G_{1}^c) + a_i  \right)  \\
&=& \frac{1}{i} \left( (a_1 -1)G^c_{i-1}  + (a_2-1) G^c_{i-2}  +  \cdots + (a_{i-1}-1) G^c_{1}  + (a_i -1)  \right)  \\
&& +  \frac{1}{i} \left( G^c_{i-1}   + G^c_{i-2}  +  \cdots + G^c_{1} + 1  \right) +\frac{1}{i} \left( a_1  c_{i-1}  + a_2  c_{i-2} +  \cdots + a_{i-1}   \right).
\end{eqnarray*}
The second term in the RHS is equal to $\frac{1}{i} c_{i}$. Using $a_1  c_{i-1}  + a_2  c_{i-2} +  \cdots + a_{i-1} = (i-1) c_{i}$, the third term in the RHS is equal to $\frac{i-1}{i} c_{i}$. Thus, by bringing these two terms in the LHS, we get  $G_i^c = \frac{1}{i} \left( (a_1 -1)G^c_{i-1}  + (a_2-1) G^c_{i-2}  +  \cdots + (a_{i-1}-1) G^c_{1}  + (a_i -1)  \right)$, for all $i\geq 2$, with $G^c_1 = a_1 -1$. Since $a_i \geq 1$ for all $i$,  we get $G^c_i \geq 0$ for all $i\geq 1$ by an induction argument .

(ii) We now strengthen the result in step (i) and show that $H_i^c := c_{i+1} - c_i \frac{\zeta+i-1}{i} \geq 0$ for all $i$, with $\zeta = \max\{ \frac{df-1}{2}, 1 \}$. Similarly as in step (i), we have 
\begin{eqnarray*}
c_{i+1} 
&=& \frac{1}{i} \left( (a_1 -\zeta)G^c_{i-1}  + (a_2-\zeta) G^c_{i-2}  +  \cdots + (a_{i-1}-\zeta) G^c_{1}  + (a_i -\zeta)  \right) \\ &&  +  \frac{\zeta}{i} \left( G^c_{i-1}   + G^c_{i-2}  +  \cdots + G^c_{1} + 1  \right) +\frac{1}{i} \left( a_1  c_{i-1}  + a_2  c_{i-2} +  \cdots + a_{i-1}   \right),
\end{eqnarray*} 
where the second term in the RHS equals $\frac{\zeta}{i} c_{i}$, and the third term equals $\frac{i-1}{i} c_{i}$. Thus, we get $H^c_i = \frac{1}{i} \left( (a_1 -\zeta)G^c_{i-1}  + (a_2-\zeta) G^c_{i-2}  +  \cdots + (a_{i-1}-\zeta) G_{1}^c  + (a_i -\zeta)  \right)$, for all $i$. By step (i), we have $G^c_i \geq 0$ for $i\geq 1$. Using the separation condition $a_i \geq \zeta$ for all $i$, we get $H_i^c  \geq 0$ for all $i$.

(iii) We show that $H_i^g := g_{i+1} - g_i \frac{\zeta+i-1}{i} \leq 0$ for all $i \geq 1$. For $df=2$ this statement follows with $\zeta=1$ since $g_{i+1}=\frac{1}{2i}(g_i+g_{i-1}+...+1)=\frac{2i-1}{2i} g_i$ and hence $(g_i)$ is decreasing. Let us now consider the case $df\geq 3$ with $\zeta = \frac{df-1}{2}$. As above we have $H_i^{g} = \frac{1}{i} \sum_{l=1}^i (b_l - \zeta)  G^g_{i-l}$, where $G^g_i := g_{i+1} - g_i$.  We plug in $b_l - \zeta = \frac{1}{2} \sum_{j=2}^{df-1} (\rho_j^l - 1)  = \frac{1}{2}\sum_{j=2}^{df-1}(\rho_j-1)(1+\rho_j + ... + \rho_j^{l-1}) = \frac{1}{2}\sum_{j=2}^{df-1}(\rho_j-1) \sum_{k=1}^l \rho_j^{k-1}$. Thus, we get:
\begin{eqnarray*}
H_i^{g} &=& \frac{1}{2 i }\sum_{j=2}^{df-1}(\rho_j-1) \sum_{l=1}^i \sum_{k=1}^l \rho_j^{k-1}  G^g_{i-l}
= \frac{1}{2 i }\sum_{j=2}^{df-1}(\rho_j-1) \sum_{k=1}^i  \rho_j^{k-1}  \sum_{l=k}^i G^g_{i-l} \\
&=& \frac{1}{2 i }\sum_{j=2}^{df-1}(\rho_j-1) \sum_{k=1}^i   \rho_j^{k-1} g_{i-k+1}
= \frac{1}{2 i }\sum_{j=2}^{df-1}(\rho_j-1) \left( g_i + \rho_j g_{i-1} + ... + \rho_j^{i-1}   \right) \leq 0.
\end{eqnarray*}

(iv) The inequalities established in steps (ii) and (iii) imply $\frac{c_{i+1}}{c_i} \geq \frac{\zeta+i-1}{i}$ and $\frac{g_{i+1}}{g_i} \leq \frac{\zeta+i-1}{i}$ for all $i$. Then, we get $\frac{c_{i+1}}{c_i} \geq \frac{g_{i+1}}{g_i}$, that is equivalent to $\frac{c_{i+1}}{g_{i+1}} \geq \frac{c_{i}}{g_i}$, for all $i$, because the sequences $c_i$ and $g_i$ are strictly positive. The conclusion follows. 

\section*{B Additional theory}
\setcounter{equation}{0}\def\theequation{B.\arabic{equation}}

\subsection*{B.1 Pseudo likelihood and PML estimator}

The FA estimator is the PML estimator based on the Gaussian likelihood function obtained from the pseudo model $y_i = \mu + F \beta_i + \varepsilon_i$ with $\beta_i \sim N ( 0 , I_k)$ and $\varepsilon_i \sim N( 0 , V_{\varepsilon})$ mutually independent and i.i.d.\ across $i=1,...,n$. Then, $y_i \sim N(  \mu , \Sigma (\theta) )$ under this pseudo model, where $\Sigma (\theta ):= F F' + V_{\varepsilon}$ and $\theta := ( vec(F)',diag(V_{\varepsilon})')' \in \mathbb{R}^r$ with $r = (k+1)T$. It yields the pseudo log-likelihood function
$
\hat{L} (\theta, \mu) = - \frac{1}{2} \log \vert \Sigma(\theta) \vert - \frac{1}{2 n} \sum_{i=1}^n (y_i - \mu)' \Sigma(\theta)^{-1} (y_i - \mu)  
= - \frac{1}{2} \log \vert \Sigma(\theta) \vert - \frac{1}{2} Tr \left( \hat{V}_y \Sigma(\theta)^{-1} \right)
- \frac{1}{2} ( \bar{y} - \mu)' \Sigma(\theta)^{-1} ( \bar{y} - \mu),
$
up to constants, where $\bar{y} = \frac{1}{n} \sum_{i=1}^n y_i$ 
and $\hat{V}_y = \frac{1}{n} \sum_{i=1}^n (y_i - \bar{y}) (y_i - \bar{y})'$. 
We concentrate out parameter $\mu$ to get its estimator  $\hat{\mu} =  \bar{y}$. Then, estimator $\hat{\theta}= ( vec(\hat{F})',diag(\hat{V}_{\varepsilon})')'$ is defined by the maximization of
\begin{equation} \label{Lhattheta}
\hat{L}(\theta) := - \frac{1}{2} \log \vert \Sigma(\theta) \vert - \frac{1}{2} Tr \left( \hat{V}_y \Sigma(\theta)^{-1} \right),
\end{equation}
subject to the normalization restriction that $F' V_{\varepsilon}^{-1} F$ is a diagonal matrix, with diagonal elements ranked in decreasing order.\footnote{If the risk-free rate vector is considered observable, we can rewrite the model as $\tilde{y}_i = F \tilde \beta_i + \varepsilon_i = \mu + F \beta_i + \varepsilon_i$, where $\tilde y_i = y_i - r_f$ is the vector of excess returns and $\mu = F \mu_{\tilde \beta}$.  It corresponds to a constrained model with parameters $\theta$ and  $\mu_{\tilde\beta}$. The maximization of the corresponding Gaussian pseudo likelihood function leads to a constrained FA estimator, that we do not consider in this paper since it does not match a standard FA formulation.}

%

\subsection*{B.2 Global identification and consistency \label{app:consistency}}

The population criterion $L_0(\theta)$ is defined in Section \ref{regularity}, with $V_y = V_y^0 = \Sigma (\theta_0) = F_0 F_0' + V_{\varepsilon}^0$.

\begin{lemma}  \label{App:lemma:Global:identification}
The following conditions are equivalent:
a) the true value $\theta_0$ is the unique maximizer of $L_0(\theta)$ for $\theta \in \Theta$;
b) $\Sigma (\theta) = \Sigma(\theta_0)$, $\theta \in \Theta$ $\Rightarrow$ $\theta = \theta_0$, up to sign changes in the columns of $F$. 
They yield the global identification in the FA model.
\end{lemma}

In Lemma \ref{App:lemma:Global:identification}, condition a) is the standard identification condition for a M-estimator with population criterion $L_0(\theta)$. Condition (b) is the global identification condition based on the variance matrix as in Anderson and Rubin (1956). 
Condition (b) corresponds to our Assumption \ref{ass:A:4}.


Let us now establish the consistency of the FA estimators in our setting. Write $\hat V_y = \frac{1}{n} \sum_{i=1}^n ( \varepsilon_i - \bar \varepsilon)( \varepsilon_i - \bar \varepsilon)' + F [ \frac{1}{n} \sum_{i=1}^n ( \beta_i - \bar \beta)( \beta_i - \bar \beta)'] F' + F [ \frac{1}{n} \sum_{i=1}^n ( \beta_i - \bar \beta) ( \varepsilon_i - \bar \varepsilon) ' ]+  [\frac{1}{n} \sum_{i=1}^n ( \varepsilon_i - \bar \varepsilon)( \beta_i - \bar \beta)'] F'$, where $\bar \varepsilon = \frac{1}{n} \sum_{i=1}^n  \varepsilon_i$ and $\bar \beta = \frac{1}{n} \sum_{i=1}^n  \beta_i$. Under the normalization in Assumption \ref{ass:A:1} we have:
\begin{equation} \label{formula:Vhaty}
\hat V_y = \frac{1}{n} \varepsilon \varepsilon'  - \bar\varepsilon \bar \varepsilon' + F F'  + F \left(\frac{1}{n} \varepsilon \beta \right)' + \left(\frac{1}{n} \varepsilon \beta \right) F'.
\end{equation}

\begin{lemma} \label{lemma:conv:espilon:beta} Under  Assumptions \ref{Ass1}, \ref{Ass2}, and \ref{ass:A:2}, \ref{ass:A:3}, as $n \rightarrow \infty$, we have: (a) $\bar{\varepsilon} = o_p( \frac{1}{n^{1/4}})$, (b) $\frac{1}{n} \varepsilon \varepsilon' \overset{p}{\rightarrow} V_{\varepsilon}^0$,  and (c) $\frac{1}{n} \varepsilon \beta  \overset{p}{\rightarrow} 0$.
\end{lemma}

\noindent From Equation (\ref{formula:Vhaty}) and Lemma \ref{lemma:conv:espilon:beta}, we have $\hat{V}_y \overset{p}{\rightarrow} V_y^0$. Thus, $\hat{L}(\theta)$ converges in probability to $L_0(\theta)$ as $n \rightarrow \infty$, uniformly over $\Theta$ compact. From standard results on M-estimators, we get consistency of $\hat{\theta}$. Moreover, from $\bar y = \mu + \bar \varepsilon$, we get the consistency of $\hat\mu$.

\begin{proposition} \label{prop:consistency:FA}
Under Assumptions \ref{Ass1}, \ref{Ass2}, and \ref{ass:A:2}-\ref{ass:A:4}, the FA estimators $\hat{F}$, $\hat{V}_{\varepsilon}$ and $\hat{\mu}$ are consistent as $n \rightarrow \infty$ and $T$ is fixed. 
\end{proposition}

Anderson and Rubin (1956) establish consistency in Theorem 12.1 (see beginning of the proof, page 145) within a Gaussian ML framework. Anderson and Amemiya (1988) provide a version of this result in their Theorem 1 for generic distribution of the data, dispensing for compacity of the parameter set but using a more restrictive identification condition. 

\subsection*{B.3 Asymptotic expansions of estimators $\hat{V}_{\varepsilon}$ and $\hat{F}$}

The FA estimators $\hat{V}_{\varepsilon}$ and $\hat{F}$ are consistent M-estimators under nonlinear constraints, and admit expansions at first order for fixed $T$ and $n\rightarrow \infty$, namely $\hat{V}_{\varepsilon} = \tilde{V}_{\varepsilon} + \frac{1}{\sqrt{n}} \Psi_{\varepsilon} + o_p(\frac{1}{\sqrt{n}})$ and $\hat{F}_j = F_j + \frac{1}{\sqrt{n}} \Psi_{F_j} + o_p(\frac{1}{\sqrt{n}})$ (see Appendix B.5.1). The next  proposition (new to the literature) characterizes the diagonal random matrix $\Psi_{\varepsilon}$ and the random vectors $\Psi_{F_j}$ by using conditions (FA1) and (FA2) in Section 2  (see proof at the end of the section).  


\begin{proposition} \label{lemma:1}  Under Assumptions \ref{Ass1}, \ref{Ass2}, and \ref{ass:A:1}-\ref{ass:A:4}, \ref{ass:A:6},  we have (a) for $j=1,...,k$
	\begin{eqnarray}
		\Psi_{F_j} &=& R_j  (\Psi_y - \Psi_{\varepsilon}) V_{\varepsilon}^{-1} F_j   + \Lambda_j \Psi_{\varepsilon} V_{\varepsilon}^{-1} F_j, \label{cond:PsiFj}
	\end{eqnarray}
	where $R_j := \frac{1}{2\gamma_j} P_{F_j,{V}_{\varepsilon}} 
	+  \frac{1}{\gamma_j}M_{F,V_{\varepsilon}}   +
	\sum_{\ell =1, \ell \neq j}^k  \frac{1}{\gamma_j - \gamma_{\ell}} P_{F_{\ell},{V}_{\varepsilon}}$ and $\Lambda_j := - \sum_{\ell =1, \ell \neq j}^k  \frac{\gamma_l}{\gamma_j - \gamma_{\ell}} P_{F_{\ell},{V}_{\varepsilon}}$ and $P_{F_j,V_{\varepsilon}} = F_j (F_j' V_{\varepsilon}^{-1} F_j)^{-1} F_j' V_{\varepsilon}^{-1} = \frac{1}{\gamma_j}  F_j F_j' V_{\varepsilon}^{-1}$ is the GLS orthogonal projection onto $F_j$. Further, (b) the diagonal matrix $\Psi_{\varepsilon}$ is such that:
	\begin{equation} \label{cond:Psie}
		diag \left( M_{F,{V}_{\varepsilon}} (\Psi_y - \Psi_{\varepsilon}) M_{F,{V}_{\varepsilon}}' \right) = 0 .
	\end{equation}
\end{proposition}

Equation (\ref{cond:PsiFj}) yields the asymptotic expansion of the eigenvectors by accounting for estimation errors of matrix $\hat{V}_y \hat{V}_{\varepsilon}^{-1}$ (first term) and of the normalization constraint (second term).
To interpret Equation (\ref{cond:Psie}), we can observe that the matrix $M_{F,{V}_{\varepsilon}} (\Psi_y - \Psi_{\varepsilon}) M_{F,{V}_{\varepsilon}}'$ yields the first-order term in the asymptotic expansion of $\sqrt{n} \hat{S}$ (up to the left- and right-multiplication by diagonal matrix $V_{\varepsilon}^{-1/2}$). Thus, Equation (\ref{cond:Psie}) is implied by the property that the diagonal terms of matrix $\hat{S}$ are equal to zero as stated in Lemma \ref{prop:1} (c).

Let us now give the explicit expression of $\Psi_{\varepsilon}$. By using $M_{F,{V}_{\varepsilon}} \Psi_y M_{F,{V}_{\varepsilon}}' = M_{F,{V}_{\varepsilon}} Z_n M_{F,{V}_{\varepsilon}}'$, we can rewrite Equation (\ref{cond:Psie}) as $diag \left( M_{F,{V}_{\varepsilon}} (Z_n - \Psi_{\varepsilon}) M_{F,{V}_{\varepsilon}}' \right) = 0$. Now, since $\Psi_{\varepsilon}$ is diagonal, we have $diag \left( M_{F,{V}_{\varepsilon}} \Psi_{\varepsilon} M_{F,{V}_{\varepsilon}}' \right) = M_{F,{V}_{\varepsilon}}^{\odot 2} diag ( \Psi_{\varepsilon} )$, where $M_{F,{V}_{\varepsilon}}^{\odot 2} = M_{F,{V}_{\varepsilon}} \odot M_{F,{V}_{\varepsilon}}$. Thus, we get:
\begin{equation} \label{les}
	M_{F,{V}_{\varepsilon}}^{\odot 2} diag ( \Psi_{\varepsilon} ) = diag( M_{F,{V}_{\varepsilon}} Z_n M_{F,{V}_{\varepsilon}}').
\end{equation}
To have a unique solution for vector $diag ( \Psi_{\varepsilon} )$, we need the non-singularity of the $T \times T$ matrix $M_{F,{V}_{\varepsilon}}^{\odot 2}$. It is the local identification condition in the FA model stated in Assumption \ref{ass:A:5}. Let us write $G=[g_1: \cdots : g_{T-k}]$. Then, we have $M_{F,V_{\varepsilon}} = G G' V_{\varepsilon}^{-1}  = \sum_{j=1}^{T-k} g_j (V_{\varepsilon}^{-1}g_j)'$, and so we get the Hadamard product
$
M_{F,V_{\varepsilon}}^{\odot 2} = \sum_{i,j=1}^{T-k} [g_i (V_{\varepsilon}^{-1}g_i)'] \odot [g_j (V_{\varepsilon}^{-1}g_j)'] =  \left[ \sum_{i,j=1}^{T-k} ( g_i \odot g_j) ( g_i \odot g_j)' \right] $ $V_{\varepsilon}^{-2} $
$=2\left(\boldsymbol{X}'\boldsymbol{X} \right) V_{\varepsilon}^{-2}.$\footnote{Let us recall the following property of the Hadamard product: $(a b') \odot (cd') = (a \odot c) (b \odot d)'$ for conformable vectors $a, b,c, d$. The last equalitiy because $\boldsymbol{X}' = \left[ \frac{1}{\sqrt{2}}g_1 \odot g_1 ~:~\cdots~:~\frac{1}{\sqrt{2}}g_{T-k}\odot g_{T-k} ~:~\{g_i \odot g_j \}_{i<j} \right]$ (see beginning of the proof of Proposition 2 (a)) .} Hence, we can state the local identification condition in Assumption \ref{ass:A:5} as a full-rank condition for matrix $\boldsymbol{X}$, analogously as in linear regression (Lemma \ref{App:lemma:Local:identification}). In Lemma \ref{App:lemma:Local:identification} in Appendix B.4 i), we also show equivalence with invertibility of the bordered Hessian, i.e., the Hessian of the Lagrangian function in a constrained M-estimation.  

Under Assumption \ref{ass:A:5}, we get from Equation (\ref{les}):
\begin{equation} \label{asy:Psie:1}
	\Psi_{\varepsilon}  = \mathcal{T}_{F,V_{\varepsilon}}( Z_n ),
\end{equation}
where $\mathcal{T}_{F,V_{\varepsilon}}(V) := diag \left( [M_{F,{V}_{\varepsilon}}^{\odot 2}]^{-1} diag( M_{F,{V}_{\varepsilon}} V M_{F,{V}_{\varepsilon}}') \right)$, for any matrix $V$. Mapping $\mathcal{T}_{F,V_{\varepsilon}}(\cdot)$ is linear and such that $\mathcal{T}_{F,V_{\varepsilon}}(V) =V$, for a diagonal matrix $V$. We have
$
diag( M_{F,V_{\varepsilon}} Z_n  M_{F,V_{\varepsilon}}') 
= diag \left( G Z^*_n G' \right) = 2 \boldsymbol{X}' vech\left( Z^*_n\right)
$,\footnote{We have $diag(GAG')=2\boldsymbol{X}'vech(A)$ for any $T\times T$ symmetric matrix $A$; see beginning of the proof of Proposition \ref{thm:asy:tests} (a).} and so \begin{equation} \label{asy:Psie:2} diag(\Psi_\varepsilon) =  V_{\varepsilon}^{2} \left(\boldsymbol{X}'\boldsymbol{X} \right)^{-1} \boldsymbol{X}'vech\left( Z^*_n\right) .\end{equation}

Anderson and Rubin (1956), Theorem 12.1, show that the FA estimator is asymptotically normal if $\sqrt{n} ( \hat{V}_y - V_y)$ is asymptotically normal. They use a linearization of the first-order conditions similar as the one of Proposition \ref{lemma:1}. Their Equation (12.16) corresponds to our Equation  (\ref{cond:Psie}). However, they only provide an implicit characterization of the $\Psi_{F_j}$ and not an explicit expression for $\Psi_{\varepsilon}$ and $\Psi_{F_j}$ in terms of asymptotically Gaussian random matrices like $Z_n$ as we do. These key developments pave the way to establishing the asymptotic distributions of estimators $\hat{F}$ and $\hat{V}_{\varepsilon}$ in general settings, that we cover in Appendix B.5.

\noindent \textbf{Proof of Proposition \ref{lemma:1}:} From (\ref{formula:Vhaty}) and Lemma \ref{lemma:conv:espilon:beta} we have $\hat{V}_{y} = \tilde{V}_y + \frac{1}{\sqrt{n}} \Psi_y + o_p(\frac{1}{\sqrt{n}})$, where $\tilde{V}_y = F F' + \tilde{V}_{\varepsilon}$ and $\Psi_y = \frac{1}{\sqrt{n}} (\varepsilon \beta F'+ F \beta' \varepsilon ') + \sqrt{n} \left( \frac{1}{n} \varepsilon \varepsilon' - \tilde{V}_{\varepsilon}\right)$. Let us substitute this expansion for $\hat{V}_y$ into (FA2) and rearrange to obtain $\hat F \hat \Gamma  - F F' \hat V_{\varepsilon}^{-1} \hat F = \frac{1}{\sqrt n} \Psi_y \hat V_{\varepsilon}^{-1} \hat F + ( \tilde V_{\varepsilon} \hat V_{\varepsilon}^{-1} - I_T) \hat F + o_p(\frac{1}{\sqrt n})$, where $\hat{\Gamma}=\hat{F}'\hat{V}^{-1}_\varepsilon \hat{F}=diag(\hat{\gamma}_1,\dots,\hat{\gamma}_k)$. From $\hat V_{\varepsilon} = \tilde V_{\varepsilon} + \frac{1}{\sqrt n} \Psi_{\varepsilon} + o_p(\frac{1}{\sqrt n})$, we have $ \tilde V_{\varepsilon} \hat V_{\varepsilon}^{-1} - I_T = - \frac{1}{\sqrt n} \Psi_{\varepsilon} \hat V_{\varepsilon}^{-1} + o_p(\frac{1}{\sqrt n})$. Substituting into the above equation and right multiplying both sides by $(F' \hat V_{\varepsilon}^{-1} \hat F)^{-1}$ gives $\hat F \hat{\mathcal{D}} - F = \frac{1}{\sqrt n} ( \Psi_y - \Psi_{\varepsilon} ) \hat V_{\varepsilon}^{-1} \hat F (F' \hat V_{\varepsilon}^{-1} \hat F)^{-1} + o_p(\frac{1}{\sqrt n})$, where $\hat{\mathcal{D}} := \hat \Gamma (F' \hat V_{\varepsilon}^{-1} \hat F)^{-1}$. By the root-$n$ convergence of the FA estimates (see Section B.5.1), we get
\begin{equation} \label{eq:exp:1}
	\hat F \hat{\mathcal{D}} - F = \frac{1}{\sqrt n} ( \Psi_y - \Psi_{\varepsilon} ) V_{\varepsilon}^{-1} F \Gamma^{-1} + o_p(\frac{1}{\sqrt n}),
\end{equation}
and $\hat{\mathcal{D}} = I_k + O_p(\frac{1}{\sqrt n})$, where $\Gamma = diag ( \gamma_1,...,\gamma_k)$. 
We can push the expansion by plugging into 
(\ref{eq:exp:1}) the expansion of $\hat{\mathcal{D}}$. We have $F' \hat V_{\varepsilon}^{-1} \hat F = [  I_k - ( \hat F - F)' \hat V_{\varepsilon}^{-1} \hat F \hat \Gamma^{-1} ] \hat\Gamma$, so that $\hat{\mathcal{D}} = [  I_k - ( \hat F - F)' \hat V_{\varepsilon}^{-1} \hat F \hat \Gamma^{-1} ]^{-1} = I_k + ( \hat F - F)'  V_{\varepsilon}^{-1}  F  \Gamma^{-1} + o_p(\frac{1}{\sqrt n})$. By plugging into (\ref{eq:exp:1}), we get:
\begin{equation} \label{eq:exp:2}
	\hat F - F + F [ (\hat F - F)'V_{\varepsilon}^{-1} F \Gamma^{-1}] =  \frac{1}{\sqrt n} ( \Psi_y - \Psi_{\varepsilon} ) V_{\varepsilon}^{-1} F \Gamma^{-1} + o_p(\frac{1}{\sqrt n}).
\end{equation}
By multiplying both sides with $M_{F,V_{\varepsilon}}$, we get $M_{F,V_{\varepsilon}}(\hat F - F) = \frac{1}{\sqrt n} M_{F,V_{\varepsilon}} ( \Psi_y - \Psi_{\varepsilon} ) V_{\varepsilon}^{-1} F \Gamma^{-1} + o_p(\frac{1}{\sqrt n})$. Then,
$\hat F - F = \frac{1}{\sqrt n} M_{F,V_{\varepsilon}} ( \Psi_y - \Psi_{\varepsilon} ) V_{\varepsilon}^{-1} F \Gamma^{-1} +  \frac{1}{\sqrt n} F A + o_p(\frac{1}{\sqrt n})$, where $A$ is a random $k\times k$ matrix to be determined next. By plugging  into (\ref{eq:exp:2}), we get $F ( A + A') = P_{F,V_{\varepsilon}}( \Psi_y - \Psi_{\varepsilon} ) V_{\varepsilon}^{-1} F \Gamma^{-1} + o_p(\frac{1}{\sqrt n})$. By multiplying both sides by $\frac{1}{2} \Gamma^{-1}F' V_{\varepsilon}^{-1}$ and using $F' V_{\varepsilon}^{-1} P_{F,V_{\varepsilon}} = F' V_{\varepsilon}^{-1}$, we get the symmetric part of matrix $A$, i.e., $\frac{1}{2}( A + A') = \frac{1}{2} \Gamma^{-1} F' V_{\varepsilon}^{-1}( \Psi_y - \Psi_{\varepsilon} ) V_{\varepsilon}^{-1} F \Gamma^{-1}$ (we include higher-order terms in the remainder $o_p(\frac{1}{\sqrt n})$). Thus, $\hat F - F = \frac{1}{\sqrt n} \Psi_F + o_p(\frac{1}{\sqrt n})$, where 
\begin{equation} \label{appendix:Psi:F}
	\Psi_F = M_{F,V_{\varepsilon}} ( \Psi_y - \Psi_{\varepsilon} ) V_{\varepsilon}^{-1} F \Gamma^{-1} +  \frac{1}{2} P_{F,V_{\varepsilon}}( \Psi_y - \Psi_{\varepsilon} ) V_{\varepsilon}^{-1} F \Gamma^{-1}  +  F \tilde A,
\end{equation}
and $\tilde A= \frac{1}{2}(A-A')$ is an antisymmetric $k\times k$ random matrix.
To find the antisymmetric matrix $\tilde{A}=(\tilde{a}_{\ell,j})$, we use that $\hat F' \hat V_{\varepsilon}^{-1} \hat F$ is diagonal. Plugging the expansions of the FA estimates, for the term at order $1/\sqrt n$ we get that the out-of-diagonal elements of matrix $\Psi_F' V_{\varepsilon}^{-1} F + F'V_{\varepsilon}^{-1} \Psi_F - F'V_{\varepsilon}^{-1}  \Psi_{\varepsilon} V_{\varepsilon}^{-1} F 
= \frac{1}{2} \Gamma^{-1} F' V_{\varepsilon}^{-1} ( \Psi_y - \Psi_{\varepsilon} ) V_{\varepsilon}^{-1} F + \frac{1}{2}  F' V_{\varepsilon}^{-1} ( \Psi_y - \Psi_{\varepsilon} ) V_{\varepsilon}^{-1} F \Gamma^{-1} +  \Gamma \tilde A - \tilde A \Gamma  - F' V_{\varepsilon}^{-1}  \Psi_{\varepsilon}  V_{\varepsilon}^{-1} F$ are nil. Setting the $(\ell,j)$ element of this matrix equal to $0$, we get $\tilde{a}_{\ell,j} = - \tilde a_{j,\ell} = \frac{1}{\gamma_j - \gamma_{\ell}} \left[ \frac{1}{2} ( \frac{1}{\gamma_j} + \frac{1}{\gamma_{\ell}})  F'_{\ell} V_{\varepsilon}^{-1} ( \Psi_y - \Psi_{\varepsilon} ) V_{\varepsilon}^{-1} F_j  -  F_{\ell}' V_{\varepsilon}^{-1}  \Psi_{\varepsilon}  V_{\varepsilon}^{-1} F_j   \right]$, for $ j \neq \ell$. Then, from Equation (\ref{appendix:Psi:F}), the $j$th column of $\Psi_F$ is
	$\Psi_{F_j} = 
	\frac{1}{\gamma_j}  M_{F,V_{\varepsilon}} ( \Psi_y - \Psi_{\varepsilon} ) V_{\varepsilon}^{-1} F_j +  \frac{1}{2 \gamma_j} P_{F_j,V_{\varepsilon}}( \Psi_y - \Psi_{\varepsilon} ) V_{\varepsilon}^{-1} F_j   +  \sum_{\ell=1:\ell \neq j}^k \frac{1}{\gamma_j - \gamma_{\ell}} P_{F_{\ell},V_{\varepsilon}} ( \Psi_y - \Psi_{\varepsilon} ) V_{\varepsilon}^{-1} F_j
	-  \sum_{\ell=1:\ell \neq j}^k \frac{\gamma_{\ell}}{\gamma_j - \gamma_{\ell}}  P_{F_{\ell},V_{\varepsilon}} \Psi_{\varepsilon}  V_{\varepsilon}^{-1} F_j$,
where we use $P_{F,V_{\varepsilon}} = \sum_{\ell=1}^k P_{F_{\ell},V_{\varepsilon}}$. Part (a) follows.

Let us now prove part (b). The asymptotic expansion of condition (FA1) yields:
\begin{equation} \label{FP}
	diag( \Psi_y) = diag \left( \sum_{j=1}^k ( F_j \Psi_{F_j}' + \Psi_{F_j} F_j' ) + \Psi_{\varepsilon} \right).
\end{equation}
From part (a) and the definition of $P_{F_j,{V}_{\varepsilon}}$ we have $
\sum_{j=1}^k\Psi_{F_j}F_j'  = \frac{1}{2} \sum_{j=1}^k P_{F_j,{V}_{\varepsilon}} ( \Psi_y - \Psi_{\varepsilon}) P_{F_j,{V}_{\varepsilon}}'
+  M_{F,V_{\varepsilon}} (\Psi_y - \Psi_{\varepsilon}) P_{F,{V}_{\varepsilon}}'  +
\sum_{\ell \neq j}  \frac{\gamma_j}{\gamma_j - \gamma_{\ell}} P_{F_{\ell},{V}_{\varepsilon}} (\Psi_y - \Psi_{\varepsilon}) P_{F_j,{V}_{\varepsilon}}'
- \sum_{\ell \neq j}^k  \frac{\gamma_{\ell} \gamma_j}{\gamma_j - \gamma_{\ell}} P_{F_{\ell},{V}_{\varepsilon}} \Psi_{\varepsilon} P_{F_j,{V}_{\varepsilon}}' $ $
=: N_1 + N_2 + N_3 + N_4,$
where $P_{F,{V}_{\varepsilon}} = \sum_{j=1}^k P_{F_j,{V}_{\varepsilon}}= I_T - M_{F,{V}_{\varepsilon}}$ and $\sum_{\ell \neq j}$ denotes the double sum over $j,\ell=1,...,k$ such that $\ell \neq j$. Matrix $N_1$ is symmetric and it contributes $2N_1$ to the RHS of (\ref{FP}). Instead, matrix $N_4$ is antisymmetric (it can be seen by interchanging indices $j$ and $\ell$ in the summation) and it does not contribute to the RHS of (\ref{FP}).  For matrix $N_3$ we have
$
N_3 + N_3' = \sum_{\ell \neq j}  \frac{\gamma_j}{\gamma_j - \gamma_{\ell}} P_{F_{\ell},{V}_{\varepsilon}} (\Psi_y - \Psi_{\varepsilon}) P_{F_j,{V}_{\varepsilon}}' 
+ \sum_{\ell \neq j}  \frac{\gamma_{\ell}}{\gamma_{\ell} - \gamma_{j}} P_{F_{\ell},{V}_{\varepsilon}} (\Psi_y - \Psi_{\varepsilon}) P_{F_j,{V}_{\varepsilon}}' 
= \sum_{\ell \neq j}  P_{F_{\ell},{V}_{\varepsilon}} (\Psi_y - \Psi_{\varepsilon}) P_{F_j,{V}_{\varepsilon}}'  = \sum_{\ell ,j}  P_{F_{\ell},{V}_{\varepsilon}} (\Psi_y - \Psi_{\varepsilon}) P_{F_j,{V}_{\varepsilon}}'   
- \sum_{ j}  P_{F_{j},{V}_{\varepsilon}} (\Psi_y - \Psi_{\varepsilon}) P_{F_j,{V}_{\varepsilon}}' 
= P_{F,{V}_{\varepsilon}} (\Psi_y - \Psi_{\varepsilon}) P_{F,{V}_{\varepsilon}}' - 2 N_1,
$
where we have interchanged $\ell$ and $j$ in the first equality  when writing $N_3'$. Thus, we get:
\begin{eqnarray}
	\sum_{j=1}^k ( F_j \Psi_{F_j}' + \Psi_{F_j} F_j') 
	&=& M_{F,{V}_{\varepsilon}} (\Psi_y - \Psi_{\varepsilon}) P_{F,{V}_{\varepsilon}}'
	+ P_{F,{V}_{\varepsilon}} (\Psi_y - \Psi_{\varepsilon}) M_{F,{V}_{\varepsilon}}'
	+ P_{F,{V}_{\varepsilon}} (\Psi_y - \Psi_{\varepsilon}) P_{F,{V}_{\varepsilon}}'  \nonumber \\
	&=& (\Psi_y - \Psi_{\varepsilon})  - M_{F,{V}_{\varepsilon}} (\Psi_y - \Psi_{\varepsilon}) M_{F,{V}_{\varepsilon}}'.  
	\label{app:eq:FPsi}
\end{eqnarray}
Then, Equation (\ref{FP}) with (\ref{app:eq:FPsi}) yields Equation (\ref{cond:Psie}).

\subsection*{B.4 Local analysis of the first-order conditions of FA estimators \label{app:C:local}}

Consider the criterion $L(\theta) = -\frac{1}{2} \log \vert \Sigma (\theta) \vert - \frac{1}{2} Tr\left( V_y \Sigma(\theta)\right)$, where $V_y$ is a p.d.\ matrix in a neighbourhood of $V_y^0$. In our Assumptions, $\theta_0$ is an interior point of $\Theta$. Let $\theta^* = (vec(F^*)',diag(V_{\varepsilon}^*)')'$ denote the maximizer of $L(\theta)$ subject to $\theta \in \Theta$. According to Anderson (2003), the first-order conditions (FOC) for the maximization of $L(\theta)$ are: (a) $diag ( V_y) = diag ( F^* (F^*)' + V_{\varepsilon}^*)$ and (b) $F^*$ is the matrix of eigenvectors of $V_y (V_{\varepsilon}^*)^{-1}$ associated to the $k$ largest eigenvalues $1+\gamma_j^*$ for $j=1,...,k$, normalized such that $(F^*)' ( V_{\varepsilon}^*)^{-1} F^* = diag (\gamma_1^*,...,\gamma_k^*)$.

\subsubsection*{i) Local identification}

Let $V_y = V_y^0$. The true values $F_0$ and $V_{\varepsilon}^0$ solve the FOC. Let $F = F_0 + \epsilon \Psi_F^{\epsilon}$ and $V_{\varepsilon} = V_{\varepsilon}^0 + \epsilon \Psi_{V_{\varepsilon}}^{\epsilon}$, where $\epsilon$ is a small scalar and $\Psi_F^{\epsilon},\Psi_{V_{\varepsilon}}^{\epsilon}$ are deterministic conformable matrices, be in a neighbourhood of $F_0$ and $V_{\varepsilon}^0$ and solve the FOC up to terms $O(\epsilon^2)$. The model is locally identified if, and only if, it implies $\Psi_{V_{\varepsilon}}^{\epsilon}=0$ and $\Psi_F^{\epsilon}=0$. 

\begin{lemma}  \label{App:lemma:Local:identification} Under Assumption \ref{Ass1},
the following four conditions are equivalent:
(a) Matrix $M_{F_0,V_{\varepsilon}^0}^{\odot 2}$ is non-singular,
(b) Matrix $\boldsymbol{X}$ is full-rank,
(c) Matrix $\Phi^{\odot 2}$ is non-singular, where $\Phi := V_{\varepsilon}^0 - F_0 (F_0' (V_{\varepsilon}^0)^{-1} F_0 )^{-1} F_0'$, 
(d) Matrix $B_0 ' J_0 B_0$ is non-singular, where $J_0 := - \frac{\partial^2 L_0 ( \theta_0)}{\partial \theta \partial \theta'}$ and $B_0$ is any full-rank $r \times (r- \frac{1}{2}k ( k-1))$ matrix such that $\frac{\partial g(\theta_0)}{\partial \theta'} B_0=0$, for $g(\theta) = \{ [F'V_{\varepsilon}^{-1} F]_{i,j} \}_{i<j}$ the $\frac{1}{2}k ( k-1)$ dimensional vector of the constraints.
They yield the local identification of our model.
\end{lemma}

In Lemma \ref{App:lemma:Local:identification}, condition (a) corresponds to Assumption \ref{ass:A:5} and is equivalent to condition (b) that $\boldsymbol{X}$ is full-rank. Condition (c) is used in Theorem 5.9 of Anderson and Rubin (1956) to show local identification. Condition (d) involves the second-order partial derivatives of the population criterion function. While the Hessian matrix $J_0$ itself is singular because of the rotational invariance of the model to latent factors, the second-order partial derivatives matrix along parameter directions, which are in the tangent plan to the contraint set, is non-singular. Condition (d)  is equivalent to invertibility of the bordered Hessian.

\subsubsection*{ii) Local misspecification}

Now, let $V_y = V_y^0 + \epsilon \Psi_y^{\epsilon}$ be in a neighbourhood of $V_y^0$. Let $F^* = F_0 + \epsilon \Psi_F^{\epsilon} + O(\epsilon^2)$ and $V_{\varepsilon}^* = V_{\varepsilon}^0 + \epsilon \Psi_{V_{\varepsilon}}^{\epsilon} + O(\epsilon^2)$ be the solutions of the FOC. Consider $V_y - \Sigma^*$, where $\Sigma^* = F^* (F^*)' + V_{\varepsilon}^*$, i.e., the difference between variance $V_y$ and its $k$-factor approximation with population FA. We want to find the first-order development of $V_y - \Sigma^*$ for small $\epsilon$. From the FOC, we have that the diagonal of such symmetric matrix is null, but not necessarily the out-of-diagonal elements.

 From the arguments in the proof of Proposition \ref{lemma:1}, Equations (\ref{FP}) and (\ref{app:eq:FPsi}), we get:
\begin{eqnarray}
\Psi_F^{\epsilon} F_0' + F_0 (\Psi_F^{\epsilon})' = \Psi_y^{\epsilon} - \Psi_{V_{\varepsilon}}^{\epsilon} - M_{F_0,V_{\varepsilon}^0} (\Psi_y^{\epsilon} - \Psi_{V_{\varepsilon}}^{\epsilon}) M_{F_0,V_{\varepsilon}^0}',  \label{app:FOC:eq1} \\
diag( M_{F_0,V_{\varepsilon}^0} (\Psi_y^{\epsilon} - \Psi_{V_{\varepsilon}}^{\epsilon}) M_{F_0,V_{\varepsilon}^0}') = 0. \qquad \qquad  \qquad \label{app:FOC:eq2}
\end{eqnarray}
As in Section B.3, Equation (\ref{app:FOC:eq2}) yields:
\begin{equation}  \label{app:psi:epsilonXXX}
diag (\Psi_{V_{\varepsilon}}^{\epsilon} ) = ( V_{\varepsilon}^0)^{2} \left(\boldsymbol{X}'\boldsymbol{X} \right)^{-1} \boldsymbol{X}' vech\left(  G_0' (V_{\varepsilon}^0)^{-1} \Psi_y^{\epsilon} (V_{\varepsilon}^0)^{-1} G_0 \right).
\end{equation}
Now, using Equation (\ref{app:FOC:eq1}), we get
$
V_y - \Sigma^* = \epsilon \left( \Psi_y^{\epsilon}  - F_0 (\Psi_F^{\epsilon})' - \Psi_F^{\epsilon} F_0' - \Psi_{V_{\varepsilon}}^{\epsilon}\right) + O(\epsilon^2)$ \break $
= \epsilon M_{F_0,V_{\varepsilon}^0} (\Psi_y^{\epsilon} - \Psi_{V_{\varepsilon}}^{\epsilon}) M_{F_0,V_{\varepsilon}^0}' + O(\epsilon^2) 
= \epsilon G_0  \Delta^* G_0' + O(\epsilon^2),
$
where $\Delta^* := G_0' (V_{\varepsilon}^0)^{-1} \Psi_y^{\epsilon} (V_{\varepsilon}^0)^{-1} G_0 
- G_0' (V_{\varepsilon}^0)^{-1} \Psi_{V_{\varepsilon}}^{\epsilon} (V_{\varepsilon}^0)^{-1} G_0$. Using that $vech( G_0' diag(a) G_0 ) =  \boldsymbol{X} a$, and Equation (\ref{app:psi:epsilonXXX}), the vectorized form of matrix $\Delta^*$ is:
$
vech(\Delta^*) = vech\left(   G_0' (V_{\varepsilon}^0)^{-1} \Psi_y^{\epsilon} (V_{\varepsilon}^0)^{-1} G_0 \right) 
$ $-  \boldsymbol{X}  (V_{\varepsilon}^0)^{-2} diag ( \Psi_{V_{\varepsilon}}^{\epsilon} ) 
= M_{\boldsymbol{X}} vech\left(  G_0' (V_{\varepsilon}^0)^{-1} \Psi_y^{\epsilon} (V_{\varepsilon}^0)^{-1} G_0 \right).
$
Thus, we have shown that, at first order in $\epsilon$, the difference between $V_y = V_y^0 + \epsilon \Psi_y^{\epsilon}$ and the FA $k$-factor approximation $\Sigma^*$ is $\epsilon G_0  \Delta^* G_0'$, with $vech(\Delta^*) = M_{\boldsymbol{X}}vech\left(  G_0' (V_{\varepsilon}^0)^{-1} \Psi_y^{\epsilon} (V_{\varepsilon}^0)^{-1} G_0 \right).$ It shows that the small perturbation $\epsilon \Psi_y^{\epsilon}$ around $V_y^0$ keeps the DGP within the $k$-factor specification (at first order) if, and only if, we have that vector $vech\left(  G_0' (V_{\varepsilon}^0)^{-1} \Psi_y^{\epsilon} (V_{\varepsilon}^0)^{-1} G_0 \right)$
 is spanned by the columns of $\boldsymbol{X}$. 

Consider $\Psi_y^{\epsilon} = H \xi \xi' H'$, where $H:=[ F_0 : G_0]$ and vector $\xi =( \xi_F',\xi_G')'$ are partitioned in $k$ and $T-k$ dimensional components, which corresponds to a local alternative with $(k+1)$th factor $H \xi$ and small loading  $\epsilon$  in the perturbation $\epsilon \Psi_y^{\epsilon}$. Then, we have $ G_0' (V_{\varepsilon}^0)^{-1} \Psi_y^{\epsilon} (V_{\varepsilon}^0)^{-1} G_0 = \xi_G \xi_G'$ since $F_0'(V_{\varepsilon}^0)^{-1} G_0 =0$ and $G_0'(V_{\varepsilon}^0)^{-1} G_0 = I_{T-k}$. Thus, $vech(\Delta^*)=M_{\boldsymbol{X}}vech\left(  \xi_G \xi_G' \right)$. Hence, it is only the component of $vech\left(  \xi_G \xi_G' \right)$ that is orthogonal to the range of $\boldsymbol{X}$, which generates a local deviation from a $k$-factor specification  through the multiplication by the projection matrix $M_{\boldsymbol{X}}$. It clarifies the role of the projector in the local power. On the contrary, the component spanned by the columns of $\boldsymbol{X}$ can be ``absorbed" in the $k$-factor specification by a redefinition of the factor $F$ and the variance $V_{\varepsilon}$ through $F^*$ and  $V_{\varepsilon}^*$. 

\subsection*{B.5 Feasible asymptotic normality of the FA estimators}

\subsubsection*{B.5.1 Asymptotic expansions}

We first establish the asymptotic expansion of $\hat{\theta}$ along the lines of pseudo maximum likelihood estimators (White (1982)). The sample criterion is $\hat{L}(\theta) $ given in Equation (\ref{Lhattheta}), where $\theta = \left( vec(F)',diag(V_{\varepsilon})' \right)'$ is subject to the nonlinear vector constraint 
$g(\theta) := \{ [F'V_{\varepsilon}^{-1} F]_{i,j} \}_{i<j} = 0$, i.e., matrix $F' V_{\varepsilon}^{-1} F$ is diagonal.  By standard methods for constrained M-estimators, we consider the FOC of the Lagrangian function: $\frac{\partial \hat{L}(\hat{\theta})}{\partial \theta} -  \frac{\partial g(\hat{\theta})'}{\partial \theta}\hat{\lambda}_L = 0$ and $g( \hat \theta ) = 0$, where $\hat \lambda_L$ is the $\frac{1}{2}k(k-1)$ dimensional vector of estimated Lagrange multipliers. Define vector $\tilde{\theta} := \left( vec(F_0)',diag(\tilde{V}_{\varepsilon})' \right)'$, which also satisfies the constraint $g(\tilde{\theta}) =0$ by the in-sample factor normalization. We apply the mean value theorem to the FOC around $\tilde \theta$ and get:
\begin{eqnarray}
\hat J (\bar\theta) \sqrt{n} ( \hat{\theta} - \tilde{\theta}) + A (\hat\theta) \sqrt{n} \hat{\lambda}_L &=& \sqrt{n} \frac{ \partial \hat{L}( \tilde{\theta} )}{\partial \theta},   \label{app:FOC1:L} \\
A (\bar \theta) '\sqrt{n} ( \hat{\theta} - \tilde{\theta}) &=& 0,  \label{app:FOC2:L}
\end{eqnarray}
where $\hat J (\theta)  := - \frac{ \partial^2 {\hat L}( {\theta} )}{\partial \theta \partial \theta'}$ is the $r \times r$ Hessian matrix, $A (\theta) := \frac{ \partial g( {\theta} )'}{\partial \theta}$ is the $r \times \frac{1}{2}k(k-1)$ dimensional  gradient matrix of the constraint function, and $\bar \theta$ is a mean value vector between $\hat \theta$ and $\tilde \theta$ componentwise. 
Matrix $A (\theta)$ is full rank for $\theta$ in a neighbourhood of $\theta_0$. For any $\theta$ define the $r \times (r-\frac{1}{2}k(k-1))$ matrix $B(\theta)$ with orthonormal columns that span the orthogonal complement of the range of $A(\theta)$. Matrix function $B(\theta)$ is continuous in $\theta$ in a neighbourhood of $\theta_0$.\footnote{Matrix $B(\theta)$ is uniquely defined up to rotation and sign changes in their columns. We can pick a unique representer such that matrix $B(\theta)$  is locally continuous, e.g., by taking $B(\theta) = \tilde B(\theta) [ \tilde B (\theta)' \tilde B(\theta )]^{-1/2}$, where matrix $\tilde B(\theta )$ consists of the first $r-\frac{1}{2}k(k-1)$ columns of $I_r - A(\theta)[A(\theta)'A(\theta)]^{-1} A(\theta)'$, if those columns are linearly independent.} Then, by multiplying Equation (\ref{app:FOC1:L}) times $B(\hat \theta)'$ to get rid of the Lagrange multiplier vector, using the identity $I_r = A(\theta) ( A(\theta)'A(\theta))^{-1} A(\theta)' + B(\theta) B(\theta)'$ for $\theta=\bar \theta$ and Equation (\ref{app:FOC2:L}), we get $[B(\hat \theta)' \hat J (\bar \theta) B(\bar \theta) ] B(\bar \theta)' \sqrt{n} ( \hat{\theta} - \tilde{\theta}) =  B(\hat\theta)' \sqrt{n}\frac{ \partial \hat{L}( \tilde{\theta} )}{\partial \theta}$. By the uniform convergence of $\hat J(\theta)$ to $J(\theta):= - \frac{ \partial^2 {L}_0( {\theta})}{\partial \theta \partial \theta'}$, and the consistency of the FA estimator $\hat\theta$ (Section B.2), matrix $B(\hat \theta)' \hat J (\bar \theta) B(\bar \theta)$ converges to $B_0'J_0 B_0$, where $J_0 := J(\theta_0)$ and $B_0 := B(\theta_0)$. Matrix $B_0' J_0 B_0$ is invertible under the local identification Assumption \ref{ass:A:5} (see Lemma \ref{App:lemma:Local:identification} condition d)). Then, $ B(\bar \theta)' \sqrt{n} ( \hat{\theta} - \tilde{\theta}) = [B(\hat \theta)' \hat J (\bar \theta) B(\bar \theta) ]^{-1} B(\hat\theta)' \sqrt{n}\frac{ \partial \hat{L}( \tilde{\theta} )}{\partial \theta}$ w.p.a.\ $1$. By using again $I_r = A(\bar \theta) ( A(\bar \theta)'A(\bar \theta))^{-1} A(\bar \theta)' + B(\bar \theta) B(\bar \theta)'$ and Equation (\ref{app:FOC2:L}), we get $\sqrt{n} ( \hat{\theta} - \tilde{\theta}) =  B(\bar \theta) [B(\hat \theta)' \hat J (\bar \theta) B(\bar \theta) ]^{-1} B(\hat\theta)' \sqrt{n}\frac{ \partial \hat{L}( \tilde{\theta} )}{\partial \theta}$. The distributional results established below imply $\sqrt{n}\frac{ \partial \hat{L}( \tilde{\theta} )}{\partial \theta} = O_p(1)$. Thus, we get $\sqrt{n}$-consistency:
\begin{equation}  \label{thetahatminusthetatilde}
\sqrt{n} ( \hat{\theta} - \tilde{\theta}) =  B_0 (B_0' J_0 B_0)^{-1}  B_0' \sqrt{n}\frac{ \partial \hat{L}( \tilde{\theta} )}{\partial \theta} + o_p(1).
\end{equation}

Let us now find the score $\frac{ \partial \hat{L}( {\theta} )}{\partial \theta}$. We have
$
\frac{ \partial \hat{L}( {\theta} )}{\partial \theta} = \left(  \frac{\partial vec( \Sigma (\theta))}{\partial \theta'} \right)' vec \left( \frac{ \partial \hat{L}(\theta)}{\partial \Sigma} \right),
$
where 
$
 vec \left( \frac{ \partial \hat{L}(\theta)}{\partial \Sigma} \right) $ $
= \frac{1}{2} \left(  \Sigma (\theta)^{-1} \otimes \Sigma (\theta)^{-1} \right) vec \left( \hat{V}_y - \Sigma (\theta) \right).
$
Moreover, by using $vec( \Sigma(\theta)) = \sum_{j=1}^k F_j \otimes F_j + [e_1 \otimes e_1~:~\cdots~:~ e_T \otimes e_T]diag( V_{\varepsilon})$, where $e_t$ is the $t$-th column of $I_T$, we get:
$
\frac{\partial vec( \Sigma (\theta))}{\partial \theta'} 
= $ \break $ \left[ (I_T \otimes F_1)+(F_1 \otimes I_T)~: ~\cdots~: ~(I_T \otimes F_k)+(F_k \otimes I_T)~:~ \right. $ $\left. e_1 \otimes e_1~:~\cdots~:~ e_T \otimes e_T \right].
$
Thus, we get:
$
\sqrt{n} \frac{ \partial \hat{L}( \tilde{\theta} )}{\partial \theta} = \frac{1}{2} \left(  \frac{\partial vec( \Sigma (\tilde{\theta}))}{\partial \theta'} \right)' \left(  \tilde{V}_y^{-1} \otimes  \tilde{V}_y^{-1} \right)  \sqrt{n} vec \left( \hat{V}_y -  \tilde{V}_y \right) $. From Equation (\ref{formula:Vhaty}) and Lemma \ref{lemma:conv:espilon:beta} we have $\hat V_y = \tilde V_y + \frac{1}{\sqrt n} ( Z_n + W_n F' + F W_n') + o_p(\frac{1}{\sqrt n})$, where $W_n := \frac{1}{\sqrt{n}} \varepsilon  \beta$. Thus, $\sqrt{n} \frac{ \partial \hat{L}( \tilde{\theta} )}{\partial \theta} =\frac{1}{2} \left(  \frac{\partial vec( \Sigma ({\theta}_0))}{\partial \theta'} \right)' \left(  {V}_y^{-1} \otimes  {V}_y^{-1} \right) vec \left( W_n F' + F W_n' + Z_n \right) + o_p(1)
$ and, from Equation (\ref{thetahatminusthetatilde}), we get:
\begin{equation} \label{asymptdistr:theta}
\sqrt{n} ( \hat{\theta} - \tilde{\theta}) = \frac{1}{2}  B_0 \left(   B_0'J_0 B_0\right)^{-1}B_0' \left(  \frac{\partial vec( \Sigma ({\theta}_0))}{\partial \theta'} \right)'  \left(  {V}_y^{-1} \otimes  {V}_y^{-1} \right)  vec \left( W_n F' + F W_n'+ Z_n \right) + o_p(1).
\end{equation}

\subsubsection*{B.5.2 Asymptotic normality}

In this subsection, we establish the asymptotic normality of estimators $\hat{F}$ and $\hat{V}_{\varepsilon}$. 
From Lemma \ref{CLT:Zn}, as $n\rightarrow \infty$ and $T$ is  fixed, we have the Gaussian distributional limit $Z_n \Rightarrow Z$ with $vech(Z) \sim N( 0, \Omega_Z)$, where the asymptotic variance $\Omega_Z$ is related to the asymptotic variance $\Omega$ of $\mathscr{Z}$ such that $Cov (Z_{ts},Z_{rp}) = \sqrt{V_{\varepsilon,tt}V_{\varepsilon,ss}
V_{\varepsilon,rr} V_{\varepsilon,pp}} Cov (\mathscr{Z}_{ts},\mathscr{Z}_{rp})$. Moreover, $Z_n^* \Rightarrow Z^* = G'V_{\varepsilon}^{-1} Z V_{\varepsilon}^{-1} G $ and $\bar{Z}_n:=Z_n-\mathcal{T}_{F,V_{\varepsilon}}(Z_n) \Rightarrow \bar{Z}$, where $\bar Z = Z - \mathcal{T}_{F,V_{\varepsilon}}(Z) = Z - V_{\varepsilon}^2 diag \left( (\boldsymbol{X}'\boldsymbol{X})^{-1} \boldsymbol{X}' vech(Z^*) \right)$ (see (\ref{asy:Psie:2})). The distributional limit of $W_n$ is given next.

\begin{lemma}  \label{lemma:asydistr:Wn} Under Assumptions \ref{Ass1}, \ref{Ass2} and \ref{ass:A:2}, \ref{ass:A:3}, \ref{ass:A:8}, as $n \rightarrow \infty$, 
(a) we have $W_n \Rightarrow \bar W$, where $vec(\bar W) \sim N(0,\Omega_W)$ with $\Omega_W = Q_{\beta} \otimes V_{\varepsilon}$, and (b) if additionally $E[w_{i,t} w_{i,r} w_{i,s}]= 0$, for all $t,r,s$ and $i$, then $Z$ and $\bar W$ are independent.
\end{lemma}

We get the following proposition from Lemmas \ref{CLT:Zn} and \ref{lemma:asydistr:Wn} (see proof at the end of the section).

\begin{proposition} \label{prop:FA:asymptdistr}
Under Assumptions \ref{Ass1}-\ref{Ass2} and \ref{ass:A:1}-\ref{ass:A:6}, \ref{ass:A:8}, as $n \rightarrow \infty$ and $T$ is fixed, for $ j=1,..,k$:
\begin{eqnarray}
&& \hspace{-0.75cm} \sqrt{n} diag ( \hat{V}_{\varepsilon} - \tilde{V}_{\varepsilon}) \Rightarrow V_{\varepsilon}^2( \boldsymbol{X}'\boldsymbol{X})^{-1} \boldsymbol{X}' vech( Z^*) ,  \label{asy:FA:estimators} \\
&& \hspace{-0.75cm} \sqrt{n} ( \hat{F}_j - F_j ) \Rightarrow R_j (\bar W F' + F \bar W' +  \bar{Z}) V_{\varepsilon}^{-1} F_j  +  \Lambda_j \{[ V_{\varepsilon} ( \boldsymbol{X}'\boldsymbol{X})^{-1} \boldsymbol{X}' vech( {Z}^*)] \odot  F_j\}, \qquad \quad \label{asy:FA:estimators2} \\
&& \hspace{-0.75cm} \sqrt n ( \hat F_j \hat{\mathcal{D}} - F_j ) \Rightarrow \frac{1}{\gamma_j} ( \bar W F' + F \bar W' + \bar Z) V_{\varepsilon}^{-1} F_j, \qquad \quad \label{asy:FA:estimators4}
\end{eqnarray}
where deterministic matrices $R_j$ and $\Lambda_j$ are defined in Proposition \ref{lemma:1}, and $\hat{\mathcal{D}} := \hat \Gamma (F' \hat V_{\varepsilon}^{-1} \hat F)^{-1}$ and $\hat \Gamma := diag ( \hat \gamma_1,...,\hat \gamma_k)$.
\end{proposition}
The joint asymptotic Gaussian distribution of the FA estimators involves the Gaussian matrices $Z^*$, $\bar{Z}$ and $\bar W$, the former two being symmetric. The asymptotic distribution of $\hat{V}_{\varepsilon}$ involves recentering around $\tilde{V}_{\varepsilon} = \frac{1}{n} \sum_{i=1}^n E[ \varepsilon_i \varepsilon_i']$, i.e., the finite-sample average cross-moments of errors, and not $V_{\varepsilon}$. For the asymptotic distribution of any functional that depends on $F$ up to one-to-one transformations of its columns, we can use the Gaussian law of (\ref{asy:FA:estimators4}) involving  $\bar W$ and $\bar{Z}$ only.

The asymptotic expansions (\ref{asy:FA:estimators})-(\ref{asy:FA:estimators2}) characterize explicitly the matrices $C_1(\theta)$ and $C_2(\theta)$ that appear in Theorem 2 in Anderson and Amemiya (1988). Their derivation is based on an asymptotic normality argument treating $\hat{\theta}$ as a M-estimator. However, neither the asymptotic variance nor a feasible CLT are given in Anderson and Amemiya (1988). We cannot use their results for our empirics.

To further compare our Proposition \ref{prop:FA:asymptdistr} with Theorem 2 in Anderson and Amemiya (1988), let $\bar{Z} = Z - \mathcal{T}_{F,V_{\varepsilon}}(Z) = \check{Z} - \mathcal{T}_{F,V_{\varepsilon}}(\check{Z})$, where $\check{Z} := Z - diag(Z)$ is the symmetric matrix of the off-diagonal elements of $Z$ with zeros on the diagonal.\footnote{Here,  $diag(Z)$ is the diagonal matrix with the same diagonal elements as $Z$.} Hence, the zero-mean Gaussian matrix $\bar{Z}$ only involves the off-diagonal elements of $Z$. Moreover, since $ V_{\varepsilon}^2 ( \boldsymbol{X}'\boldsymbol{X})^{-1} \boldsymbol{X}' vech( \Delta_n^*)  = V_{\varepsilon}^2 diag ( V_{\varepsilon}^{-1} \Delta_n V_{\varepsilon}^{-1} ) = diag (\Delta_n)$ for a diagonal matrix $\Delta_n$ and $\Delta_n^* := G' V_{\varepsilon}^{-1} \Delta_n V_{\varepsilon}^{-1} G$, we can write the asymptotic expansion of $\hat{V}_{\varepsilon}$ as $\sqrt{n} diag ( \hat{V}_{\varepsilon} - \tilde{V}_{\varepsilon}) =   V_{\varepsilon}^2( \boldsymbol{X}'\boldsymbol{X})^{-1} \boldsymbol{X}' vech( \check{Z}_n^*) 
+ diag( Z_n) + o_p(1)$, 
where $\check{Z}_n^* = G' V_{\varepsilon}^{-1} \check{Z}_n V_{\varepsilon}^{-1} G$ and $\check{Z}_n := Z_n - diag(Z_n)$. Thus, we get:
$
\sqrt{n} diag ( \hat{V}_{\varepsilon} - \tilde{V}_{\varepsilon}) \Rightarrow  V_{\varepsilon}^2( \boldsymbol{X}'\boldsymbol{X})^{-1} \boldsymbol{X}' vech( \check{Z}^*) 
+ diag( Z),
$
where $\check{Z}^* = G' V_{\varepsilon}^{-1} \check{Z} V_{\varepsilon}^{-1} G$. Hence, the asymptotic distribution of the FA estimators depends on the diagonal elements of $Z$ via term $diag(Z)$ in the asymptotic distribution of $\hat{V}_{\varepsilon}$. In Theorem 2 in Anderson and Amemiya (1988), this term does not appear because in their results the asymptotic distribution of $\hat{V}_{\varepsilon}$ is centered around $diag( \frac{1}{n}\varepsilon \varepsilon')$ instead of $\tilde{V}_{\varepsilon}$. Our recentering around $\tilde{V}_{\varepsilon}$ avoids a random bias term.

Finally, by applying the CLT to (\ref{asymptdistr:theta}), the asymptotic distribution of vector $\hat \theta$ is:
\begin{eqnarray}  
\sqrt{n} ( \hat{\theta} - \tilde{\theta})  \Rightarrow  \frac{1}{2}  B_0 \left( B_0'J_0 B_0\right)^{-1} B_0' \left(  \frac{\partial vec( \Sigma ({\theta}_0))}{\partial \theta'} \right)'  \left(  {V}_y^{-1} \otimes  {V}_y^{-1} \right) vec \left( \bar W F' + F \bar W'+ Z \right). \quad \label{asy:FA:estimators3}
\end{eqnarray}
The Gaussian asymptotic distribution in (\ref{asy:FA:estimators3}) matches those in (\ref{asy:FA:estimators}) and (\ref{asy:FA:estimators2}) written for the components, and its asymptotic variance yields the `sandwich formula''. The result in (\ref{asy:FA:estimators3}) is analogue to Theorem 2 in Anderson and Amemiya (1988), for different factor normalization and recentering of the variance estimator.

\noindent \textbf{Proof of Proposition  \ref{prop:FA:asymptdistr}:}
 From (\ref{asy:Psie:2}), we have the asymptotic expansion: $\sqrt{n} diag ( \hat{V}_{\varepsilon} - \tilde{V}_{\varepsilon}) = diag (\Psi_{\varepsilon}) + o_p(1) = V_{\varepsilon}^2( \boldsymbol{X}'\boldsymbol{X})^{-1} \boldsymbol{X}' vech( Z_n^*) + o_p(1).$ Moreover, from Proposition \ref{lemma:1} (a) and using $\Psi_y - \Psi_{\varepsilon} = W_n F' + F W_n' + \bar{Z}_n$, we have:
$
\sqrt{n} ( \hat{F}_j - F_j ) = R_j (\Psi_y - \Psi_{\varepsilon}) V_{\varepsilon}^{-1} F_j  + \Lambda_j \Psi_{\varepsilon} V_{\varepsilon}^{-1} F_j + o_p(1) 
=  R_j (W_n F' + F W_n' +  \bar{Z}_n) V_{\varepsilon}^{-1} F_j  + \Lambda_j [ diag( \Psi_{\varepsilon}) \odot (V_{\varepsilon}^{-1} F_j)] + o_p(1)
 = R_j (W_n F' + F W_n' +  \bar{Z}_n) V_{\varepsilon}^{-1} F_j  +  \Lambda_j \{ [ V_{\varepsilon} ( \boldsymbol{X}'\boldsymbol{X})^{-1} \boldsymbol{X}' vech( Z_n^*)] \odot  F_j \} + o_p(1).
$
Lemmas \ref{CLT:Zn} and \ref{lemma:asydistr:Wn} yield  (\ref{asy:FA:estimators})-(\ref{asy:FA:estimators2}), together with (\ref{asy:FA:estimators4}) from  (\ref{eq:exp:1}) since $\Psi_y - \Psi_{\varepsilon} \Rightarrow \bar W F' + F \bar W' + \bar Z$.

\subsubsection*{B.5.3 Feasible CLT for the FA estimators}

\noindent \textbf{i) Feasible CLT for $Z_n$ via a parametric estimator of the asymptotic variance} 

We first show that, under strengthening of Assumption \ref{Ass2}, we get a parametric structure for the variance $V[vech(Z)] = \Omega_Z(V_{\varepsilon},\vartheta)$ with a vector of unknown parameters $\vartheta$ of dimension $T+1$. 
\begin{assumption} \label{block:structure:reinforced}
The standardized errors processes $w_{i,t}$ in Assumption \ref{Ass2} are (a) stationary martingale difference sequences (mds), and (b) $E[ w_{i,t}^2  w_{i,r} w_{i,s}] =0$, for $t > r > s$.
\end{assumption}
Assumption \ref{block:structure:reinforced} holds e.g.\ for conditionally homoskedastic mds, and for ARCH processes (see below). The stationarity condition is used only to ensure that $Cov(w^2_{i,t}, w^2_{i,s})$ depends on the difference $t-s$ rather than on $t$ and $s$ individually. This assumption could be replaced by this weaker condition without affecting the results. This condition allows us  to significantly reduce the number of parameters to estimate, making estimation more tractable. Let $\mathscr{Z} :=  V_{\varepsilon}^{-1/2} Z V_{\varepsilon}^{-1/2}$.
Then, using Lemma \ref{CLT:Zn}, under Assumptions \ref{Ass2} and \ref{block:structure:reinforced}, we have $V[\mathscr{Z}_{t,t}] =    \psi(0) +2 \kappa$, $V[\mathscr{Z}_{t,s}]=   \psi (t-s)+q+\kappa$ and $Cov( \mathscr{Z}_{t,t}, \mathscr{Z}_{s,s}) =  \psi (t-s)$, where $\psi(t-s):= \underset{n\rightarrow\infty}{\lim}\frac{1}{n} \sum_{i} Cov( w_{i,t}^2, w_{i,s}^2) \sigma_{ii}^2$. Quantity $\psi (t-s)$ depends on the difference $t-s$ only, by stationarity. The other covariance terms between elements of $\mathscr{Z}$ vanish. Then, we have $\Omega = [\psi(0) - 2 q] D(0) + \sum_{h=1}^{T-1} \psi(h) D(h)  + (q +\kappa) I_{T(T+1)/2}$, where $D(0) = \sum_{t=1}^T vech(E_{t,t})vech(E_{t,t})'$ and $D(h)= \tilde D(h) + \bar D(h)$ with $\tilde D(h) = \sum_{t=1}^{T-h} [ vech(E_{t,t})vech(E_{t+h,t+h})'  + vech(E_{t+h,t+h})vech(E_{t,t})']$ and $\bar D( h) = \sum_{t=1}^{T-h}  vech(E_{t,t+h} + E_{t+h,t} )vech(E_{t,t+h} + E_{t+h,t})'$ for $h=1,...,T-1$,
and where $E_{t,s}$ denote the $T \times T$ matrix with entry $1$ in position $(t,s)$ and $0$ elsewhere. Hence, with $Z = V_{\varepsilon}^{1/2} \mathscr{Z}V_{\varepsilon}^{1/2}$, we get a parametrization $\Omega_Z(V_{\varepsilon},\vartheta)$ for $V[vech(Z)]$ with $\vartheta = (q+\kappa, \psi(0)-2q,\psi(1),...,\psi(T-1))'$.

Then, we obtain a parametric structure for $M_{\boldsymbol{X}}\Omega_{Z^*}M_{\boldsymbol{X}}=M_{\boldsymbol{X}}\boldsymbol{R}'\Omega_{Z}\boldsymbol{R}M_{\boldsymbol{X}}$.
\begin{lemma} \label{lemma:parametric:Omega:Z:bar:star}
Under Assumptions \ref{Ass1}-\ref{ass:A:6}, we have:
\begin{equation} \label{parametric:Omega:Z:bar:star}
M_{\boldsymbol{X}}\Omega_{Z^*}M_{\boldsymbol{X}} =  \sum_{h=1}^{T-1} [\psi(h) + q +\kappa] M_{\boldsymbol{X}}\boldsymbol{R}' \bar D(h) \boldsymbol{R} M_{\boldsymbol{X}}.
\end{equation}
\end{lemma}
Hence, the parametric structure $M_{\boldsymbol{X}}\Omega_{Z^*}M_{\boldsymbol{X}}( V_{\varepsilon},G,\tilde\vartheta)$ depends linearly on vector $\tilde \vartheta$ that stacks the $T-1$ parameters $\psi(h) + q +\kappa$, for $h=1,...,T-1$. It does not involve parameter $\psi(0)$, i.e., the quartic moment of errors, because the asymptotic expansion of the LR statistic does not involve the diagonal terms of $Z$. Moreover, the unknown parameters appear through the linear combinations $\psi(h) + q +\kappa$ that are the scaled variances of the out-of-diagonal elements of $Z$.
We can estimate the unknown parameters in $\tilde \vartheta$ by least squares applied on (\ref{parametric:Omega:Z:bar:star}), using the nonparametric estimator $M_{\hat{\boldsymbol{X}}}\hat{\Omega}_{Z^*}M_{\hat{\boldsymbol{X}}}$ defined in Proposition \ref{thm:asy:tests}, after half-vectorization and replacing $V_{\varepsilon}$ and $G$ by their FA estimates. It yields a consistent estimator of $M_{\boldsymbol{X}}\Omega_{Z^*}M_{\boldsymbol{X}}$ incorporating the restrictions implied by Assumption \ref{block:structure:reinforced}.

To get a feasible CLT for the FA estimates, we need to estimate the additional parameters $\psi(0)-2q$ and $q+\kappa$. We consider the matrix $\hat{\Omega}_{Z^*}$ from Proposition  \ref{thm:asy:tests}, that involves fourth-order moments of residuals. 

\begin{lemma} \label{lemma:pnTau}
Under Assumptions \ref{Ass1}-\ref{ass:A:6}, and $\sqrt{n} \sum_{m=1}^{J_n} B_{m,n}^2= o(1)$, up to pre- and post-multiplication by an orthogonal matrix and its transpose, we have $\hat{\Omega}_{Z^*} =\boldsymbol{R}'\tilde \Xi_n \boldsymbol{R} + o_p(1)$, where $\tilde \Xi_n = [\psi_n(0)-2q_n]D(0) +\sum_{h=1}^{T-1}\psi_n(h)D(h)+(q_n+\kappa_n)I_{T(T+1)/2}+(q_n+\xi_n)vech(I_T)vech(I_T)'$ and $\displaystyle \xi_n :=  \frac{1}{n} \sum_{m=1}^{J_n} \sum_{i\neq j \in I_m} \sigma_{ii}\sigma_{jj}$.
\end{lemma}

With blocks of equal size, the condition $\sqrt{n} \sum_{m=1}^{J_n} B_{m,n}^2= o(1)$ holds if $J_n = n^{\bar \alpha}$ and $\bar \alpha >1/2$. Now, we have the relation
$3D(0)+\sum_{h=1}^{T-1}D(h)-vech(I_T)vech(I_T)'=I_{T(T+1)/2}$, 
which implies
$3\boldsymbol{R}'D(0)\boldsymbol{R}+\sum_{h=1}^{T-1}\boldsymbol{R}'D(h)\boldsymbol{R}-vech(I_{T-k})vech(I_{T-k})'=I_{p}$. Hence, matrix
\begin{eqnarray} \label{system:Xi}
\boldsymbol{R}' \tilde \Xi_n\boldsymbol{R} &=&[\psi_n(0)+q_n+3\kappa_n]\boldsymbol{R}'D(0)\boldsymbol{R}+\sum_{h=1}^{T-1}[\psi_n(h)+q_n+\kappa_n]\boldsymbol{R}'D(h)\boldsymbol{R} \nonumber \\
&& +(\xi_n-\kappa_n)vech(I_{T-k})vech(I_{T-k})'
\end{eqnarray}
 depends on $T+1$ linear combinations of the elements of $\vartheta_n = (q_n+\kappa_n, \psi_n(0)-2q_n,\psi_n(1),...,\break \psi_n(T-1))'$ and $\xi_n - \kappa_n$. Thus, the linear system (\ref{system:Xi}) is rank-deficient to identify $\vartheta_n$. Moreover, in Assumption \ref{ass:A:3} (b), $\kappa_n$ is defined as a double sum over squared covariances scaled by $n$, and  is assumed to converge to a constant $\kappa$.  Such a convergence is difficult to assume for $\xi_n$ since $\xi_n$ is a double sum over products of two variances scaled by $n$.

We apply half-vectorization on (\ref{system:Xi}), replace the LHS by its consistent estimate $\hat \Xi$, and plug-in the FA estimates in the RHS. From Lemma \ref{lemma:pnTau}, least squares estimation on such a linear regression yields consistent estimates of linear combinations $\psi(0)+q+3\kappa$ and $\psi(h)+q+\kappa$ for $h=1,\dots,T-1$. Consistency of those parameters applies independently of $\xi_n-\kappa_n$ converging as $n\rightarrow \infty$, or not.\footnote{To see this, write the half-vectorization of the RHS of (\ref{system:Xi}) as $\chi \eta_n$, where $\chi$ is the $\frac{p(p+1)}{2} \times (T+1)$ matrix of regressors and $\eta_n$ the $(T+1)\times 1$ vector of unknown parameters. Then, $vech( \hat \Omega_{Z^*} ) = \hat \chi \eta_n + o_p(1)$, by Lemma \ref{lemma:pnTau}, the consistency of the FA estimates, and the last column of $\chi$ not depending on unknown parameters. Thus, $\hat \eta_n := ( \hat \chi' \hat \chi)^{-1} \hat \chi ' vech( \hat \Omega_{Z^*}) = \eta_n + o_p(1)$. In particular, we also have $\widehat{\xi_n-\kappa_n } = \xi_n-\kappa_n + o_p(1)$. } 
In order to identify the components of $\vartheta$, we need an additional condition. We use the assumption $\psi(T-1)=0$. That condition is implied by serial uncorrelation in the squared standardized errors after lag $T-1$, that is empirically relevant in our application with monthly returns data. Then, parameter $q+\kappa$ is estimated by $\widehat{\psi_n(T-1)+q_n+\kappa_n }$, and by difference we get the estimators of $\psi(0)-2q$ and $\psi(h)$, for $h=1,...,T-2$.

Let us now discuss the case of ARCH errors.
Suppose the $w_{i,t}$ follow independent ARCH(1) processes with Gaussian innovations that are independent across assets, i.e.,
$
w_{i,t} = h_{i,t}^{1/2} z_{i,t}, \  z_{i,t} \sim IIN(0,1), \ h_{i,t} = c_i + \alpha_i w_{i,t-1}^2
$
with $c_i = 1 - \alpha_i$. Then $E[w_{i,t}]=0$, $E[w_{i,t}^2]=1$, $\eta_i := V[w_{i,t}^2] = \frac{2}{1-3\alpha_i^2}$, $Cov ( w_{i,t}^2, w_{i,t-h}^2 ) = \eta_i \alpha_i^h$. Moreover, $E[ w_{i,t} w_{i,r} w_{i,s} w_{i,p}] =0$ if one index among $t,r,s,p$ is different from all the others. Indeed, without loss of generality, suppose $t$ is different from $s,p,r$. By the law of iterated expectation:
$
E[\varepsilon_{i,t} \varepsilon_{i,s} \varepsilon_{i,p} \varepsilon_{i,r} ]
= E[ E[\varepsilon_{i,t} \vert \{ z_{i,\tau}^2\}_{\tau=-\infty}^{\infty}, \{ z_{i,\tau} \}_{\tau \neq t}] \varepsilon_{i,s} \varepsilon_{i,p} \varepsilon_{i,r} ]
= E[ h_{i,t}^{1/2} E[ z_{i,t} \vert z_{i,t}^2] \varepsilon_{i,s} \varepsilon_{i,p} \varepsilon_{i,r} ] =0$. 
Then, Assumption \ref{block:structure:reinforced} holds. 
The explicit formula of $\Omega$ involves $\psi(h) = \underset{n\rightarrow\infty}{\lim} \frac{1}{n}\sum_{i=1}^n \frac{2\alpha_i^h}{1-3\alpha_i^2} \sigma_{ii}^2$, for $h =0,1,...,T-1$. Hence, setting $\psi(T-1)=0$ is a mild assumption for identification purpose  since $\alpha_i^{T-1}$ is small.
If $\alpha_i= 0$ for all $i$, i.e., no ARCH effects, we have $\psi(0)= 2q$ and $\psi(h) = 0$ for $h>0$, so that $\Omega=(q+\kappa) I_{\frac{T(T+1)}{2}}$.

\noindent \textbf{ii) Feasible CLT for $W_n$}

Let us now establish a feasible CLT for $W_n$.  
In order to estimate matrix $Q_{\beta}$ in the asymptotic variance $\Omega_{W}$ in Lemma \ref{lemma:asydistr:Wn}, we use the estimated betas and residuals, and combine them with a temporal sample splitting approach to cope with the EIV problem caused by the fixed $T$ setting. Specifically, let us split the time spell into two consecutive sub-intervals with $T_1$ and $T_2$ observations, with $T_1 + T_2 = T$ and such that $T_1 >k$ and $T_2 \geq k$. The factor model in the two sub-intervals reads $y_{1,i} = \mu_1 +  F_1 \beta_i +  \varepsilon_{1,i}$ and $y_{2,i} = \mu_2 +  F_2 \beta_i +  \varepsilon_{2,i}$, and let $V_{1,\varepsilon}$ and $V_{2,\varepsilon}$ denote the corresponding diagonal matrices of error average unconditional variances.\footnote{We can take the two sub-intervals as the halves of the time span. If this choice does not meet conditions $T_1 >k$ and $T_2 \geq k$ in a subperiod, we take the second sub-interval such that $T_2=k$, and add to the first sub-interval a sufficient number of dates from the preceeding subperiod in order to get $T_1 = k+1$.}	The conditions $T_1 >k$ and $T_2 \geq k$ are needed because we estimate residuals and betas in the first and the second sub-intervals, namely $\hat\varepsilon_{1,i} = M_{\hat F_1,\hat V_{1,\varepsilon}} ( y_{1,i} - \bar y_1)$ and $\hat \beta_i = ( \hat F_2' \hat V_{2,\varepsilon}^{-1} \hat F_2)^{-1}\hat F_2' \hat V_{2,\varepsilon}^{-1} ( y_{2,i} - \bar y_2)$. Here, $\hat F_{j}$ and $\hat V_{j,\varepsilon}$ for $j=1,2$ are deduced from the FA estimates in the full period of $T$ observations. 
Define $\hat\Psi_{\beta} = \frac{1}{n} \sum_m \sum_{i,j\in I_m} (\hat\beta_i \hat\beta_j') \otimes (\hat\varepsilon_{1,i} \hat\varepsilon_{1,j}')$. By using $\hat\varepsilon_{1,i} = ( M_{\hat F_1,\hat V_{1,\varepsilon}} F_1) \beta_i + M_{\hat F_1,\hat V_{1,\varepsilon}} (\varepsilon_{1,i} - \bar \varepsilon_{1})$, $M_{\hat F_1,\hat V_{1,\varepsilon}} F_1 = O_p(\frac{1}{\sqrt n})$ and $\frac{1}{n^2} \sum_m  b_{m,n}^2 = \sum_m  B_{m,n}^2 = o(1)$, we get $\hat\Psi_{\beta} = (I_k \otimes M_{\hat F_1,\hat V_{1,\varepsilon}}) \left( \frac{1}{n} \sum_m \sum_{i,j\in I_m} (\hat\beta_i \hat\beta_j') \otimes [(\varepsilon_{1,i} - \bar \varepsilon_{1}) (\varepsilon_{1,j} - \bar \varepsilon_{1})'] \right) (I_k \otimes M_{\hat F_1,\hat V_{1,\varepsilon}}') + o_p(1)$. Now, we use $\hat \beta_i = \left[ ( \hat F_2' \hat V_{2,\varepsilon}^{-1} \hat F_2)^{-1}\hat F_2' \hat V_{2,\varepsilon}^{-1}  F_2 \right] \beta_i  + ( \hat F_2' \hat V_{2,\varepsilon}^{-1} \hat F_2)^{-1}\hat F_2' \hat V_{2,\varepsilon}^{-1} (\varepsilon_{2,i} - \bar \varepsilon_2)$, and $\bar{\varepsilon}_1 = o_p(n^{-1/4})$, $\bar{\varepsilon}_2 = o_p(n^{-1/4})$ from Lemma \ref{lemma:conv:espilon:beta} (a), as well as the the mds condition in Assumption \ref{block:structure:reinforced}. We get $\hat\Psi_{\beta} =\hat\Psi_{\beta,1} + \hat\Psi_{\beta,2} + o_p(1)$,  where $\hat\Psi_{\beta,1} = (I_k \otimes M_{F_1,V_{1,\varepsilon}}) \left( \frac{1}{n} \sum_m \sum_{i,j\in I_m} (\beta_i \beta_j') \otimes (\varepsilon_{1,i} \varepsilon_{1,j}') \right) (I_k \otimes M_{ F_1, V_{1,\varepsilon}}')$ and $\hat\Psi_{\beta,2} = \left( [( F_2'V_{2,\varepsilon}^{-1} F_2)^{-1}F_2'V_{2,\varepsilon}^{-1}] \otimes M_{F_1,V_{1,\varepsilon}} \right) \left( \frac{1}{n} \sum_m \sum_{i,j\in I_m} (\varepsilon_{2,i} \varepsilon_{2,j}') \otimes (\varepsilon_{1,i} \varepsilon_{1,j}') \right) \break \left( [( F_2'V_{2,\varepsilon}^{-1} F_2)^{-1}F_2'V_{2,\varepsilon}^{-1}] \right.  \left.\otimes M_{F_1,V_{1,\varepsilon}} \right)'$. We use $\frac{1}{n} \sum_m \sum_{i,j\in I_m} (\beta_i \beta_j') \otimes (\varepsilon_{1,i} \varepsilon_{1,j}') = Q_{\beta} \otimes V_{1,\varepsilon} + o_p(1)$, and $\frac{1}{n} \sum_m \sum_{i,j\in I_m} (\varepsilon_{2,i} \varepsilon_{2,j}') \otimes (\varepsilon_{1,i} \varepsilon_{1,j}')  = \Omega_{21} + o_p(1)$, where $\Omega_{21}$ is the sub-block of matrix $\Omega_Z$ that is the asymptotic variance of $\frac{1}{\sqrt n} \sum_{i=1}^n \varepsilon_{2,i} \otimes \varepsilon_{1,i} \Rightarrow N(0,\Omega_{21})$. Then, $\hat\Psi_{\beta} = Q_{\beta} \otimes ( M_{F_1,V_{1,\varepsilon}} V_{1,\varepsilon}) + \left( [( F_2'V_{2,\varepsilon}^{-1} F_2)^{-1}F_2'V_{2,\varepsilon}^{-1}] \otimes M_{F_1,V_{1,\varepsilon}} \right)\Omega_{21} \left( [( F_2'V_{2,\varepsilon}^{-1} F_2)^{-1}F_2'V_{2,\varepsilon}^{-1}] \otimes M_{F_1,V_{1,\varepsilon}} \right)' + o_p(1)$. Thus, we get a consistent estimator of $Q_{\beta} \otimes (V_{1,\varepsilon}^{-1/2} M_{F_1,V_{1,\varepsilon}}V_{1,\varepsilon}^{1/2})$ by subtracting to $\hat \Psi_{\beta}$ a consistent estimator of the second term on the RHS (bias term),\footnote{Sample splitting makes the estimation of the bias easier, but we can avoid such a splitting  at the expense of a more complicated debiasing procedure.} and then by pre- and post-multiplying times $(I_k \otimes \hat V_{1,\varepsilon}^{-1/2})$. To get a consistent estimator of $Q_{\beta}$, we apply a linear transformation that amounts to computing the trace of the second term of a Kronecker product, and divide by $Tr(V_{1,\varepsilon}^{-1/2} M_{F_1,V_{1,\varepsilon}}V_{1,\varepsilon}^{1/2}) = T_1 - k$. Thus:
	$\hat Q_{\beta} = \frac{1}{n (T_1 - k)} \sum_m \sum_{i,j\in I_m} (\hat\beta_i \hat\beta_j') (\hat\varepsilon_{1,j}' \hat V_{1,\varepsilon}^{-1}\hat\varepsilon_{1,i})- \frac{1}{T_1 - k} \sum_{j=1}^{T_1} (I_k \otimes e_j') \left\{  \left( [( \hat F_2' \hat V_{2,\varepsilon}^{-1} \hat F_2)^{-1} \hat F_2'\hat V_{2,\varepsilon}^{-1}] \otimes \right.\right. \break \left.\left. [ \hat V_{1,\varepsilon}^{-1/2} M_{\hat F_1,\hat V_{1,\varepsilon}}] \right)\hat \Omega_{21} \left( [\hat V_{2,\varepsilon}^{-1} \hat F_2 ( \hat F_2'\hat V_{2,\varepsilon}^{-1} \hat F_2)^{-1} ] \otimes [ M_{\hat F_1,\hat V_{1,\varepsilon}}'  \hat V_{1,\varepsilon}^{-1/2}] \right) \right\}(I_k \otimes e_j)$,
where the $e_j$ are $T_1$-dimensional unit vectors, and $\hat \Omega_{21}$ is obtained from Subsection B.5.3 i). If estimate $\hat Q_{\beta}$ is not positive definite, we regularize it by deleting the negative eigenvalues.

\noindent \textbf{iii) Joint feasible CLT}

To get a feasible CLT for the FA estimators from (\ref{asy:FA:estimators})-(\ref{asy:FA:estimators2}), we need the joint distribution of the Gaussian matrix variates $Z$ and $W$. Under the condition of Lemma \ref{lemma:asydistr:Wn} (b), the estimates of the asymptotic variances of $vech(Z)$ and $vec(W)$ are enough, since these vectors are independent. Otherwise, to estimate the covariance $Cov( vech(Z), vec(W))$, we need to extend the approaches of the previous subsections.

\subsubsection*{B.5.4  Special cases}

In this subsection, we particularize the asymptotic distributions of the FA estimators for three special cases along the lines of Section \ref{discu}, plus a fourth special case that allows us to further discuss the link with Anderson and Amemiya (1988).

\noindent \textbf{i) Gaussian errors} 

When the errors admit a Gaussian distribution $\varepsilon_i\overset{ind}{\sim} N(0, \sigma_{ii} V_{\varepsilon})$ with diagonal $V_{\varepsilon}$, matrix $\frac{1}{\sqrt{q}} V_{\varepsilon}^{-1/2} Z V_{\varepsilon}^{-1/2}$ is in the GOE for dimension $T$, i.e., $\frac{1}{\sqrt{q}} vech( V_{\varepsilon}^{-1/2} Z V_{\varepsilon}^{-1/2} ) \sim N ( 0, I_{T(T+1)/2})$, where $q = \underset{n  \rightarrow \infty}{\lim} \frac{1}{n} \sum_i \sigma_{ii}^2$. Moreover, $vec(W) \sim N(0, Q_{\beta}  \otimes V_{\varepsilon})$, where $Q_{\beta}  = \underset{n  \rightarrow \infty}{\lim} \frac{1}{n} \sum_i \sigma_{ii} \beta_i \beta_i'$, mutually independent of $Z$ because of the symmetry of the Gaussian distribution.

\noindent \textbf{ii) Quasi GOE errors} 

As an extension of the previous case, here let us suppose that the errors meet Assumption \ref{Ass2}, the Conditions (a) and (b) in Proposition \ref{prop:gen:Gaussian} plus additionally (c) $\underset{n\rightarrow\infty}{\lim} \frac{1}{n} \sum_{i=1}^n V(\varepsilon_{i,t}^2) = \eta V_{\varepsilon,tt}^2$, for a constant $\eta > 0$, and (d) $\underset{n\rightarrow\infty}{\lim} \frac{1}{n} \sum_{i=1}^n E[ \varepsilon_{i,t}^2 \varepsilon_{i,r} \varepsilon_{i,p}] = 0$ for $r \neq p$. This setting allows e.g.\ for conditionally homoskedastic mds processes in the errors, but excludes ARCH effects. Then, the arguments in Lemma \ref{CLT:Zn} imply $vech(V_{\varepsilon}^{-1/2} Z V_{\varepsilon}^{-1/2}) \sim N(0,\Omega)$ with $\Omega = \left( \begin{array}{cc} (\eta/2+\kappa)I_T & 0 \\ 0 & (q+\kappa)I_{\frac{1}{2}T(T-1)} \end{array} \right)$. The distribution of $V_{\varepsilon}^{-1/2} Z V_{\varepsilon}^{-1/2}$ is similar to (scaled) GOE holding in the Gaussian case up to the variances of diagonal and of out-of-diagonal elements being different when $\eta \neq 2 q$. Hence, contrasting with test statistics, the asymptotic distributions of FA estimates differ in cases i) and ii) beyond scaling factors. It is because the asymptotic distributions of FA estimates involve diagonal elements of $Z$ as well.

\noindent \textbf{iii) Spherical errors} 

Let us consider the case $\varepsilon_i\overset{ind}{\sim} (0, \sigma_{ii} V_{\varepsilon})$ where $V_{\varepsilon} = \bar\sigma^2 I_T$, with independent components across time and the normalization $\underset{n  \rightarrow \infty}{\lim} \frac{1}{n} \sum_i \sigma_{ii}=1$. By repeating the arguments of Section B.3 for the constrained FA estimators (see Section 5.3), we get $Tr(M_F(\Psi_y-\Psi_\varepsilon)M_F)=0$ instead of equation (\ref{cond:Psie}). It yields the asymptotic expansions $\sqrt{n} ( \hat{\sigma}^2 - \tilde{\sigma}^2 ) = \frac{1}{T-k} Tr ( M_F Z_n ) + o_p(1) = \frac{\bar \sigma^2}{T-k} Tr ( Z_n^* ) + o_p(1)$, and 
$
\sqrt{n} ( \hat{F}_j - F_j ) = \frac{1}{\bar\sigma^2} R_j (\Psi_y - \Psi_{\varepsilon})  F_j  - \frac{1}{\bar \sigma^2} \Lambda_j \Psi_{\varepsilon}  F_j + o_p(1) 
=  \frac{1}{\bar \sigma^2}  R_j (W_n F' + F W_n' +  \bar{Z}_n) F_j  + o_p(1),
$
where we use $\Psi_y - \Psi_{\varepsilon} = W_n F' + F W_n' +  \bar{Z}_n$, $\Psi_{\varepsilon} = \frac{1}{T-k} Tr ( M_F Z_n ) I_T$ and $\Lambda_j F_j = 0$, and $\bar{Z}_n = Z_n - \frac{1}{T-k} Tr ( M_F Z_n ) I_T $. Moreover, by sphericity, we have $R_j = \frac{1}{2\gamma_j} P_{F_j} + 
  \frac{1}{\gamma_j}M_{F}  +
\sum_{\ell =1, \ell \neq j}^k  \frac{1}{\gamma_j - \gamma_{\ell}} P_{F_{\ell}}$. Thus, we get $\sqrt{n} ( \hat{\sigma}^2 - \tilde{\sigma}^2 )\Rightarrow \frac{\bar \sigma^2}{T-k} Tr ( Z^* )$ and $\sqrt{n} ( \hat{F}_j - F_j )  \Rightarrow \frac{1}{\bar \sigma^2}  R_j (W F' + F W' +  \bar{Z}) F_j $.\footnote{The asymptotic distribution of estimator $\hat{\sigma}^2$ coincides with that derived in FGS with perturbation theory methods. The asymptotic distribution of the factor estimates slightly differs from that given in FGS, Section 5.1, because of the different factor normalization adopted by FA compared to PCA even under sphericity.} The Gaussian matrix $Z$ is such that $Z_{tt} \sim N(0,\eta)$ and $Z_{t,s} \sim N(0,q)$ for $t\neq s$, mutually independent, where $\eta = \underset{n  \rightarrow \infty}{\lim} \frac{1}{n} \sum_i V[\varepsilon_{i,t}^2]$, and $vec(W) \sim N(0, Q_{\beta} \otimes I_T)$. Variables $Z$ and $W$ are independent if $E[\varepsilon_{i,t}^3]=0$. FGS, Section 4.3.1, explain how we can estimate $q$ and $\eta$ by solving a system of two linear equations based on estimated moments of $\hat \varepsilon_{i,t}$.

\noindent \textbf{iv) Cross-sectionally homoskedastic errors and link with Anderson and Amemiya (1988)}

Let us now make the link with the distributional results in Anderson and Amemiya (1988). In our setting, the analogous conditions as those in their Corollary 2 would be: (a) random effects for the loadings that are i.i.d.\ with $E[\beta_i] = 0$, $V[\beta_i]=I_k$, (b) error terms are i.i.d.\ $\varepsilon_i \sim (0 , V_{\varepsilon})$ with $V_{\varepsilon}= diag(V_{\varepsilon,11},...,V_{\varepsilon,TT})$ such that $E[ \varepsilon_{i,t} \varepsilon_{i,r} \varepsilon_{i,s} \varepsilon_{i,p} ] = V_{\varepsilon,tt} V_{\varepsilon,ss}$, for $t = r > s = p$, and $=0$, otherwise, and (c) $\beta_i$ and $\varepsilon_i$ are mutually independent. Thus, $\sigma_{ii}=1$ for all $i$, i.e., errors are cross-sectionally homoskedastic. Under the aforementioned Conditions (a)-(c), the Gaussian distributional limits $Z$ and $W$ are such that  $V[Z_{tt}] = \eta_t V_{\varepsilon,tt}^2$, for $\eta_t := V[ \varepsilon_{i,t}^2]/V_{\varepsilon,tt}^2$, $V[ Z_{ts} ] = V_{\varepsilon,tt} V_{\varepsilon,ss}$, for $t \neq s$, all covariances among different elements of $Z$ vanish, and $V[ vec(W) ] = I_k \otimes V_{\varepsilon}$. Equations (\ref{asy:FA:estimators})-(\ref{asy:FA:estimators2}) yield the asymptotic distributions of the FA estimates. In particular, they do not depend on the distribution of the $\beta_i$. Moreover, the distribution of the out-of-diagonal elements of $Z$ does not depend on the distribution of the errors, while, for the diagonal term,  we have $\eta_t = 2$ for Gaussian errors. As remarked in Section B.5.2, if the asymptotic distribution of estimator $\hat{V}_{\varepsilon}$ is centered around the realized matrix $\frac{1}{n} \sum_i \varepsilon_i \varepsilon_i'$ instead of its expected value, that distribution involves the out-of-diagonal elements of $Z$, and the elements of $W$. Hence, in that case, the asymptotic distribution of the FA estimates is the same independent of the errors being Gaussian or not, and depends on $F$ and ${V}_{\varepsilon}$ only, as found in Anderson and Amemiya (1988).


\subsection*{B.6 Orthogonal transformations and maximal invariant statistic}

In this subsection, we consider the transformation $\mathcal{O}$ that maps matrix $\hat G$ into $\hat G O$, where $O$ is an orthogonal matrix in $\mathbb{R}^{(T-k)\times (T-k)}$, and the transformation $\mathcal{O}_D$ that maps matrix $\hat{\boldsymbol{D}}$ into $ \hat{\boldsymbol{D}} O_D$, where $O_D$ is an orthogonal matrix in $\mathbb{R}^{df \times df}$. These transformations are induced from the freedom in chosing the orthonormal bases spanning the orthogonal complements of $\hat F$ and $\hat{\boldsymbol{X}}$. We show that they imply a group of orthogonal transformations on the vector $\hat{W} = \sqrt{n} \hat{\boldsymbol{D}}' vech(\hat{S}^*)$, with $\hat{S}^* = \hat{G}' \hat V_{\varepsilon}^{-1}  ( \hat V_y - \hat V_{\varepsilon})   \hat V_{\varepsilon}^{-1} \hat{G}$, and establish the maximal invariant.

Under the transformation $\mathcal{O}$, matrix $\hat{S}^*$ is mapped into $ O^{-1} \hat{S}^* O$. This transformation is mirrored by a linear mapping at the level of the half-vectorized form $vech(\hat{S}^*)$. In fact, this mapping is norm-preserving, since $\Vert vech(S) \Vert^2 =  \frac{1}{2} \Vert S \Vert^2$ and $\Vert O^{-1} S O \Vert = \Vert S \Vert$ for any conformable symmetric matrix $S$ and orthogonal matrix $O$. This mapping is characterized in the next lemma. 

\begin{lemma} \label{lemma:transformation:vech}
For any symmetric matrix $S$ and orthogonal matrix $O$ in $\mathbb{R}^{m \times m}$, we have \break $vech(O^{-1} S O) = \mathscr{R} (O) vech(S)$, where $\mathscr{R}(O) = \frac{1}{2} A_m' ( O' \otimes O' ) A_m$ is an orthogonal matrix, and $A_m$ is the duplication matrix defined in Appendix A. Transformations $\mathscr{R}(O)$ with orthogonal $O$ have the structure of a group: (a) $\mathscr{R}(I_m) = I_{\frac{1}{2} m(m+1)}$, (b) $\mathscr{R}(O_1) \mathscr{R}(O_2) = \mathscr{R}(O_2 O_1)$, and (c) $[\mathscr{R}(O)]^{-1} = \mathscr{R}(O^{-1})$.
\end{lemma}

With this lemma, we can give the transformation rules under $\mathcal{O}$ for a set of relevant statistics in the next proposition. We denote generically with $\widetilde{\cdot}$ a quantity computed with $\hat G O$ instead of $\hat{G}$. 

\begin{proposition} \label{prop:transformation:invariance} Under Assumptions \ref{Ass1} and \ref{ass:A:5},
(a) $vech(\widetilde{\hat{S}^*}) = \mathscr{R}(O) vech(\hat{S}^*)$, (b) $\widetilde{\boldsymbol{X}} =  \mathscr{R}(O) {\boldsymbol{X}} $, \break (c) $ I_p - \widetilde{\boldsymbol{X}} (\widetilde{\boldsymbol{X}}'\widetilde{\boldsymbol{X}})^{-1} \widetilde{\boldsymbol{X}}'  = \mathscr{R}(O) [ I_p -  \boldsymbol{X} (\boldsymbol{X}'\boldsymbol{X})^{-1} \boldsymbol{X}'] \mathscr{R}(O)^{-1}$, (d) $\widetilde{\boldsymbol{R}} = \boldsymbol{R} \mathscr{R}(O)^{-1}$, \break (e) $\widetilde{\boldsymbol{R}} (I_p - \widetilde{\boldsymbol{X}} (\widetilde{\boldsymbol{X}}'\widetilde{\boldsymbol{X}})^{-1} \widetilde{\boldsymbol{X}}') 
= {\boldsymbol{R}} (I_p - {\boldsymbol{X}} ({\boldsymbol{X}}'{\boldsymbol{X}})^{-1} {\boldsymbol{X}}') \mathscr{R}(O)^{-1}$. 
\end{proposition}

From Proposition \ref{prop:transformation:invariance} (c), under transformation $\mathcal{O}$, matrix $\hat{\boldsymbol{D}}$ is mapped into $\mathscr{R}(O)\hat{\boldsymbol{D}}$. Combining with transformation $\mathcal{O}_D$, we have $\widetilde{\hat{\boldsymbol{D}}} = \mathscr{R}(O)\hat{\boldsymbol{D}} O_D$. Thus, using Proposition \ref{prop:transformation:invariance} (a), under $\mathcal{O}$ and $\mathcal{O}_D$, vector $\hat W$ is mapped into $\widetilde{\hat W} = \sqrt{n} \widetilde{\hat{\boldsymbol{D}}}'vech(\widetilde{\hat S^*})=  O_D' \hat{W}_D$. Thus, statistic $\hat{W}$ is invariant under $\mathcal{O}$, while $\mathcal{O}_D$ operates as the group of orthogonal transformations. The maximal invariant under this group of transformations is the squared norm $\Vert \hat W \Vert^2=\hat W ' \hat W$. 

\noindent \textbf{Proof of Proposition \ref{prop:transformation:invariance}:} With $\widetilde{\hat{S}^*} = O^{-1} {\hat{S}^*} O$, part (a) follows from Lemma \ref{lemma:transformation:vech}. Let $\tilde{G} = G O$. Then, for any diagonal matrix $\Delta$, on the one hand, we have $vech(\tilde G' \Delta \tilde G) = \tilde{\boldsymbol{X}}diag (\Delta)$, and on the other hand, we have $vech(\tilde G' \Delta \tilde G) = vech( O^{-1} G' \Delta G O) = \mathscr{R}(O) vech( G' \Delta G ) = \mathscr{R}(O){\boldsymbol{X}}diag (\Delta)$. By equating the two expressions for any diagonal matrix $\Delta$, part (b) follows. Statement (c) is a consequence thereof and $\mathscr{R}(O)$ being orthogonal. Moreover, with $\tilde Q = Q O$ and using $vech(\tilde Q ' Z \tilde Q) = vech( O^{-1} Q' Z Q O) = \mathscr{R}(O) \boldsymbol{R}' vech(Z)$, we deduce part (d). Statement (e) is a consequence of (c) and (d).

\subsection*{B.7 Proofs of Lemmas \ref{App:lemma:Global:identification}-\ref{lemma:transformation:vech}}

\noindent \textbf{Proof of Lemma  \ref{App:lemma:Global:identification}:} The equivalence of conditions (a) and (b) is a consequence of the fact that function $\mathscr{L}(A) = - \frac{1}{2} \log \vert A \vert - \frac{1}{2} Tr ( V_y^0 A^{-1})$, where $A$ is a p.d.\ matrix, is uniquely maximized for $A = V_y^0$ (see Magnus and Neudecker (2007), p.\ 410),  and $L_0(\theta)=\mathscr{L}(\Sigma(\theta))$.


\noindent \textbf{Proof of Lemma  \ref{lemma:conv:espilon:beta}:} (a) From Assumption \ref{Ass2}, we have $E[\bar \varepsilon] = 0$ and $V[ \bar \varepsilon]=  V \left[\frac{1}{n} \sum_{i,k=1}^n s_{i,k} V_{\varepsilon}^{1/2} w_k \right]\break= V_{\varepsilon}^{1/2} \frac{1}{n^2} \sum_{i,j,k,l=1}^ns_{i,k}s_{j,l} E[ w_k w_l'] V_{\varepsilon}^{1/2} =  (\frac{1}{n^2} \sum_{i,j}^n \sigma_{i,j}) V_{\varepsilon}$ where the $s_{i,k}$ are the elements of $\Sigma^{1/2}$. Now,  $\frac{1}{n^2} \sum_{i,j=1}^n \sigma_{i,j} \leq C \frac{1}{n^2} \sum_{m=1}^{J_n} b_{m,n}^{1+\delta} = O( n^{\delta-1} \sum_{m=1}^{J_n} B_{m,n}^{1+\delta} )  = O ( n^{\delta-1} J_n^{1/2} (\sum_{m=1}^{J_n} B_{m,n}^{2(1+\delta)})^{1/2} ) \break= o( n^{-1} J_n^{1/2} )=o( n^{-1/2})$ from the Cauchy-Schwarz inequality and Assumptions \ref{Ass2} (c) and  (d). Part (a) follows. To prove part (b), we use $E[\frac{1}{n} \varepsilon \varepsilon'] \rightarrow V_{\varepsilon}^0$ and $V[ vech( (V_{\varepsilon}^0)^{-1/2}( \frac{1}{n} \varepsilon \varepsilon')(V_{\varepsilon}^0)^{-1/2}) ] = \frac{1}{n} \Omega_n$ from the proof of Lemma \ref{CLT:Zn}, and $\frac{1}{n} \Omega_n = o(1)$ by Assumption \ref{ass:A:3}. Finally, to show part (c), write $\frac{1}{n} \sum_{i=1}^n \varepsilon_i \beta_i' = (V_{\varepsilon}^0)^{1/2}  \frac{1}{n} \sum_{i,j=1}^n s_{i,j} w_j  \beta_i'$. Then, $E[\frac{1}{n} \sum_{i=1}^n \varepsilon_i \beta_i' ]=0$ while the variance of $vec( \frac{1}{n} \sum_{i=1}^n \varepsilon_i \beta_i')$ vanishes asymptotically since $V[ vec ( \frac{1}{n} \sum_{i,j=1}^n s_{i,j} w_j \beta_i' )] =\frac{1}{n^2} \sum_{i,j,m,l=1}^n s_{i,j} s_{m,l} \break (\beta_i \beta_l') \otimes E[w_j w_m'] = \frac{1}{n^2} \sum_{i,l=1}^n \sigma_{i,l} (\beta_i \beta_l') \otimes I_T = o(1)$ under Assumptions \ref{Ass2} and \ref{ass:A:2}.

\noindent \textbf{Proof of Lemma  \ref{App:lemma:Local:identification}:} From the arguments in the proof of Proposition \ref{lemma:1} with $\Psi_y = 0$, the solution of the FOC is such that $\Psi_{F,j}^{\epsilon} = (\Lambda_j^0 - R_j^0) \Psi_{V_{\varepsilon}}^{\epsilon} (V_{\varepsilon}^0)^{-1}F_j$ for $j=1,...,k$, and $diag( M_{F_0,V_{\varepsilon}^0} \Psi_{V_{\varepsilon}}^{\epsilon} M_{F_0,V_{\varepsilon}^0}')\break =0$. Since $\Psi_{V_{\varepsilon}}^{\epsilon}$ is diagonal,  the latter equation yields $M_{F_0,V_{\varepsilon}^0}^{\odot 2} diag (\Psi_{V_{\varepsilon}}^{\epsilon} ) =  0$. Under condition (a) of Lemma \ref{App:lemma:Local:identification}, we get $\Psi_{V_{\varepsilon}}^{\epsilon}=0$, which in turn implies $\Psi_F^{\epsilon} = 0$. Thus, condition (a) is sufficient for local identification. It is also necessary to get uniqueness of the solution $\Psi_{V_{\varepsilon}}^{\epsilon}=0$.  Moreover, conditions (a) and (b) of Lemma \ref{App:lemma:Local:identification} are equivalent as shown in Appendix B.3.  Further, conditions (a) and (c) are equivalent since $\Phi^{\odot 2} = M_{F_0,V_{\varepsilon}^0}^{\odot 2} (V_{\varepsilon}^0)^2$. Finally, let us show that condition (d) of Lemma \ref{App:lemma:Local:identification} is both sufficient and necessary for local identification. The FOC for the Lagrangian problem are $\frac{\partial L_0(\theta)}{\partial \theta} - \frac{\partial g(\theta)'}{\partial  \theta} \lambda_L =0$ and $g(\theta)=0$, where $\lambda_L$ is the Lagrange multiplier vector. By expanding at first-order around $\theta_0$ and $\lambda_0=0$, we get $H_0 
\left(
\begin{array}{cc} \theta - \theta_0 \\ \lambda \end{array} 
\right) = 0$, where $H_0 := \left(
\begin{array}{cc} J_0 &  A_0 \\ A_0' & 0 \end{array} 
\right)$, with
 $A_0 = \frac{\partial g(\theta_0)'}{\partial  \theta}$, is the bordered Hessian. The parameters are locally identified if, and only if,  $H_0$ is invertible. The latter condition is equivalent to $B_0'J_0 B_0$ being invertible.\footnote{Indeed, we can show $\vert H_0 \vert = (-1)^{\frac{1}{2}k(k-1)} \vert A_0'A_0 \vert \vert B_0'J_0 B_0 \vert$ by using $J_0 A_0 =0$, where the latter equality follows since the criterion is invariant to rotations of the latent factors.}

\noindent \textbf{Proof of Lemma \ref{lemma:asydistr:Wn}:} By Assumption \ref{Ass2},  $vec(W_n) = (I_k \otimes V_{\varepsilon}^{1/2}) \frac{1}{\sqrt{n}} \sum_{m=1}^{J_n} x_{m,n}$ where the $x_{m,n} := \sum_{i,j \in I_m} s_{i,j} (\beta_i \otimes w_j)$ are independent across $m$. Now, we apply the Liapunov CLT to show $\frac{1}{\sqrt{n}} \sum_{m=1}^{J_n} x_{m,n} \Rightarrow N(0,Q_{\beta} \otimes I_T)$. We have $E[ x_{m,n}]=0$ and $E[ x_{m,n} x_{m,n}'] = \left( \sum_{i,j \in I_m} \sigma_{i,j} \beta_i \beta_j' \right) \otimes I_T$ and, by Assumption \ref{ass:A:8}, $\Omega_{W,n} := \frac{1}{n} \sum_{m=1}^{J_n} E[ x_{m,n} x_{m,n}']$ converges to the positive definite matrix $Q_{\beta} \otimes I_T$. Let us now check the multivariate Liapunov condition $\Vert \Omega_{W,n}^{-1/2} \Vert^4 \frac{1}{n^2} \sum_{m=1}^{J_n} E[ \Vert x_{m,n} \Vert^4] = o(1)$. Since $\Vert \Omega_{W,n}^{-1/2} \Vert = O_p(1)$, it suffices to prove $ \frac{1}{n^2} \sum_{m=1}^{J_n} E[ (x_{m,n}^{p,t})^4] =o(1)$, for any $p=1,...,k$ and $t=1,...,T$, where $x_{m,n}^{p,t} := \sum_{i,j \in I_m} s_{i,j} \beta_{i,p} w_{j,t}$. For this purpose, Assumptions \ref{ass:A:1} and \ref{ass:A:2} yield $E[ (x_{m,n}^{p,t})^4] \leq C (\sum_{i,j \in I_m} \sigma_{i,j} )^2$. Then, we get $ \frac{1}{n^2} \sum_{m=1}^{J_n} E[ (x_{m,n}^{p,t})^4] \leq C \frac{1}{n^2} \sum_{m=1}^{J_n} b_{m,n}^{2(1+\delta)} \leq C n^{2\delta}  \sum_{m=1}^{J_n} B_{m,n}^{2(1+\delta)} =o(1)$ by Assumptions \ref{Ass2} (c) and (d). Part (a) of Lemma \ref{lemma:asydistr:Wn} follows. Moreover, $E[vech(\zeta_{m,n}) x_{m,n}'] =0$ and the proof of Lemma \ref{CLT:Zn} imply part (b).

\noindent \textbf{Proof of Lemma \ref{lemma:parametric:Omega:Z:bar:star}:} From the proof of Proposition \ref{thm:asy:tests} we have $M_{\boldsymbol{X}} \Omega_{Z^*} M_{\boldsymbol{X}} = M_{\boldsymbol{X}} \boldsymbol{R}' \Omega \boldsymbol{R} M_{\boldsymbol{X}}$, where  $\Omega = D + \kappa I_{T(T+1)/2}  = [\psi(0) - 2 q] D(0) + \sum_{h=1}^{T-1} \psi(h) [\tilde D(h) + \bar D (h)] + (q +\kappa) I_{T(T+1)/2}$. Then, since the columns of $\boldsymbol{R}$ are orthonormal, we get $M_{\boldsymbol{X}} \Omega_{Z^*} M_{\boldsymbol{X}} = [\psi(0) - 2 q] M_{\boldsymbol{X}} \boldsymbol{R}' D(0) \boldsymbol{R} M_{\boldsymbol{X}} + \sum_{h=1}^{T-1} \psi(h) M_{\boldsymbol{X}} \boldsymbol{R}' \tilde D(h)  \boldsymbol{R} M_{\boldsymbol{X}} + \sum_{h=1}^{T-1} \psi(h) M_{\boldsymbol{X}} \boldsymbol{R}' \bar D (h) \boldsymbol{R} M_{\boldsymbol{X}} + (q +\kappa) M_{\boldsymbol{X}}$. Now, we show that the the first two terms in this sum are nil. We have $G' E_{t,t} G  = Q' V_{\varepsilon}^{1/2} E_{t,t}  V_{\varepsilon}^{1/2} Q= V_{\varepsilon,tt} Q' E_{t,t} Q$ and thus $vech(G' E_{t,t} G ) = V_{\varepsilon,tt} vech( Q' E_{t,t} Q ) = V_{\varepsilon,tt}\boldsymbol{R}' vech( E_{t,t} )$ (see the proof of Proposition \ref{thm:asy:tests}). Hence, the kernel of matrix $M_{\boldsymbol{X}}$ is spanned by vectors $\boldsymbol{R}' vech( E_{t,t} )$, for $t=1,...,T$. We deduce that $M_{\boldsymbol{X}} \boldsymbol{R}' D(0) = 0$ and $ M_{\boldsymbol{X}} \boldsymbol{R}' \tilde D(h) \boldsymbol{R} M_{\boldsymbol{X}} =  0$.
Furthermore, from $I_{T(T+1)/2} = 2 \sum_{t=1}^T vech( E_{t,t}) vech( E_{t,t})' + \sum_{t< s} vech( E_{t,s} + E_{s,t}) vech( E_{t,s} + E_{s,t})' = 2 D(0) + \sum_{h=1}^{T-1} \bar D ( h)$, we get $M_{\boldsymbol{X}} = M_{\boldsymbol{X}} \boldsymbol{R}' I_{T(T+1)/2} \boldsymbol{R} M_{\boldsymbol{X}} 
=   \sum_{h=1}^{T-1}  M_{\boldsymbol{X}} \boldsymbol{R}' \bar D ( h) \boldsymbol{R} M_{\boldsymbol{X}}$. The conclusion follows. 

\noindent \textbf{Proof of Lemma \ref{lemma:pnTau}:} By the root-$n$ consistency of the FA estimators, $\hat{z}^*_{m,n} = z^*_{m,n} + O_p( \frac{b_{m,n}}{\sqrt{n}} )$, uniformly in $m$, where $z^*_{m,n}= \sum_{i\in I_m}G'V_{\varepsilon}^{-1} \varepsilon_i \varepsilon_i' V_{\varepsilon}^{-1} G= \sum_{i\in I_m} Q' e_i e'_i Q$. Under the condition $\frac{1}{n^{3/2}} \sum_{m=1}^{J_n} b_{m,n}^2 = \sqrt n \sum_{m=1}^{J_n} B_{m,n}^2 =o(1)$, we have $\hat{\Omega}_{Z^*}=\frac{1}{n}\sum_{m=1}^{J_n} E[vech(z^*_{m,n})vech(z^*_{m,n})']+o_p(1)$, up to pre- and post-multiplication by an orthogonal matrix. Moreover, $vech(z^*_{m,n}) = \boldsymbol{R}'[\sum_{i\in I_m} vech( e_i e_i')] = \frac{1}{2} \boldsymbol{R}' A_T'[\sum_{i\in I_m} ( e_i \otimes e_i)]$, and $\sum_{i\in I_m} ( e_i \otimes e_i) = \sum_{a,b} \sigma_{a,b} (w_a \otimes w_b)$. Thus, we get $
E \left[ vech(z^*_{m,n})vech(z^*_{m,n})'\right] = \frac{1}{4} \boldsymbol{R}' A_T'\left\{ \sum_{a,b,c,d \in I_m } \sigma_{a,b} \sigma_{c,d} E[ (w_a \otimes w_b) (w_c \otimes w_d)' ] \right\} A_T \boldsymbol{R}$. The non-zero contributions to the term in the curly brackets come from the combinations with $a=b=c=d$, $a=b\neq c=d$, $a=c\neq b=d$ and $a=d\neq b=c$, yielding:
$
 \sum_{a,b,c,d} \sigma_{a,b} \sigma_{c,d} E[ (w_a \otimes w_b) (w_c \otimes w_d)' ]   
= $ $\sum_a  \sigma_{a,a}^2 E[ (w_a w_a') \otimes (w_a w_a') ]   + ( \sum_{a \neq c}\sigma_{a,a} \sigma_{c,c} ) vec(I_T) vec(I_T)'  + ( \sum_{a \neq b}\sigma_{a,b}^2  ) ( I_{T^2} + K_{T,T}) = \sum_a  [\sigma_{a,a}^2 V( w_a \otimes w_a )]  
+ ( \sum_{a}\sigma_{a,a} )^2 vec(I_T) vec(I_T)'  + ( \sum_{a \neq b}\sigma_{a,b}^2  ) ( I_{T^2} + K_{T,T})$. Then, using $w_a \otimes w_a =  A_T vech( w_a w_a')$, we get $\frac{1}{4}  A_T'\left\{ \sum_{a,b,c,d \in I_m } \sigma_{a,b} \sigma_{c,d} E[ (w_a \otimes w_b) (w_c \otimes w_d)' ] \right\} A_T = \sum_a  [\sigma_{a,a}^2 V( vech( w_a w_a') )]  
+ ( \sum_{a}\sigma_{a,a} )^2 vech(I_T) vech(I_T)'  + ( \sum_{a \neq b}\sigma_{a,b}^2  ) I_{\frac{T(T+1)}{2}}$. Then, since \break $\frac{1}{n}\sum_{i=1}^n  \sigma_{i,i}^2 V[ vech( w_i w_i' ) ] = D_n$, where matrix $D_n$ is defined in Assumption \ref{ass:A:6} we get $\hat{\Omega}_{Z^*}= \boldsymbol{R}' \tilde \Xi_n \boldsymbol{R} +o_p(1)$, where $\tilde \Xi_n  = D_n + (q_n + \xi_n) vech(I_T) vech(I_T)' + \kappa_n I_{\frac{T(T+1)}{2}}$. Moreover, under Assumption \ref{block:structure:reinforced}, and singling out parameter $q_n$ along the diagonal, we have $D_n = [\psi_n(0) - 2 q_n] D(0) + \sum_{h=1}^{T-1} \psi_n(h) [\tilde D(h) + \bar D(h)]  + q_n I_{T(T+1)/2}$. The conclusion follows.

\noindent \textbf{Proof of Lemma \ref{lemma:transformation:vech}:}   We use  $vec(S) = A_m vech(S)$, where the $m^2 \times \frac{1}{2} m(m+1)$ matrix $A_m$ is such that: (i) $A_m'A_m = 2 I_{\frac{1}{2}m(m+1)}$, (ii) $K_{m,m} A_m = A_m$, where $K_{m,m}$ is the commutation matrix for order $m$, and (iii) $A_m A_m' = I_{m^2} + K_{m,m}$ (see the proof of Proposition \ref{thm:asy:tests} and also Theorem 12 in Magnus, Neudecker (2007) Chapter 2.8). Then, $vech(S) = \frac{1}{2} A_m' vec ( S)$ by property (i), and
$vech(O^{-1} S O) = \frac{1}{2} A_m' vec( O^{-1}  S O ) = \frac{1}{2} A_m ' ( O' \otimes O' ) vec(S) 
= \frac{1}{2} A_m' ( O' \otimes O' ) A_m vech(S),
$
for all symmetric matrix $S$. 
It follows $\mathscr{R}(O) = \frac{1}{2} A_m' ( O'\otimes O') A_m$. Moreover, by properties (i)-(iii), we have (a) $\mathscr{R}(I_m) = I_{\frac{1}{2}m(m+1)}$, (b) 
$\mathscr{R}(O_1) \mathscr{R}(O_2) = \frac{1}{4} A_m' ( O_1' \otimes O_1' ) A_m A_m' ( O_2' \otimes O_2' ) A_m  
= \frac{1}{4} A_m' ( O_1' \otimes O_1' ) (I_{m^2} + K_{m,m} ) ( O_2'\otimes O_2' ) A_m
= \frac{1}{4} A_m' ( O_1' O_2'\otimes O_1' O_2' ) (I_{m^2} + K_{m,m} )  A_m
= \frac{1}{2} A_m' [ (O_2 O_1)' \otimes (O_2 O_1)' ]  A_m  = \mathscr{R}(O_2 O_1),
$
and thus (c) $[\mathscr{R}(O)]^{-1} = \mathscr{R}(O^{-1})$. 



\section*{C  Numerical checks of conditions (\ref{cond:MLR}) of Proposition   \ref{thm:AUMPI}}
\setcounter{equation}{0}\def\theequation{F.\arabic{equation}}

In this section, we check numerically the validity of Inequalities (\ref{cond:MLR}) for given $df$, $\lambda_j$, $\nu_j$, and $m=3,....,M$, for a large bound $M$. The idea is to compute the frequency of the LHS of (\ref{cond:MLR}) becoming strictly negative over a large number of potential values of $\lambda_j$ and $\nu_j$, $j=1,...,df$, for any given $df >1$.\footnote{From Footnote \ref{footnote19}, we know that Inequalities (\ref{cond:MLR}) are automatically met with $df=1$. A given value of $df$ may result from several different combinations of $T$ and $k$, while a given $T$ implies different values of $df$ depending on $k$. For instance, $df=2$ applies with $(T,k)=(4,1)$, $(8,4)$, and $(13,8)$, among other combinations. For $T=20$, the tests for $k=1,2,...,14$ yield $df=170,151,133,116,100,85,71,58,46,35,25,16,8,1$, respectively.} Table \ref{numcheck}  provides those frequencies for $m = 3, ...,16$ (cumulatively), with $\lambda_j$ uniformly drawn in $[\underline \lambda, \bar \lambda]$ for  $j=1, ...,df$, and with $\nu_1=0$ \footnote{This normalization results from ranking the eigenvalues $\mu_j$, so that $\mu_1$ is the smallest one.} and $\nu_j$  uniformly drawn in $[0,\bar \nu]$, for  $j=2, ...,df$, and different combinations of bounds $\underline\lambda$, $\bar \lambda$, $\bar\nu$, and degrees of freedom $df=2,...,12$. Each frequency is computed from $10^8$ draws of $\lambda_j$ and $\nu_j$, $j=1,...,df$. In the SMC, we also report  a table of frequencies for large grids of equally-spaced points in $[\underline{\lambda},\bar \lambda]^{df}\times [0,\bar \nu]^{df-1}$, which corroborate the findings of this section.

\subsection*{C.1 Calibration of $\bar \nu$, $\underline{\lambda}$ and $\bar \lambda$}

To calibrate the bounds $\bar \nu$, $\underline{\lambda}$ and $\bar \lambda$  with realistic values, we run the following numerical experiment. For $T=20$ and $k=7$, we simulate $10,000$ draws from random $T\times k$ matrix $\tilde F$ such that  $vec(\tilde F) \sim N(0,I_{Tk})$ and set $F = V_{\varepsilon}^{1/2} U \Gamma^{1/2}$, $U = \tilde F  ( \tilde F' \tilde F)^{-1/2}$, $G = V_{\varepsilon}^{1/2} Q$, $Q = \tilde Q ( \tilde Q ' \tilde Q)^{-1/2}$, $\tilde{Q}$ are the first $T-k$ columns of $I_T - U U'$, for $V_{\varepsilon} = diag(V_{\varepsilon,11},...,V_{\varepsilon, TT})$, with 
$V_{\varepsilon,tt}=1.5$ for $t=1,...,10$, and $V_{\varepsilon,tt} = 0.5$ for $t=11,...,20$, and $\Gamma = T diag (4,3.5,3,2.5,2,1.5,1)$, $c_{k+1}= 10T$, and $\xi_{k+1}=e_1$. With these choices, the ``signal-to-noise" $\frac{1}{T} F_j'V_{\varepsilon}^{-1} F_j$ for the seven factors $j=1,...,7$ are $4,3.5,3,2.5,2,2.5,1$, and the ``signal-to-noise" for the weak factor is $\frac{1}{T} F_{k+1}'V_{\varepsilon}^{-1} F_{k+1} = 10 n^{-1/2}$. Moreover, the errors follow the ARCH model of Section B.5.3 (i) with ARCH parameters  either (a) $\alpha_i =0.2$ for all $i$, or (b) $\alpha_i =0.5$ for all $i$, and $q=4$, and $\kappa=0$ (cross-sectional independence). The choices $\alpha_i  = 0.2$, $0.5$ both meet the condition $3 \alpha_i^2 < 1$ ensuring the existence of fourth-order moments. Moreover, with $q-1 = 3$, we have a cross-sectional variance of the $\sigma_{ii}$ that is three times larger than the mean (normalized to $1$). For each draw, we compute the $df=71$ non-zero eigenvalues and associated eigenvectors of $\Omega_{\bar Z^*}$, and the values of parameters $\nu_j$ and $\lambda_j$. In our simulations (a) with $\alpha_i=0.2$, the draws of $\max_{j=1,...,df} \nu_j$ range between $0.21$ and $0.30$, with $95\%$ quantile equal to $0.28$, while the $5\%$ and $95\%$ quantiles of the $\lambda_j$ are $0.13$ and $7.65$. Instead, (b) with $\alpha_i=0.5$, the $\max_{j=1,...,df} \nu_j$ range between $0.70$ and $0.79$, with $95\%$ quantile equal to $0.77$, and the $5\%$ and $95\%$ quantiles of the $\lambda_j$ are $0.12$ and $6.64$. To get further insights in the choice of parameters $\bar \nu, \underline{\lambda}, \bar{\lambda}$, we also consider the values implied by the FA estimates in our empirical analysis. Here, when testing for the last retained $k$ in a given subperiod, the median across subperiods of $\max_{j=1,...,df} \nu_j$ is $0.76$, and smaller than about $0.90$ in most subperiods. Similarly, assuming $c_{k+1}= 10T$ and $\xi_{k+1}=e_1$ as above,  the median values of the smallest and the largest estimated $\lambda_j$ are $0.0024$ and $5.84$. Inspired by these findings, we set $\bar \lambda =7$, and consider $\bar \nu = 0.2$, $0.7$, $0.9$, $0.99$, and $\underline{\lambda}=0$, $0.1$, $0.5$, $1$, to get realistic settings with different degrees of dissimilarity from the case with serially uncorrelated squared errors (increasing with $\bar \nu$), and separation of the alternative hypothesis from the null hypothesis (increasing with $\underline{\lambda}$).


\subsection*{C.2 Results with Monte Carlo draws}

In Table \ref{numcheck}, the entries are nil for $\bar \nu$ sufficiently small and $\underline \lambda$ sufficiently large, suggesting that the AUMPI property holds for those cases that are closer to the setting with uncorrelated squared errors and sufficiently separated from the null hypothesis. Violations of Inequalities  (\ref{cond:MLR}) concern $df = 2, 3, 4, 5$.\footnote{A given number of simulated draws become increasingly sparse when considering larger values of $df$, which makes the exploration of the parameter space more challenging in those cases. However, unreported theoretical considerations show via an asymptotic approximation that the monotone likelihood property holds for $df \rightarrow \infty$ since the limiting distribution is then Gaussian. This finding resonates with the absence of violations in Table \ref{numcheck} for the larger values of $df$.} Let us focus on the setting with $\bar \nu =0.7$ and $\underline{\lambda}=0.1$. We find $3752$ violations of Inequalities  (\ref{cond:MLR}) out of $10^8$ simulations, all occurring for $df=2$, except $65$ for $df=3$. For those draws violating Inequalities  (\ref{cond:MLR}) for $df=2$, a closer inspection shows that (a) they feature values $\nu_2$ close to upper bound $\bar \nu=0.7$, and values of $\lambda_2$ close to lower bound $\underline{\lambda}=0.1$, and (b) several of them yield non-monotone density ratios $\frac{f(z;\lambda_1,\lambda_2)}{f(z;0,0)}$, with the non-monotonicity region corresponding to large values of $z$. As an illustration,  let us take the density ratio for $df=2$ with $\nu_2=0.666$, $\lambda_1 = 1.372$, and $\lambda_2=   0.130$. Here, the eigenvalues of the covariance matrix are $\mu_1=1$ (by normalization) and $\mu_2 =(1-\nu_2)^{-1}=2.994$, and the non-centrality parameter $\lambda_2$ is small. The quantiles of the asymptotic distribution under the null hypothesis for asymptotic size $\alpha = 20\%, 10\%, 5\%, 1\%, 0.1\%$ are $9.3, 12.8, 16.2, 24.5, 36.5$. Non-monotonicity applies for $z \geq 16$. The optimal rejection regions $\{ \frac{f(z;\lambda_1,\lambda_2)}{f(z;0,0)} \geq C \}$ correspond to those of the LR test $\{ z \geq \tilde{C} \}$, e.g., for asymptotic levels such as $\alpha=20\%$, but not for $\alpha=5\%$ or smaller. Indeed, in the latter cases, because of non-monotonicity of the density ratio, the optimal rejection regions are finite intervals in argument $z$. With $\bar \nu = 0.7$, we do not find violations with $\underline{\lambda}=0.5$ or larger. 

\begin{table}[H]
\begin{center}
	
	
	\begin{tabular}{||c | c||c|c|c|c|c| c| c| c| c| c| c| c|} 
		\hline
	& 	$df$ &  2 & 3 & 4 & 5 & 6 & 7 & 8 & 9 & 10 & 11 & 12\\ \hline\hline
$\bar \nu = 0.2$ & $\underline{\lambda}= 0$      & 0.002  & 0  & 0  & 0  & 0 &  0 & 0  & 0 & 0 & 0 & 0 \\ \hline
&  $\underline{\lambda} =0.1$       &  0.000 &  0& 0  & 0  & 0 & 0 &  0& 0& 0 & 0 & 0 \\ \hline	
&   $\underline{\lambda} =0.5$  & 0 &0  &0  & 0 & 0& 0&0 &0 &0 &0 & 0 \\ \hline
&   $\underline{\lambda} =1$       &0  &0  &0  &0  &0 & 0& 0& 0& 0& 0& 0 \\ \hline
\hline
$\bar \nu = 0.7$ &  $\underline{\lambda}= 0$       & 0.051 &  0.000 & 0.000 & 0 & 0 & 0 & 0 & 0 & 0 & 0 & 0\\ \hline
&$\underline{\lambda} =0.1$       &  0.037 & 0.000 & 0 & 0  & 0 & 0 & 0 & 0 & 0 & 0 & 0 \\ \hline
&$\underline{\lambda} =0.5$      &  0 &0  &0  & 0 & 0& 0&0 &0 &0 &0 & 0\\ \hline
&$\underline{\lambda} =1$       & 0 &0  &0  & 0 & 0& 0&0 &0 &0 &0 & 0 \\ \hline
\hline
$\bar \nu = 0.9 $ & $\underline{\lambda}= 0$       & 0.151 & 0.004  & 0.000  & 0.000  & 0 & 0 & 0 & 0 & 0 & 0 & 0 \\ \hline
&$\underline{\lambda} =0.1$	     & 0.134 &  0.004 & 0.000 & 0 & 0& 0 & 0 & 0 & 0 & 0 & 0\\ \hline
&$\underline{\lambda} =0.5$	     & 0.007 & 0.000 & 0 & 0 &0 & 0& 0& 0& 0& 0&0\\ \hline
&$\underline{\lambda} =1$	     & 0 &0  &0  & 0 & 0& 0&0 &0 &0 &0 & 0\\ \hline
\hline
 $\bar \nu = 0.99$ & $\underline{\lambda}= 0$      & 0.426  & 0.015  &  0.000  &  0.000  &  0  & 0  & 0  & 0  & 0  & 0  & 0  \\ \hline
&$\underline{\lambda} =0.1$        & 0.411  & 0.014   & 0.000   & 0.000   & 0   & 0  & 0  & 0  & 0  & 0  & 0 \\ \hline
&$\underline{\lambda} =0.5$       &  0.218   & 0.007   & 0.000   & 0   & 0   & 0  & 0  & 0  & 0  & 0  & 0\\ \hline
&$\underline{\lambda} =1$       &  0.078  & 0.001  & 0  & 0   & 0   & 0  &  0 &  0 &  0&  0 & 0 \\ \hline
\hline
			\end{tabular}
			\end{center}
	\caption{Numerical check of Inequalities (\ref{cond:MLR}) by Monte Carlo.  We display the cumulative frequency of violations in $\permil$ of Inequalities (\ref{cond:MLR}), for $m=3,...,16$, over $10^8$ random draws of the parameters $\lambda_j \sim Unif[\underline\lambda,\bar \lambda]$ and $\nu_j \sim Unif[0,\bar\nu]$, for $\bar \lambda =7$, and different combinations of bounds $\underline\lambda$, $\bar\nu$, and degrees of freedom $df$. An entry 0.000 corresponds to less than $100$ cases out of $10^8$ draws.} \label{numcheck} 
	
\end{table}




		


\section*{D Maximum value of $k$ as a function of $T$}

In Table \ref{maxk},  we report the maximal values for the number of latent factors $k$ to have $df \geq 0$, or $df>0$.

\begin{table}[h]
\begin{center}
	
	
	\begin{tabular}{||c||c|c|c|c|c| c| c| c| c| c| c| c|} 
		\hline
		$T$ & 1    & 2 & 3 & 4 & 5 & 6 & 7 & 8 & 9 & 10 & 11 & 12\\ \hline \hline
$df\geq 0$  &  0  & 0  & 1 & 1 & 2 & 3 & 3 & 4 & 5 & 6 & 6 & 7 \\ \hline
$df>0$  & NA  & 0 & 0 & 1 & 2 & 2 & 3 & 4 & 5 & 5 & 6 & 7 \\ \hline \hline
$T$ & 13 & 14& 15 & 16 & 17 & 18 & 19 & 20 & 21 & 22 & 23 & 24 \\ \hline \hline
$df\geq 0$ &  8  & 9 & 10 & 10& 11 & 12 & 13 & 14 & 15 & 15 & 16 & 17 \\ \hline
$df>0$   & 8 & 9 & 9 & 10 & 11 & 12 & 13 & 14 & 14 & 15 & 16 & 17\\ \hline
		   \hline
			\end{tabular}
			\end{center}
	\caption{Maximum value of $k$. We give the maximum admissible value $k$ of latent factors so that the order conditions $df \geq 0$ and $df > 0$ are met, with  $df =  \frac{1}{2}[(T-k)^2 - T - k]$, for different values of the sample size $T=1, ...,24$. Condition $df \geq 0$ is required for FA estimation, and condition $df >0$ is required for testing  the number of latent factors. 
	} \label{maxk} 
	
\end{table}

\end{document}